\documentclass[numberedappendix]{emulateapj}
\newcommand{\myemail}{toshiki.saito@nao.ac.jp}
\usepackage[breaklinks,colorlinks,urlcolor=blue,citecolor=blue,linkcolor=blue]{hyperref}
\defcitealias{ion13}{paper~I}

\shorttitle{Molecular gas  in VV~114}
\shortauthors{TOSHIKI SAITO et al.}

\begin{document}

\title{ALMA Multi-line Observations of the IR-bright Merger VV~114}

\author{Toshiki Saito\altaffilmark{1, 2}, Daisuke Iono\altaffilmark{2, 3}, Min S. Yun\altaffilmark{4}, Junko Ueda\altaffilmark{2}, Kouichiro Nakanishi\altaffilmark{2, 3, 5}, Hajime Sugai\altaffilmark{6}, Daniel Espada\altaffilmark{2, 5}, Masatoshi Imanishi\altaffilmark{2, 3, 7}, Kentaro Motohara\altaffilmark{8}, Yosiaki Hagiwara\altaffilmark{2}, Ken Tateuchi\altaffilmark{8}, Minju Lee\altaffilmark{1,2} and Ryohei Kawabe\altaffilmark{1, 2, 3}}
\email{\myemail}
\affil{$^1$Department of Astronomy, Graduate school of Science, The University of Tokyo, 7-3-1 Hongo, Bunkyo-ku, Tokyo 133-0033, Japan}
\affil{$^2$National Astronomical Observatory of Japan, 2-21-1 Osawa, Mitaka, Tokyo, 181-8588, Japan}
\affil{$^3$The Graduate University for Advanced Studies (SOKENDAI), 2-21-1 Osawa, Mitaka, Tokyo 181-0015, Japan}
\affil{$^4$Department of Astronomy, University of Massachusetts, Amherst, MA 01003, USA}
\affil{$^5$Joint ALMA Observatory, Alonso de C\'{o}rdova 3107, Vitacura, Casilla 19001, Santiago 19, Chile}
\affil{$^6$Kavli Institute for the Physics and Mathematics of the Universe (WPI), The University of Tokyo, 5-1-5 Kashiwanoha, Kashiwa, Chiba 277-8583, Japan}
\affil{$^7$Subaru Telescope, 650 North A'ohoku Place, Hilo, HI 96720, USA}
\affil{$^8$Institute of Astronomy, The University of Tokyo, 2-21-1 Osawa, Mitaka, Tokyo 181-0015, Japan}

\begin{abstract}

We present ALMA cycle 0 observations of the molecular gas and dust in the IR-bright mid-stage merger VV~114 obtained at 160 -- 800 pc resolution. The main aim of this study is to investigate the distribution and kinematics of the cold/warm gas and to quantify the spatial variation of the excitation conditions across the two merging disks. The data contain 10 molecular lines, including the first detection of extranuclear CH$_3$OH emission in interacting galaxies, as well as continuum emission. We map the $^{12}$CO~(3--2)/$^{12}$CO~(1--0) and the $^{12}$CO~(1--0)/$^{13}$CO~(1--0) line ratio at 800 pc resolution (in the units of K km s$^{-1}$), and find that these ratios vary from 0.2 -- 0.8 and 5 -- 50, respectively. Conversely, the 200 pc resolution HCN~(4--3)/HCO$^+$~(4--3) line ratio shows low values ($<$ 0.5) at a filament across the disks except for the unresolved eastern nucleus which is three times higher (1.34 $\pm$ 0.09). We conclude from our observations and a radiative transfer analysis that the molecular gas in the VV~114 system consists of five components with different physical and chemical conditions; i.e., 1) dust-enshrouded nuclear starbursts and/or AGN, 2) wide-spread star forming dense gas, 3) merger-induced shocked gas, 4) quiescent tenuous gas arms without star formation, 5) H$_2$ gas mass of (3.8 $\pm$ 0.7) $\times$ 10$^7$~M$_{\odot}$ (assuming a conversion factor of $\alpha_{\rm{CO}}$ = $0.8~\rm{M}_{\odot}~\rm{(K~km~s^{-1}~pc^2)^{-1}}$) at the tip of the southern tidal arm, as a potential site of tidal dwarf galaxy formation.

\end{abstract}


\section{INTRODUCTION}

Galaxy interactions and mergers play important roles in triggering star formation and/or fueling the nuclear activity in the merging host galaxies \citep{hop06}. Recent high resolution simulations of major mergers show that large scale tidal forces as well as small scale turbulence and stellar feedback can significantly influence the distribution of gas, forming massive clumps of dense gas with $M_{\rm{H_2}}$ = 10$^6$ -- 10$^8$~M$_{\odot}$ \citep[e.g.,][]{tey10, hop13}. These simulations also predict that the star formation not only increases as the galaxies first collide, but it also persists at a higher rate throughout the merger process, peaking at the final coalescence.

(Ultra-)Luminous Infrared Galaxies \citep[U/LIRGs;][]{soi87} at low redshifts are almost exclusively strongly interacting and merging systems \citep{kar10}, often found at the mid to final stages of the merger.  The elevated level of infrared luminosity originates from the reprocessed emission from the dust particles surrounding the starburst or the Active Galactic Nuclei (AGNs), both of which are likely triggered by the tidal interaction.  The highest gas surface densities ($\Sigma_{\rm{H_2}}$ = 5.4 $\times$ 10$^4$ -- 1.4 $\times$ 10$^5$~M$_{\odot}$~pc$^{-2}$) and consequently the highest star formation activities ($\Sigma_{\rm{SFR}}$ = $\sim$ 1000~M$_{\odot}$~yr$^{-1}$~kpc$^{-2}$) are usually found near the compact nuclear region \citep[e.g., Arp220, NGC 6240;][]{d&s98, eng10, wil14}. Dense molecular gas ($n$ $\sim$ 10$^5$ -- 10$^7$~cm$^{-3}$) in U/LIRGs directly shows nuclear gas distribution and kinematics \citep[e.g.,][]{ion04, sak14}. They are often surrounded by diffuse gas ($n$ $\sim$ 10$^2$ -- 10$^3$~cm$^{-3}$) that may or may not be directly associated with star formation activities.

It has been demonstrated that the HCN~(4--3) and HCO$^+$~(4--3) emission lines, whose critical densities are 8.5 $\times$ 10$^6$ and 1.8 $\times$ 10$^6$ cm$^{-3}$, respectively, can be used as tracers of the dense gas \citep[e.g.,][]{ion13, gar14, ima14}. On the other hand, CO~(1--0) and $^{13}$CO~(1--0) line emission, whose critical densities are 4.1 $\times$ 10$^2$ and 1.5 $\times$ 10$^3$ cm$^{-3}$, respectively, have been used extensively for tracing the global gas distribution and kinematics in merging U/LIRGs \citep[e.g.,][]{yun94, ion04, ued14}. In addition, the ratio of these lines (e.g. $^{12}$CO/$^{13}$CO and HCN/HCO$^+$) have been used to investigate the properties of the ISM \citep{cas92, aal97} or to search for buried AGNs \citep[e.g.,][]{ima07, ima14}. Limitations in sensitivity and angular resolution have been the major obstacles in understanding the detailed distribution and kinematics of both dense and diffuse gas, and investigating the spatial variation of the line ratios and the physical condition of gas.

\begin{figure}
\begin{center}
\includegraphics[scale=.35]{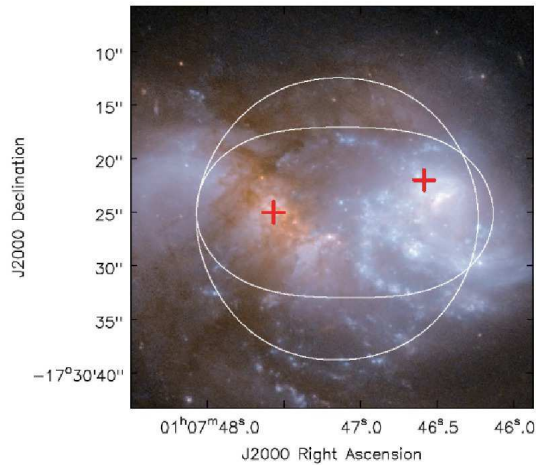}
\caption{The HST/ACS image of VV~114 \citep{eva08}. There is a dust lane from north to south in front of the eastern galaxy. The red crosses show the positions of the nuclei defined by the peak positions of the Ks-band observation \citep{tat12}. The white ellipse shows a field of view of 3-point mosaic observation with band 7, while the white circle shows a field of view of 7-point mosaic observation with band 7 (see \S\ref{obs}).
}
\label{fig_HST}
\end{center}
\end{figure}

In this paper, we present \textit{Atacama Large Millimeter/submillimeter Array} (ALMA) cycle 0 observations of the IR-bright merging galaxy VV~114. VV~114 is one of the best samples for studying the gas response during the critical stage when the two gas disks merge \citep{ion05, wil08}. The target molecular lines include $^{12}$CO~(1--0), $^{13}$CO~(1--0), $^{12}$CO~(3--2), HCN~(4--3) and HCO$^+$~(4--3), and we also present the maps of CH$_3$OH~(2$_k$--1$_k$), CS~(2--1), CN~(1$_{1/2}$--0$_{1/2}$), CN~(1$_{3/2}$--0$_{1/2}$), and CS~(7--6) lines which were observed simultaneously within the same band. The main aim of this study is to investigate the distribution and kinematics of the diffuse and dense molecular gas and to quantify the spatial variation of the excitation conditions across the two merging disks.

VV~114 is a gas-rich \citep[$M_{\rm{H_{2}}}$ = 5.1 $\times$ 10$^{10}$~M$_{\odot}$;][]{yun94} nearby (D = 82~Mpc; 1\farcs0 = 400~pc) interacting system (Figure~\ref{fig_HST}) with high-infrared luminosity \citep[$L_{\rm{IR}}$ = 4.7 $\times$ 10$^{11}$~$L_{\odot}$;][]{arm09}. The projected nuclear separation between the two optical galaxies (VV 114E and VV 114W) is about 6~kpc. \citet{fry99} found a large amount of dust ($M_{\rm{dust}}$ = 1.2 $\times$ 10$^8$~M$_{\odot}$) distributed across the system with a dust temperature of 20 -- 25~K. About half of the warmer dust traced in the mid-IR (MIR) is associated with the eastern galaxy, where both compact (nuclear region) and extended emission is found \citep{lfl02}. \citet{ric11} found a bimodal distribution of velocity dispersions of several atomic forbidden lines and emission line ratios indicative of composite activity explained by a combination of wide-spread shocks and star formation. The wide-spread star formation is also revealed by Pa$\alpha$ observation using ANIR camera mounted on miniTAO \citep[see also Appendix~\ref{A1}]{tat12}. \citet{ion13} (hereafter \citetalias{ion13}) identified a highly obscured AGN and compact starburst clumps using sub-arcsecond resolution ALMA cycle 0 observations of HCN~(4--3) and HCO$^+$~(4--3) emission.

This paper is organized as follows. We describe our observations and data reduction in \S\ref{obs}, and results in \S\ref{result}. In \S\ref{ratio} and \S\ref{der}, we provide molecular line ratios and physical parameters, such as the gas/dust mass, the gas temperature, and the gas density. In \S\ref{dis}, we present the properties of ``dense" gas (\S\ref{dense}), the comparison between molecular gas and star formation (\S\ref{SFR}), the discussions of the CO isotope enhancement (\S\ref{isotope}), the gas-to-dust mass ratio (\S\ref{G/D}), the fractional abundances of CS, CH$_3$OH, and CN relative to H$_2$ (\S\ref{chemical}), and a potential tidal dwarf galaxy formations at the tip of the tidal arms of VV~114 (\S\ref{TDG}). We summarize and conclude this paper in \S\ref{conclusion}. Throughout this paper, we adopt H$_0$ = 73 km s$^{-1}$ Mpc$^{-1}$, $\Omega_{\rm{M}}$ = 0.27, and $\Omega_{\rm{\Delta}}$ = 0.73.

\section{OBSERVATIONS AND DATA REDUCTION} \label{obs}

Observations toward VV~114 were carried out as an ALMA cycle 0 program (ID = 2011.0.00467.S; PI = D. Iono) using fourteen -- twenty 12~m antennas. The band~3 and band~7 receivers were tuned to the $^{12}$CO~(1--0), $^{13}$~CO~(1--0), $^{12}$CO~(3--2), HCN~(4--3), and HCO$^+$~(4--3) line emissions in the upper side band (see Table~\ref{table_obs}). The $^{12}$CO~(1--0) data were obtained on November 6, 2011 and May 4, 2012 in the compact and extended configurations, respectively. The $^{13}$CO~(1--0) data were obtained on May 27 and July 2, 2012 in the compact configuration. The $^{12}$CO~(3--2) emission was observed on November 5, 2011 in the compact configuration (7-point mosaic). The HCN~(4--3) and HCO$^+$~(4--3) data were obtained on July 1, 2, and 3, 2012 in the extended configuration (3-point mosaic), simultaneously. Each spectral window had a bandwidth of 1.875~GHz with 3840~channels, and two spectral windows were set to each sideband to achieve a total frequency coverage of $\sim$ 7.5~GHz in these observations. The spectral resolution was 0.488~MHz per channel. J1924-292, J0132-169, Uranus (Neptune for band 3 observations) were used for bandpass, phase, and flux calibrations. Detailed observational parameters are shown in Table~\ref{table_obs}.

\begin{figure*}[tbh]
\begin{center}
\includegraphics[scale=.5]{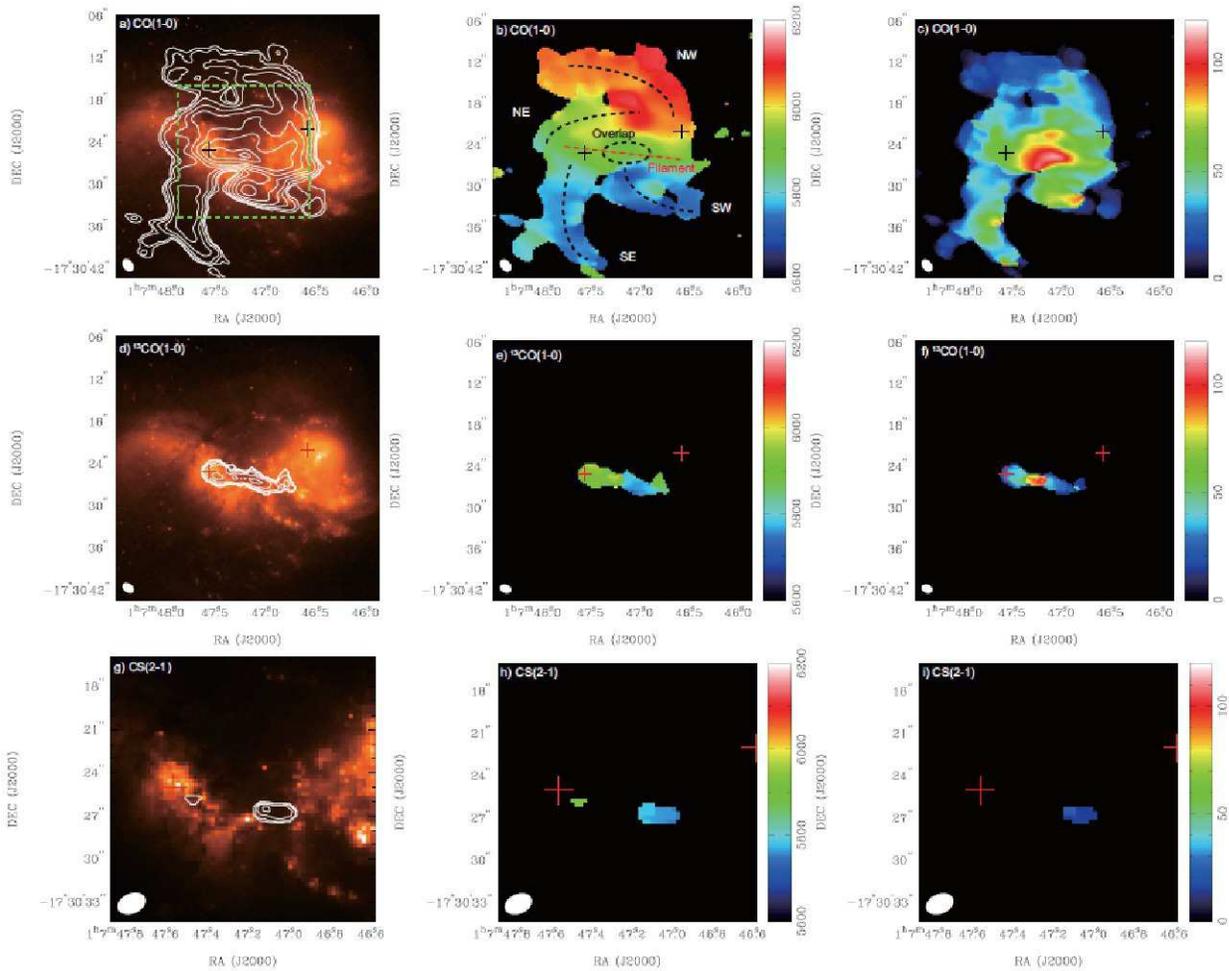}
\caption{(a) $^{12}$CO~(1--0) integrated intensity image overlaid on the HST/ACS/F435W image of VV~114. The contours are 0.2, 0.4, 0.8, 1.6, 3.2, 6.4, 12.8, 25.6, and 33.0~Jy~km~s$^{-1}$. The dashed green box shows an imaging field of other lines and continuum of this work except for the CO and $^{13}$CO lines. (b) $^{12}$CO~(1--0) velocity field image. The velocity field image in color scale ranges from 5600~km~s$^{-1}$ to 6200~km~s$^{-1}$. The dashed black lines represent tidal arms of VV~114. The dashed red line tracks the filamentary structure detected in images of other lines and dust continuum, and the dashed circle shows the overlap region. (c) $^{12}$CO~(1--0) velocity dispersion image. The velocity dispersion image in color scale ranges from 0~km~s$^{-1}$ to 120~km~s$^{-1}$. (d) The same as (a) but for $^{13}$CO~(1--0). The contours are 0.02, 0.04, 0.08, 0.16, 0.32, and 0.64~Jy~km~s$^{-1}$. (e/f) The same as (b/c), respectively, but for $^{13}$CO~(1--0). (g) The same as (a) but for CS~(2--1). The contours are 0.04, 0.08, 0.16, and 0.28~Jy~km~s$^{-1}$. (h/i) The same as (b/c), respectively, but for CS~(2--1). The beam size of each line is shown in the bottom-left of the images (Table~\ref{table_data}). The red crosses show the positions of the nuclei defined by the peak positions of the Ks-band observation \citep{tat12}.
}
\label{fig_mom1}
\end{center}
\end{figure*}

We used the delivered calibrated data and mapping was accomplished using the {\tt clean} task in {\tt CASA} \citep{CASA}. We made the data cubes with a velocity width of 5 km s$^{-1}$ for the $^{12}$CO line and 30 km s$^{-1}$ for the other lines. All maps in this paper, except for $^{12}$CO~(3--2), are reconstructed with a Briggs weighting \citep[robust = 0.5;][]{b&c92} and analyzed with {\tt MIRIAD} and {\tt AIPS}. The $^{12}$CO~(3--2) images are created with uniform weighting (see \S\ref{co32}). The synthesized beam size of the $^{12}$CO~(1--0), $^{13}$CO~(1--0), $^{12}$CO~(3--2), and HCN~(4--3) were 1\farcs97 $\times$ 1\farcs35 (P.A. = 82.3~deg.), 1\farcs77 $\times$ 1\farcs20 (P.A. = 85.8~deg.), 1\farcs64 $\times$ 1\farcs17 (P.A. = 112.6~deg.), and 0\farcs46 $\times$ 0\farcs38 (P.A. = 51.5~deg.), respectively. We also detected CN~(1$_{3/2}$--0$_{1/2}$), CN~(1$_{1/2}$--0$_{1/2}$), CS~(2--1), CH$_3$OH~(2$_k$-1$_k$), and CS~(7--6) line emission for the first time in VV~114. The properties of these molecular lines are summarized in Table~\ref{table_data}. All images which we constructed are corrected for primary beam attenuation. The on-source times of band~3 and band~7 were about 40 minutes and 80 minutes, and the rms noise levels of the channel maps with 30~km~s$^{-1}$ resolution are 1.0~mJy~beam$^{-1}$ and 0.8~mJy~beam$^{-1}$, respectively. Furthermore, we made continuum maps at each observing frequency by adding the line-free channels. The rms level of the continuum images were 0.05~mJy~beam$^{-1}$, 0.11~mJy~beam$^{-1}$, and 0.07~mJy~beam$^{-1}$ for band~3, band~7 in the compact configuration, and band~7 in the extended configuration, respectively. The continuum emission was subtracted in the $uv$-plane before making the line images. Throughout this paper, the pixel scales of the band~3 and the band~7 images are set to 0\farcs3/pixel and 0\farcs08/pixel, respectively, and only the statistical error is considered unless mentioned otherwise. The systematic error on the absolute flux is estimated to be $\sim$ 5\% and $\sim$ 10\% for both sidebands in band 3 and band 7, respectively.

In the following sections, we estimate the missing flux of each molecular line for which the single dish data are available in literature. Although the effect of missing flux becomes critical when we evaluate the global gas properties and the corresponding line ratios, the effect is negligible when we discuss structures that are smaller than the ``maximum recoverable scale" (MRS) of each configuration of ALMA. This is estimated from the minimum baseline lengths of the assigned antenna configurations and the observed frequencies. The MRS of our observations are $\sim$ 8\arcsec~and $\sim$ 7\arcsec~in band 3 and band 7, respectively (Table \ref{table_obs}). Therefore the missing flux effect in this paper is negligible, since we derive physical parameters (e.g., molecular gas mass) only for structures smaller than $\sim$ 2\arcsec.

\begin{figure*}
\begin{center}
\includegraphics[scale=.5]{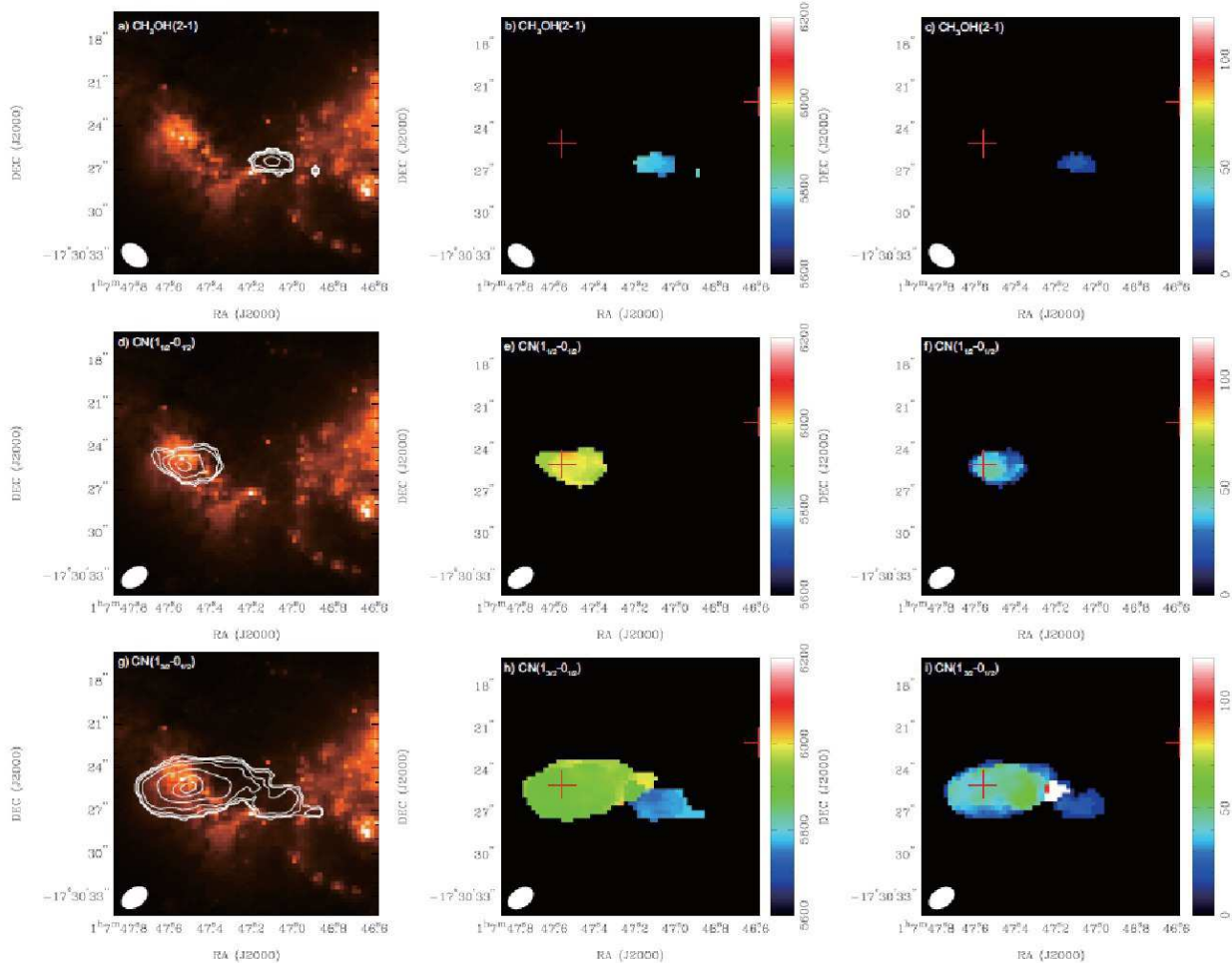}
\caption{The same as Figure \ref{fig_mom1} but for (a, b, and c) CH$_3$OH~(2$_k$--1$_k$), (d, e, and f) CN~(1$_{1/2}$--0$_{1/2}$), and (g, h, and i) CN~(1$_{3/2}$--0$_{1/2}$). (a) The contours are 0.02, 0.04, 0.08, 0.16, and 0.32~Jy~km~s$^{-1}$ (d) The contours are 0.04, 0.08, 0.16, 0.32, and 0.50~Jy~km~s$^{-1}$. (g) The contours are 0.04, 0.08, 0.16, 0.32, 0.64, 1.00, and~1.20 Jy~km~s$^{-1}$.
}
\label{fig_mom2}
\end{center}
\end{figure*}

\section{RESULTS} \label{result}

Molecular line and continuum images are shown in Figures~\ref{fig_mom1}, \ref{fig_mom2}, \ref{fig_mom3}, \ref{fig_VV114E}, and \ref{fig_contin}. The channel maps and the spectra of all line emissions are shown in Appendix~\ref{A2} and \ref{A3}.

\subsection{Line Emissions in Band 3}

\subsubsection{$^{12}$CO~(1--0)}

The integrated intensity, velocity field, and velocity dispersion maps of VV~114 are shown in Figures~\ref{fig_mom1}a, \ref{fig_mom1}b, and  \ref{fig_mom1}c, respectively. The total $^{12}$CO~(1--0) integrated intensity of VV~114 is 594.6 $\pm$ 1.6~Jy~km~s$^{-1}$, which is 1.3 times larger than that detected using the NRAO 12~m telescope \citep[461~Jy~km~s$^{-1}$;][]{san91}. This is because the pointing center for the NRAO 12~m observation was 25\farcs0 southwest of the CO centroid identified from the ALMA map (NRAO 12~m: 01h07m45.7s, -17d30m36.5s; CO centroid: 01h07m47.2s, -17d30m25.8s). At the adopted distance of VV~114 (86~Mpc), the 1\farcs97 $\times$ 1\farcs35 beam of the $^{12}$CO~(1--0) observation gives us a resolution of 790~pc $\times$ 540~pc. The two crosses shown in all images represent the peaks obtained from the miniTAO/ANIR $K$s-band observation, and we regard them as the progenitor's nuclei.

The integrated $^{12}$CO~(1--0) intensity map of VV~114 (Figure~\ref{fig_mom1}a) shows that the diffuse/cold gas forms two arm-like structures and a filamentary structure located at the center of the image. The global gas distribution is consistent with the previous $^{12}$CO~(1--0) observations \citep{yun94}. The southeastern (SE) arm clearly follows the tidal arm seen in the HST/ACS image \citep[Figure~\ref{fig_HST};][]{eva08}, while the northwestern (NW) arm has no counterpart in any other wavelengths. The region from the center of VV~114 to the eastern nucleus shows a strong concentration of molecular gas ($\simeq$ 5\farcs5 west of the eastern nucleus), and we refer to this region as the ``overlap" region with a molecular ``filament" (see Figure~\ref{fig_mom1}).

\begin{figure*}
\begin{center}
\includegraphics[scale=.5]{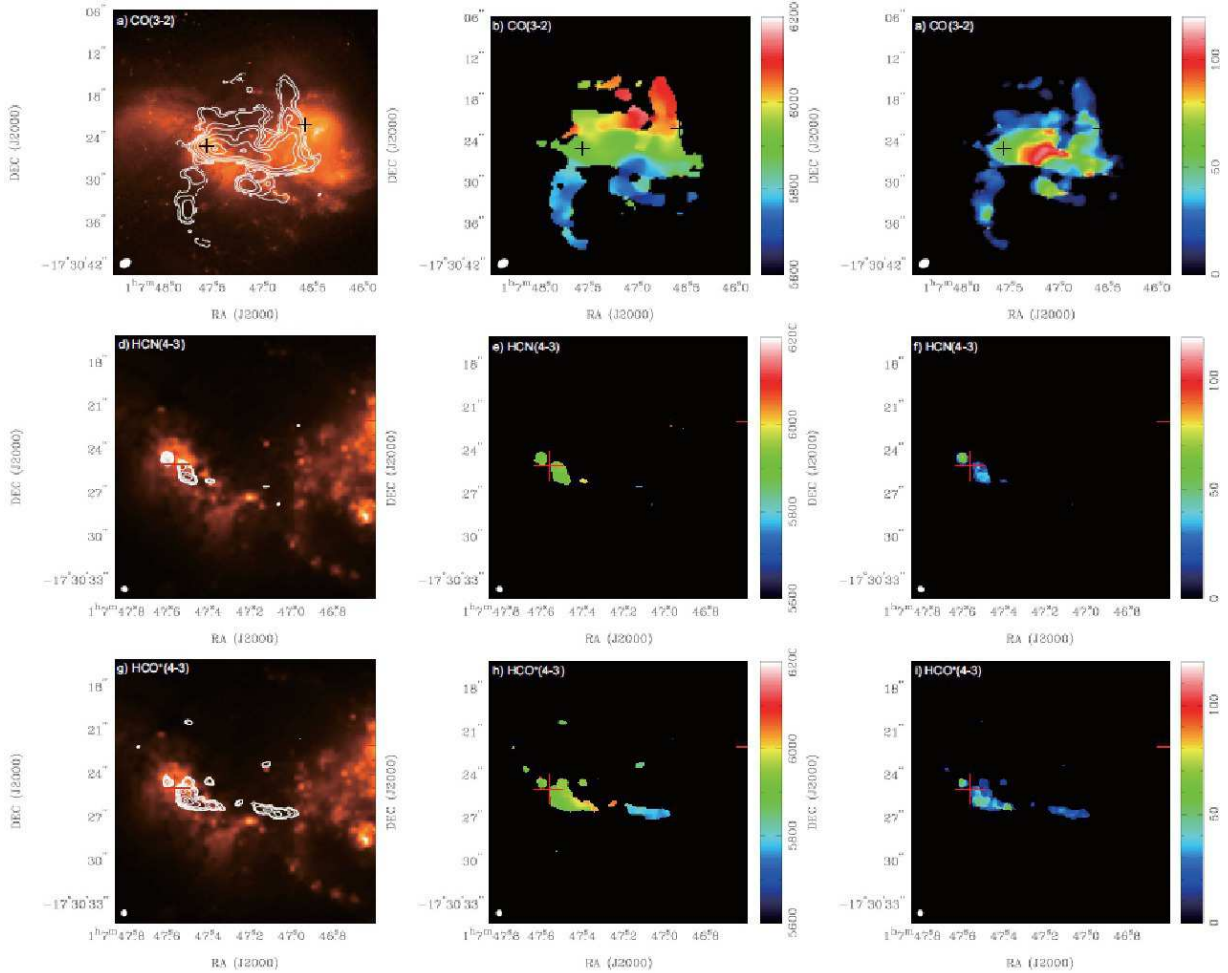}
\caption{The same as Figure \ref{fig_mom1} but for (a, b, and c) $^{12}$CO~(3--2), (d, e, and f) HCN~(4--3), and (g, h, and i) HCO$^+$~(4--3). (a) The contours are 2, 4, 8, 16, 32, 64, 128, and 170~Jy~km~s$^{-1}$ (d) The contours are 0.04, 0.08, 0.16, 0.32, 0.64, 1.28, and 1.80~Jy~km~s$^{-1}$. (g) The contours are 0.04, 0.08, 0.16, 0.32, 0.64, 1.28, and 2.40~Jy~km~s$^{-1}$.
}
\label{fig_mom3}
\end{center}
\end{figure*}

The $^{12}$CO~(1--0) velocity field map of VV~114 (Figure~\ref{fig_mom1}b) shows a significantly broad velocity range across the galaxy disks ($\simeq$ 600~km~s$^{-1}$). The SE arm has a blue-shifted velocity from 5650~km~s$^{-1}$ to 5920~km~s$^{-1}$, while the NW arm has a red-shifted velocity from 5950~km~s$^{-1}$ to 6160~km~s$^{-1}$. One possibility for the larger velocity width in the SE arm may be a highly inclined tidal arm. Two other arm-like features are also detected in the $^{12}$CO~(1--0) observations. One arm is located $\simeq$ 4\farcs0 northeast of the eastern nucleus and shows an arc around the eastern nucleus in the velocity range of 5810~km~s$^{-1}$ to 6180~km~s$^{-1}$. The other arm is located $\simeq$ 10\farcs0 west of the SE arm and has a strong peak ($\simeq$ 262.5 $\pm$ 1.0~Jy~km~s$^{-1}$) in the velocity range of 5610~km~s$^{-1}$ to 5900~km~s$^{-1}$.

The overlap region has the highest velocity dispersion ($\simeq$ 110~km~s$^{-1}$) (Figure~\ref{fig_mom1}c). The NW arm has an average velocity dispersion of $\simeq$ 30~km~s$^{-1}$, while the SE arm has $\simeq$ 40~km~s$^{-1}$. These values are significantly higher than the dispersions seen in Giant Molecular Clouds (GMCs) in the LMC \citep[2 -- 14~km~s$^{-1}$;][]{mnm08, fji14} and slightly higher than that in Giant Molecular Associations (GMAs) in the Antennae galaxy \citep[6 -- 36~km~s$^{-1}$;][]{ued12}. We suggest that the main contribution to the $^{12}$CO~(1--0) velocity dispersion is inter cloud turbulent medium along the tidal arm, and/or shocked region induced by the tidal interaction, rather than the velocity dispersion of the GMCs/GMAs.

\subsubsection{$^{13}$CO~(1--0)}

The integrated intensity, velocity field, and velocity dispersion maps of $^{13}$CO~(1--0) are shown in Figures~\ref{fig_mom1}d, \ref{fig_mom1}e, and \ref{fig_mom1}f, respectively.
The integrated $^{13}$CO~(1--0) intensity map of VV~114 (Figure~\ref{fig_mom1}d) shows a filamentary structure across the galaxy disks, which is consistent with the region where the $^{12}$CO~(1--0) filament is detected. The total $^{13}$CO~(1--0) integrated intensity is 5.9 $\pm 0.4$~Jy~km~s$^{-1}$.  The strongest peak is located $\simeq$ 4\farcs2 southwest of the eastern nucleus.
The $^{13}$CO~(1--0) velocity field map of VV~114 (Figure~\ref{fig_mom1}e) shows a narrower velocity range (5670 -- 6000~km~s$^{-1}$) than that of the $^{12}$CO~(1--0) emission (5600 -- 6200~km~s$^{-1}$). This suggests that the $^{13}$CO~(1--0) emission mainly comes from two components, the eastern galaxy and the blue-shifted component of the overlap region.
The $^{13}$CO~(1--0) velocity dispersion map of VV~114 (Figure~\ref{fig_mom1}f) shows a high velocity dispersion component ($\sim$ 100~km~s$^{-1}$) between the eastern nucleus and the overlap region. This significant velocity dispersion may be caused by a superposition of clouds (see the double-peak spectrum at R39 shown in Appendix~\ref{A3}).

\subsubsection{CS~(2--1) and CH$_3$OH~(2$_k$--1$_k$)}

The CS~(2--1) and CH$_3$OH~(2$_k$--1$_k$) lines are only detected at the overlap region (Figures~\ref{fig_mom1}g, \ref{fig_mom1}h, \ref{fig_mom1}i, \ref{fig_mom2}a, \ref{fig_mom2}b, and \ref{fig_mom2}c). This is the first detection of the CH$_3$OH~(2$_k$--1$_k$) emission in a merger-induced overlap region. We observed the blended set of 2$_1$ -- 1$_1$ ($\nu_{\rm{rest}}$ = 96.756~GHz, $E_{\rm{up}}/k$ = 28.0~K), 2$_0$ -- 1$_0$ \textit{E} ($\nu_{\rm{rest}}$ = 96.745~GHz, $E_{\rm{up}}/k$ = 20.1~K), 2$_0$ -- 1$_0$ \textit{A}$^+$ ($\nu_{\rm{rest}}$ = 96.741~GHz, $E_{\rm{up}}/k$ = 7.0~K), and 2$_{-1}$ -- 1$_{-1}$ \textit{E} ($\nu_{\rm{rest}}$ = 96.739~GHz, $E_{\rm{up}}/k$ = 12.5~K), thermal transitions of CH$_3$OH (hereafter designated the 2$_k$ -- 1$_k$ transition). The distribution of these molecular lines is clearly different from the other dense gas tracers detected in the current program. The peaks of CS~(2--1) and CH$_3$OH~(2$_k$--1$_k$) are coincident with one of the peaks of $^{13}$CO~(1--0) to within 0\farcs5. The total CS~(2--1) and CH$_3$OH~(2$_k$--1$_k$) integrated intensities are 0.4 $\pm$ 0.1~Jy~km~s$^{-1}$ and 0.5 $\pm$ 0.1~Jy~km~s$^{-1}$, respectively. The signal to noise is too low to resolve the velocity structure.

\begin{figure}
\begin{center}
\includegraphics[scale=.2]{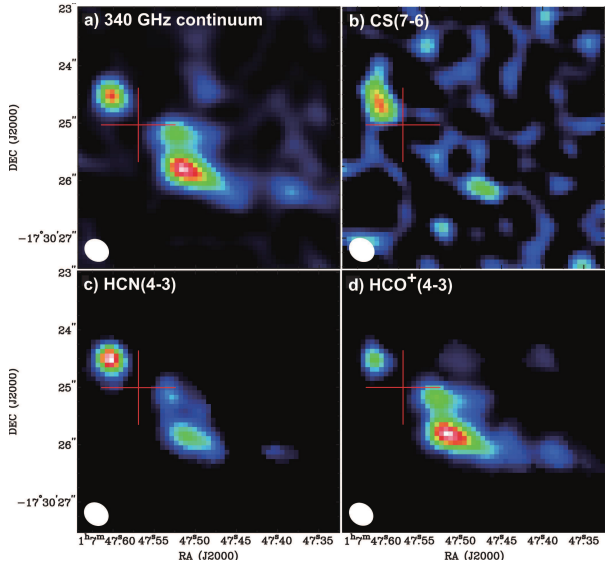}
\caption{(a) 340~GHz continuum flux image of VV~114E. The flux in color scale ranges from 0~mJy~beam$^{-1}$ to 3.4~mJy~beam$^{-1}$. (b) CS~(7--6) integrated intensity image of VV~114E. The flux in color scale ranges from 0~Jy~km~s$^{-1}$ to 0.45~Jy~km~s$^{-1}$. (c) HCN~(4--3) integrated intensity image of VV~114E. The flux in color scale ranges from 0~Jy~km~s$^{-1}$ to 2.0~Jy~km~s$^{-1}$. (d) HCO$^+$~(4--3) integrated intensity image of VV~114E. The flux in color scale ranges from 0~Jy~km~s$^{-1}$ to 2.8~Jy~km~s$^{-1}$. The beam size of each line and continuum is shown in the bottom-left of the images (Table~\ref{table_data}). The red cross shows the position of the eastern nucleus defined by the peak position of the Ks-band observation \citep{tat12}.
}
\label{fig_VV114E}
\end{center}
\end{figure}

\subsubsection{CN~(1$_{3/2}$--0$_{1/2}$) and CN~(1$_{1/2}$--0$_{1/2}$)}

Two radical CN rotational transitions \textit{N} = 1 -- 0 (\textit{J} = 3/2 -- 1/2 and 1/2 -- 1/2) are detected at the eastern nucleus. The $J$ = 3/2 -- 1/2 transition is extended toward the overlap region (Figures~\ref{fig_mom2}d, \ref{fig_mom2}e, \ref{fig_mom2}f, \ref{fig_mom2}g, \ref{fig_mom2}h, and \ref{fig_mom2}i). We can not resolve their multiplet because of the coarse frequency resolution (11.5~MHz $\simeq$ 30~km~s$^{-1}$). Because the critical density of CN is high ($\sim$ 10$^6$~cm$^{-3}$), the CN emission mainly comes from denser gas regions than regions traced by $^{12}$CO~(1--0). The \textit{J} = 3/2 -- 1/2 transition shows a similar distribution to the $^{13}$CO~(1--0) emission, but it is less extended over the overlap region. The total CN~(1$_{1/2}$--0$_{1/2}$) and CN~(1$_{3/2}$--0$_{1/2}$) integrated intensities are 2.0 $\pm$ 0.1~Jy~km~s$^{-1}$ and 5.4 $\pm$ 0.3~Jy~km~s$^{-1}$, respectively. The highest velocity dispersion in the CN~(1$_{3/2}$--0$_{1/2}$) image is also detected between the eastern nucleus and the overlap region, and this is likely caused by a superposition of clouds similar to the case of the $^{13}$CO~(1--0) image (see Appendix ~\ref{A3}).

\subsection{Line Emission in Band 7} \label{co32}

\subsubsection{$^{12}$CO~(3--2)}

The $^{12}$CO~(3--2) emission maps are presented in Figure~\ref{fig_mom3}. The estimated missing flux in our ALMA observation is 21 $\pm$ 1~\% \citep[\textit{James Clerk Maxwell Telescope} (JCMT): 2956 $\pm$ 133~$\rm{Jy~km~s^{-1}}$ and ALMA: 2343.7 $\pm$ 4.7~$\rm{Jy~km~s^{-1}}$;][]{wil08, sai13}. Although our $^{12}$CO~(3--2) observation recovers more flux than the \textit{Submillimeter Array} (SMA) observation \citep[1530 $\pm$ 16~$\rm{Jy~km~s^{-1}}$; the missing flux = 48 $\pm$ 15~\%; ][]{wil08}, there are significant negative sidelobes at the north and south of the image which is likely the cause of missing flux. We made the CLEANed image with a uniform \textit{uv} weighting to minimized the sidelobe level \citep{thomp}.

The $^{12}$CO~(3--2) integrated intensity map of VV~114 (Figure~\ref{fig_mom3}a) shows two arm-like structures and a filamentary structure similar to the $^{12}$CO~(1--0) image, and the strongest peak is at $\simeq$ 5\farcs5 west of the eastern nucleus. The global gas distribution is consistent with the previous $^{12}$CO~(3--2) observations \citep{ion04, wil08}.
The $^{12}$CO~(3--2) velocity field map of VV~114 (Figure~\ref{fig_mom3}b) also shows significant broad velocity range across the galaxy disks ($\simeq$ 600~km~s$^{-1}$), similar to the $^{12}$CO~(1--0) velocity field map. The SE arm has a blue-shifted velocity from 5650~km~s$^{-1}$ to 5920~km~s$^{-1}$, while the NW arm has a red-shifted velocity from 5950~km~s$^{-1}$ to 6160~km~s$^{-1}$. Other two arm-like features are also detected. One located $\simeq$ 4\farcs0 northeast of the eastern nucleus shows an arc around the eastern nucleus and has red-shifted velocities from 5810~km~s$^{-1}$ to 6180~km~s$^{-1}$. This arm coincides with the NE arm detected in the $^{12}$CO~(1--0). The other one located at $\simeq$ 10\farcs0 west of the SE arm has a strong peak ($\simeq$ 262.5 $\pm$ 0.9 ~Jy~km~s$^{-1}$) and blue-shifted velocities from 5610~km~s$^{-1}$ to 5900~km~s$^{-1}$. This arm also coincide with the SW arm detected in the $^{12}$CO~(1--0).
From the $^{12}$CO~(3--2) velocity dispersion map of VV~114 (Figure~\ref{fig_mom3}c), we find that the overlap region has the highest velocity dispersion ($\simeq$ 110~km~s$^{-1}$). The velocity dispersion of the NW arm is $\simeq$ 30~km~s$^{-1}$, while the SE arm is $\simeq$ 60~km~s$^{-1}$.

\begin{figure*}
\begin{center}
\includegraphics[scale=.5]{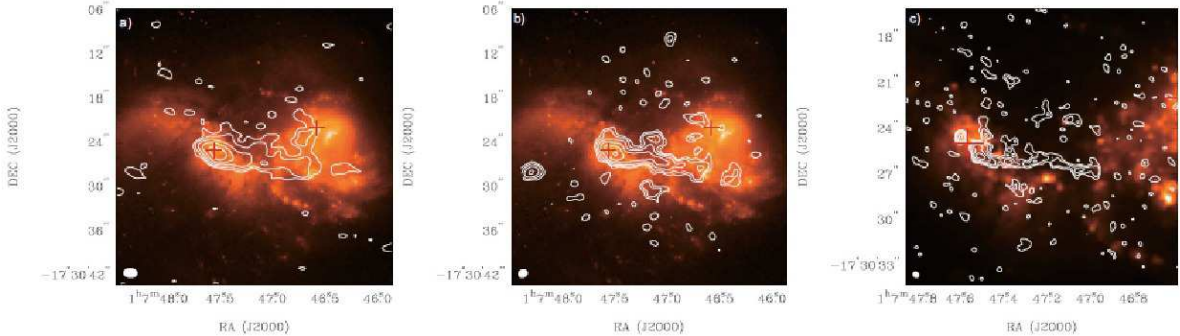}
\caption{(a) The 110~GHz continuum flux image overlaid on the HST/ACS/F435W image of VV~114. The contours are 0.10, 0.20, 0.40, 0.80, and 1.60~mJy~beam$^{-1}$. The red crosses show the positions of the nuclei defined by the peak positions of the Ks-band observation \citep{tat12}. (b) The low resolution 340~GHz continuum flux image overlaid on the HST/ACS/F435W image of VV~114. The contours are 0.22, 0.44, 0.88, 1.76, 3.52, and 7.04~mJy~beam$^{-1}$. A strong point source at the eastern edge of the image is a distant star-forming galaxy, ALMA J010748.3 -- 173028 \citep[see][]{tam14}.  (c) The high resolution 340~GHz continuum flux image overlaid on the HST/ACS/F435W image of VV~114. The contours are 0.14, 0.28, 0.56, 1.12, and 2.24~mJy~beam$^{-1}$. The red cross shows the position of the eastern nucleus defined by the peak position of the Ks-band observation \citep{tat12}. The beam size of each continuum is shown in the bottom-left of the images (Table~\ref{table_data}).
}
\label{fig_contin}
\end{center}
\end{figure*}

\subsubsection{HCN~(4--3) and HCO$^+$~(4--3)}

The HCN~(4--3) and HCO$^+$~(4--3) images are shown in Figures~\ref{fig_mom3}d, \ref{fig_mom3}e, \ref{fig_mom3}f, \ref{fig_mom3}g, \ref{fig_mom3}h, and \ref{fig_mom3}i. While the HCN~(4--3) emission is only seen near the eastern nucleus of VV~114 and is resolved into four peaks, the HCO$^+$~(4--3) emission is more extended and has at least 10 peaks in the integrated intensity map. The total integrated intensities of HCO$^+$~(4--3) and HCN~(4--3) are 15.3 $\pm$ 0.4~Jy~km ~s$^{-1}$ and 4.4 $\pm$ 0.2~Jy~km~s$^{-1}$, respectively. The higher HCO$^+$~(4--3) flux observed with the SMA \citep[17 $\pm$ 2~mJy,][]{wil08} using a 2\farcs8 $\times$ 2\farcs0 beam is likely attributed to missing flux by the ALMA observation. A compact component in the eastern nucleus is unresolved with the current resolution, and the upper limit to the size is 200~pc. The HCN~(4--3) emission is not detected in the overlap region, where both the high $^{12}$CO~(1--0) velocity dispersion and the significant CH$_3$OH~(2$_k$--1$_k$) and HCO$^+$~(4--3) detection suggest the presence of shocked gas \citep{krp08}. We concluded in \citetalias{ion13} from their source size, line widths, and the relative strengths of HCN~(4--3) and HCO$^+$~(4--3) that the unresolved eastern nucleus harbors an obscured AGN, and the dense clumps in the western galaxy are related to extended starbursts.

\subsubsection{CS~(7--6)}

The CS~(7--6) emission has the highest critical density ($n_{\rm{cr}}$ $\simeq$ 10$^7$~cm$^{-3}$) of all of the lines detected in our observations. The CS~(7--6) emission is marginally (S/N $\sim$ 4) detected at the eastern nucleus (Figure~\ref{fig_VV114E}), and the total flux is 0.5 $\pm$ 0.1~Jy~km~s$^{-1}$.

\subsection{Continuum Emission}

The continuum image at 110~GHz shows a filamentary structure similar to the molecular line image (Figure~\ref{fig_contin}a). We construct low resolution (1\farcs33 $\times$ 1\farcs12) and high resolution (0\farcs45 $\times$ 0\farcs38) images of the 340~GHz continuum (Figures~\ref{fig_contin}b, and \ref{fig_contin}c) using the combined data (compact + extended) and the extended configuration data, respectively. We find that the filamentary structure and the unresolved eastern nucleus are both present in dust continuum. The total flux of the 110~GHz and the low resolution 340~GHz continuum emission are 10.3 $\pm$ 0.2~mJy and 38.6 $\pm$ 0.3~mJy, respectively. The estimated missing flux relative to the JCMT 340~GHz observation \citep{wil08} is 75 $\pm$ 4~\% (SMA: 79 $\pm$ 7~\%). The difference in the recovered flux between $^{12}$CO~(3--2) and 340~GHz continuum emission may be caused by the difference in the distribution. The 110~GHz and 340~GHz continuum emission is detected at the eastern nucleus (S/N $\sim$ 50 and 70) and the filamentary structure (S/N $\sim$ 8 and 24) identified in the $^{13}$CO~(1--0) image, both with high significance.

\section{Spatially resolved line ratios} \label{ratio}

\begin{figure*}
\begin{center}
\includegraphics[scale=.5]{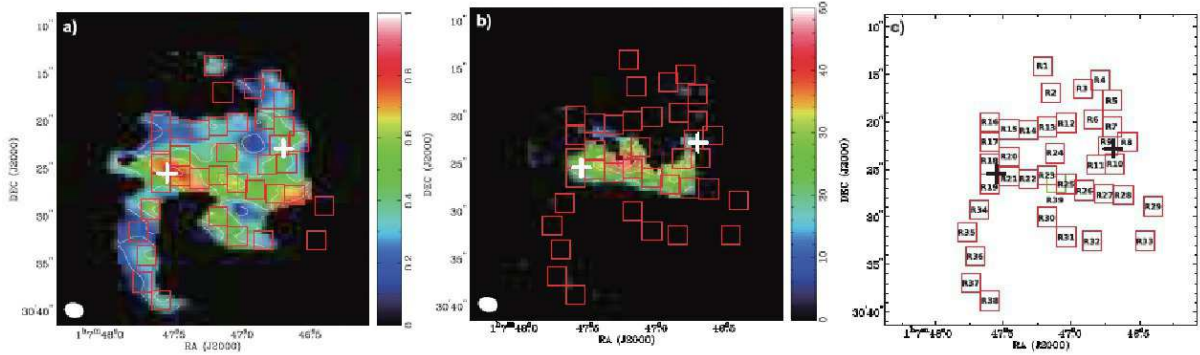}
\caption{(a) The $R_{3-2/1-0}$ image. The ratio in color scale ranges from 0 to 1. The white crosses show the positions of the nuclei defined by the peak positions of the Ks-band observation \citep{tat12}. (b) The $R_{12/13}$ image. The ratio in color scale ranges from 0 to 40. (c) Locations of 39 boxes (R1 -- R39) that are used to calculate the line ratios and physical parameters. For each ratio, the two integrated intensity images were convolved to the same resolution and expressed in units of K~km~s$^{-1}$ before calculating the ratio at locations where both lines are detected above 3~$\sigma$. The black crosses show the positions of the nuclei defined by the peak positions of the Ks-band observation \citep{tat12}. The beam size of each line ratio is shown in the bottom-left of the images.
}
\label{fig_ratio}
\end{center}
\end{figure*}

We assign 39 ``R" boxes (2\farcs0 $\times$ 2\farcs0; R1 -- R39; see Figure~\ref{fig_ratio}) for the band 3 and $^{12}$CO~(3--2) data and 15 smaller ``S" boxes (1\farcs2 $\times$ 1\farcs2; S0 -- S14; see Figure~\ref{figratio_HCN}) for the rest of the data to estimate the physical parameters, such as the molecular gas mass ($M_{\rm{H_2}}$), dense gas mass ($M_{\rm{dense}}$), dust mass ($M_{\rm{dust}}$), star formation rate (SFR), kinetic temperature ($T_{\rm{kin}}$), gas density ($n_{\rm{H_2}}$), gas column density ($N(\rm{H_2})$), and molecular abundance relative to H$_2$ ([$X$]/[H$_2$]). The positions of the boxes are chosen to cover the CO~(3--2) emission (R1 -- R39) and the HCO$^+$~(4--3) emission (S0 -- S14). The sizes of the boxes are chosen such that they are comparable to the beam size. 
Before deriving the parameters and line ratios at each box, we first matched the $uv$ range between our data set and reconstructed the integrated intensity image of each line. The shortest baseline lengths are set to 13.5 k$\lambda$ and 40.0 k$\lambda$ for the molecular lines in the band 3 and the band 7, respectively, and the images are convolved into the same resolution (2\farcs0 $\times$ 1\farcs5 with a P.A. of 83~deg, 1\farcs2 $\times$ 1\farcs0 with a P.A. of 119~deg). For each ratio, the two integrated intensity images were expressed in the units of K~km~s$^{-1}$ before calculating the ratio at locations where both lines are detected above 3~$\sigma$. The derived box-summed spectra are listed in Appendix~\ref{A3}. We carried out a multi Gaussian fit (one - three components) to reproduce the box-summed spectra, and labeled the components as ``a", ``b", and ``c" from the bluest peak (e.g., the bluest peak at R21 is labeled as R21a).

\subsection{$^{12}$CO~(3--2)/$^{12}$CO~(1--0), $R_{3-2/1-0}$}

The $^{12}$CO~(3--2)/$^{12}$CO~(1--0) ratio, $R_{3-2/1-0}$, can be used as an indicator of the dense/warm gas content relative to the total molecular gas. The $R_{3-2/1-0}$ of VV~114 varies from 0.2 to 0.8, as shown in Figure~\ref{fig_ratio} (left) and Table~\ref{table_T_R}. This range is larger than the same ratios derived for normal spirals, which is typically 0.15 -- 0.5 when observed with a similar linear resolution \citep{war10}. At the edge and the center of the filament, $R_{3-2/1-0}$ is higher (0.53 -- 0.69) than the highest peaks of each arm ($\sim$ 0.4). This suggests that the CO emitting gas at the filament have higher excitation conditions than normal spirals, while the conditions of each arm of VV~114 are consistent with arms and nuclei of normal spirals. The  $R_{3-2/1-0}$ at the eastern nucleus is 0.76 $\pm$ 0.01. It is suggested that the $R_{3-2/1-0}$ is much higher \citep[3.12 $\pm$ 0.03 in NGC 1068;][]{tsa12} for gas surrounding an AGN, and the low $R_{3-2/1-0}$ in VV~114 may be due to the difference in filling factor (160 $\times$ 140~pc beam averaging for NGC~1068, while 800~pc box averaging for VV~114). It is possible, however, that the nuclear excitation conditions are different from source to source.

\subsection{$^{12}$CO~(1--0)/$^{13}$CO~(1--0), $R_{12/13}$}

In general, the $^{12}$CO lines has higher optical depths than the $^{13}$CO~(1--0) line. Therefore, the measured $^{12}$CO~(1--0)/$^{13}$CO~(1--0) line intensity ratio, $R_{12/13}$, gives a lower limit to the CO/$^{13}$CO abundance ratio (hereafter [CO]/[$^{13}$CO]). We present the $R_{12/13}$ image of VV~114 in Figure~\ref{fig_ratio} (center). The $R_{12/13}$ increases from the arms ($<$ 17) to the filament (15 -- 32). Observationally, $R_{12/13}$ increases towards the central region of galaxies \citep{aal95}, where the gas is generally warmer and denser. \citet{aal95} suggest that the moderate optical depth of $^{12}$CO~(1--0) emission and/or the high [CO]/[$^{13}$CO] environment can increase the $R_{12/13}$ in nuclei of U/LIRGs. In order to understand which of the two (optical depths or abundances) is dominant, we calculated and mapped the optical depth of the $^{12}$CO~(1--0) and the $^{13}$CO~(1--0) as shown in Table~\ref{table_LTE} and Figure~\ref{fig_RADEX}. We provide an interpretation of these results in \S\ref{chi}.

\subsection{HCN~(4--3)/HCO$^+$~(4--3), $R_{\rm{HCN/HCO^+}}$}

\begin{figure*}
\begin{center}
\includegraphics[scale=.5]{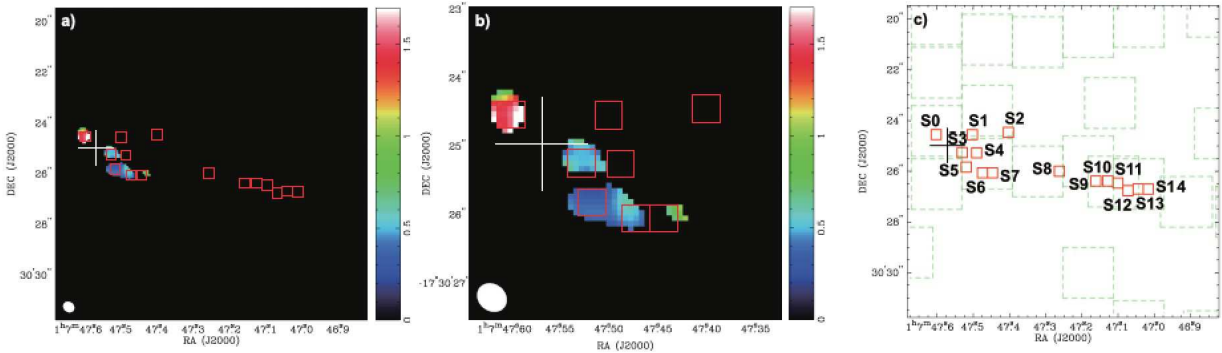}
\caption{(a) The $R_{\rm{HCN/HCO^+}}$ image. The ratio in color scale ranges from 0 to 1. The white cross shows the position of the eastern nucleus defined by the peak position of the Ks-band observation \citep{tat12}. (b) The $R_{\rm{HCN/HCO^+}}$ image near the nucleus of VV~114E. The ratio in color scale also ranges from 0 to 1. (c) Locations of 15 boxes (S0 -- S14) that are used to calculate the line ratios and physical parameters. For each ratio, the two integrated intensity images were convolved to the same resolution and expressed in units of K~km~s$^{-1}$ before calculating the ratio at locations where both lines are detected above 3~$\sigma$. The black cross shows the position of the eastern nucleus defined by the peak position of the Ks-band observation \citep{tat12}. The green open squares are the ``R" boxes shown in Figure \ref{fig_ratio}c.
}
\label{figratio_HCN}
\end{center}
\end{figure*}

In \citetalias{ion13}, the HCN~(4--3) and HCO$^+$~(4--3) maps of VV~114 allowed us to investigate the central region at 200 pc resolution for the first time, and we find that both the HCN~(4--3) and HCO$^+$~(4--3) in the eastern nucleus are compact ($<$ 200~pc), and broad [290~km~s$^{-1}$ for HCN~(4--3)]. We present the HCN~(4--3)/HCO$^+$~(4--3), $R_{\rm{HCN/HCO^+}}$, image of VV~114 in Figure~\ref{figratio_HCN}. From the higher $R_{\rm{HCN/HCO^+}}$ along with the past X-ray and NIR observations, we suggest the presence of an obscured AGN in the eastern nucleus. We also detect a 3 -- 4~kpc long filament of dense gas, which is likely to be tracing the active star formation triggered by the ongoing merger, and this is consistent with the results from the numerical model by \citet{tey10} who predict that the fragmentation and turbulent motion of dense gas across the merging disk is responsible for forming dense gas clumps with masses of 10$^6$ -- 10$^8$~M$_{\odot}$.

We present the $R_{\rm{HCN/HCO^+}}$ image in Figure~\ref{figratio_HCN}. The overlap region does not show significant HCN~(4--3) emission, and we provide the 3~$\sigma$ upper limit in Table~\ref{table_T_S}. Three out of the four boxes (i.e., S1 -- S3) in the eastern nucleus have low $R_{\rm{HCN/HCO^+}}$ ($<$ 0.5) whereas S0 has a high $R_{\rm{HCN/HCO^+}}$ (1.34 $\pm$ 0.09). It is suggested that such a high value is only produced around AGN environments \citep[e.g.,][]{khn01, har13, ion13, izm13, ima14}.

\section{Derivation of physical parameters} \label{der}

In this section, we derive the molecular gas mass (\S\ref{X}), and the physical parameters using the radiative transfer code RADEX (\S\ref{chi}) for each box defined in \S\ref{ratio}. The column density is derived using the optically thin $^{13}$CO line under the LTE assumption. We estimate the beam filling factor $\Phi_{\rm{A}}$ and the relative molecular abundance of molecule $X$ (hereafter expressed as [$X$]/[H$_2$]) (\S\ref{LTE}). Finally, we calculate the dust mass using the 340~GHz continuum emission (\S\ref{dust}).

\subsection{Molecular Gas Mass Derivation} \label{X}

The molecular gas mass M$_{\rm{X}}$ is derived by;

\begin{equation}
M_{\rm{X}} = \alpha_{\rm{X}}\:L'_{\rm{X}}\:[\rm{M}_{\odot}],
\end{equation}
where $\alpha_{\rm{X}}$ is the molecular line luminosity-to-H$_2$ mass conversion factor and $L\arcmin_{\rm{X}}$ is the velocity integrated flux \citep{s&v05}. We use the conversion factor known to be appropriate for U/LIRGs \citep[$\alpha_{\rm{CO}}$ = 0.8~M$_{\odot}~\rm{(K~km~s^{-1}~pc^2)^{-1}}$;][]{d&s98}. This is consistent with the value derived by \citet{slw13} in VV~114 ($\alpha_{\rm{CO}}$ = $0.5^{+0.6}_{-0.3}$~M$_{\odot}~\rm{(K~km~s^{-1}~pc^2)^{-1}}$). The molecular gas mass derived at the boxes defined in \S\ref{ratio} ranges between 0.2 $\times$ 10$^8$~$\left(\frac{\alpha_{\rm{CO}}}{0.8}\right)$ and 4.8 $\times$ 10$^8$~$\left(\frac{\alpha_{\rm{CO}}}{0.8}\right)$~M$_{\odot}$ (Table~\ref{table_SF_CO}). We also calculate the dense gas mass $M_{\rm{dense}}$ using $\alpha_{\rm{HCN}}$ = 10~M$_{\odot}~\rm{(K~km~s^{-1}~pc^2)^{-1}}$ \citep{g&s04} and the HCN~(4--3) luminosity which is converted to the HCN~(1--0) luminosity using HCN~(4--3)/HCN~(1--0) = 0.63 \citep[\citetalias{ion13};][]{ima07}.  The dense gas mass ranges between 1.8 $\times$ 10$^6$~$\left(\frac{\alpha_{\rm{HCN}}}{10}\right)$ and 3.8 $\times$ 10$^7$~$\left(\frac{\alpha_{\rm{HCN}}}{10}\right)$~M$_{\odot}$ (Table~\ref{table_SF_HCN}).

We note that the CO luminosity-to-H$_2$ mass conversion factor, $\alpha_{\rm{CO}}$, is very uncertain, and varies significantly from source to source \citep[0.4 -- 0.8 for LIRGs;][]{d&s98, yao03, pap12, bol13}. It may be possible that $\alpha_{\rm{CO}}$ varies from region to region within a galaxy. While one would ideally adopt a spatially varying $\alpha_{\rm{CO}}$ for a better quantification of the H$_2$ mass, such a study is beyond the scope of this present paper. For simplicity, here we adopt a constant $\alpha_{\rm{CO}}$ across all regions in VV~114, bearing in mind that the uncertainties could be as large as a factor of two. The same applies to $\alpha_{\rm{HCN}}$ \citep{g&s04}.

\subsection{Radiative Transfer Analysis using $\rm{RADEX}$} \label{chi}

\begin{figure*}
\begin{center}
\includegraphics[scale=.5]{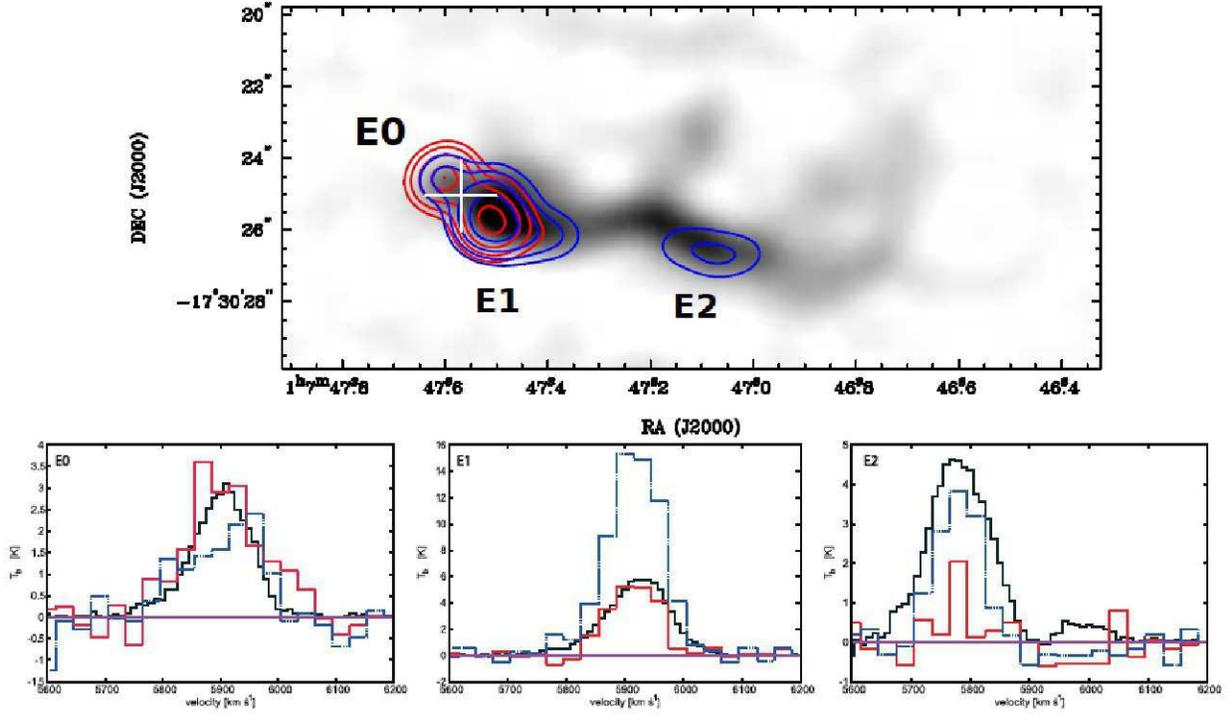}
\caption{(top) The $uv$- and beam-matched (grey color) $^{12}$CO~(3--2), (red contour) HCN~(4--3), and (blue contour) HCO$^+$~(4--3) images. The integrated intensity of the $^{12}$CO~(3--2) in color scale ranges from 0~Jy~km~s$^{-1}$ to 100~Jy~km~s$^{-1}$. The contours are 5, 10, 20, 40, and 50~Jy~km~s$^{-1}$ for HCN~(4--3), and 12, 24, 48, and 96~Jy~km~s$^{-1}$ for HCO$^+$~(4--3). The white cross shows the position of the eastern nucleus defined by the peak position of the Ks-band observation \citep{tat12}. (bottom) 1\farcs2 box-summed spectra of (black line) $^{12}$CO~(3--2), (red dashed line) HCN~(4--3) $\times$ 10, and (blue dashed line) HCO$^+$~(4--3) $\times$ 10 at the each box, labeled E0 -- E2. The spectra are taken from the ALMA data cubes after correcting the cubes for the primary beam attenuation and convolving them to 1\farcs2 $\times$ 1\farcs0 resolution (P.A. = 119 deg.).}
\label{fig_spec_RADEX}
\end{center}
\end{figure*}

We used the non-LTE radiative transfer code RADEX \citep{vdt07} and varied the parameters until the residuals between the observed line fluxes and the modeled line fluxes are minimized in a $\chi^2$ sense. We assumed a uniform spherical geometry ($dv$ = 1.0~$\rm{km~s^{-1}}$), and derived the physical conditions of molecular gas ($T_{\rm{kin}}$, $n_{\rm{H_2}}$, and $N(\rm{H}_2)$). RADEX uses an escape probability approximation to solve the non-LTE excitation assuming that all lines are from the same region. Since the molecular lines in the band~7 have significantly higher critical densities than that in the~band 3, we used two sets of molecular lines; (case 1) 2\farcs0 box-summed $^{12}$CO~(1--0), $^{13}$CO~(1--0), and $^{12}$CO~(3--2), and (case 2) 1\farcs2 box-summed HCN~(4--3), HCO$^+$~(4--3), $^{12}$CO~(3--2), and $^{12}$CO~(1--0), to solve for the degeneracy of the physical parameters. In case 2, we made the \textit{uv} and beam-matched HCN~(4--3), HCO$^+$~(4--3), and $^{12}$CO~(3--2) images (1\farcs2 $\times$ 1\farcs0 resolution with the P.A. = 119~deg.), and we defined three HCO$^+$~(4--3) peaks as E0, E1, and E2 (Figure~\ref{fig_spec_RADEX}). We also use the $uv$ and beam-matched $^{12}$CO~(1--0) data to constrain the $N(\rm{H}_2)$, allowing us to vary the [HCN]/[HCO$^+$] in case 2. All line parameters, such as the upper state energies and the Einstein coefficients, were taken from the \textit{Leiden Atomic and Molecular Database} \citep[LAMDA;][]{sco05}. In order to find the set of physical parameters that can reproduce the observed line intensities, we run RADEX by varying $T_{\rm{kin}}$, $n_{\rm{H_2}}$, and $N(\rm{H_2})$ for case 1, and $T_{\rm{kin}}$, $n_{\rm{H_2}}$, and [HCN]/[HCO$^+$] for case 2. The adopted $N(\rm{H_2})$ are 10$^{21.2}$, 10$^{21.6}$, and 10$^{21.5}$~cm$^{-2}$, at E0, E1, and E2, respectively.

We varied the gas kinetic temperature within a range of $T_{\rm{kin}}$ = 5 -- 300~K using steps of d$T_{\rm{kin}}$ = 5~K, and a gas density of $n_{\rm{H_2}}$ = $10^2 $ -- $ 10^5~\rm{cm^{-3}}$ using steps of d$n_{\rm{H_2}}$ = $10^{0.1}~\rm{cm^{-3}}$. For case 1, we fixed [$^{13}$CO]/[H$_2$] = 1.4 $\times$ 10$^{-6}$ \citep{dav13} and [CO]/[$^{13}\rm{CO}$] = 70, which are the Galactic values \citep{w&r94}. In case 2, we changed the parameters, $T_{\rm{kin}}$ = 5 -- 400~K using steps of d$T_{\rm{kin}}$ = 5~K, $n_{\rm{H_2}}$ = $10^3$ -- $10^7~\rm{cm^{-3}}$ using steps of d$n_{\rm{H_2}}$ = $10^{0.1}~\rm{cm^{-3}}$, and fixed [CO]/[H$_2$] = 1.0 $\times$ 10$^{-4}$ and [HCO$^+$]/[H$_2$] = 1.0 $\times$ 10$^{-9}$, which are the standard values observed in Galactic molecular clouds \citep{blk87}. We varied [HCN]/[HCO$^+$] from 1 -- 10, in steps of one. The parameters we used are summarized in Table~\ref{table_radex_parm}. We list the results that are within the 95~\% confidence level with 3-degree of freedom ($\chi^2 < 7.81$) (Tables~\ref{table_RADEX} and \ref{table_RADEX_HCN}). Finally, we created velocity-averaged channel maps of $n_{\rm{H_2}}$ and the optical depth of the transitions (Figure~\ref{fig_RADEX}).

We note that the uncertainty of the $N(\rm{H_2})$ for case 2 did not strongly affect the results, while that of the [CO]/[$^{13}\rm{CO}$] for case 1 changed. The effect of varying the [CO]/[$^{13}\rm{CO}$] will be discussed in \S\ref{1}. Future multi-transition HCN/HCO$^+$/CO/$^{13}$CO imaging will help us to derive these parameters directly.

\subsubsection{Case 1} \label{1}

The (box-averaged) kinetic temperature near the eastern nucleus (R21a) is constrained to within 25 -- 90~K (the best fit is 50~K), as shown in Table~\ref{table_RADEX}. The $T_{\rm{ex}}$ (58.8 $\pm$ 2.9~K) obtained from the LTE assumption at R21a (see \S\ref{LTE}) is higher than the best-fitted $T_{\rm{kin}}$. In fact, we also find five regions (R10b, R11b, R14, R16, and R25a) that show similarly high excitation temperatures. Four out of five regions are in the central filament. In general, spontaneous emission dominates over collisional excitation in sub-thermally excited conditions, and hence $T_{\rm{ex}}$ should be lower than $T_{\rm{kin}}$. One reason for this discrepancy could be attributed to the incorrect assumption of [CO]/[$^{13}$CO].  By varying this abundance ratio, we find that the temperature reversal (i.e. $T_{\rm{kin}} > T_{\rm{ex}}$) occurs only when [CO]/[$^{13}$CO] $>$ 150.  This is consistent with the results obtained by \citet{slw13} who used RADEX along with their multi CO and $^{13}$CO line data to find evidence of a cold/dense molecular gas component with extremely high [CO]/[$^{13}$CO] of 229, which is 3 times higher than that of the Galactic value \citep{w&r94}.

The derived $T_{\rm{kin}}$ at the other regions are generally higher than 100~K. The derived $T_{\rm{ex}}$ in each region are typically 10 -- 40~K, which may suggest sub-thermal conditions. The kinetic temperatures derived at the SE and NW arms are estimated to be $<$ 90 K, with higher temperature at the NW arm. The NW arm is also associated with relatively strong Pa$\alpha$ emission and $K$s-band emission, which is consistent with the higher relative temperature due to star-forming activities \citep{mnm08}. However, this is inconsistent with the general understanding that strong tidal shear in tidal arms prevents active star formation to occur \citep{aal10}.

The derived $n_{\rm{H_2}}$ in most of the boxes are less than 10$^{3.0}$~cm$^{-3}$, which is consistent with the critical densities of the low-$J$ CO lines observed here. The highest density of 10$^{3.4}$ -- 10$^{5.0}$~cm$^{-3}$ is estimated at R21a, and this is consistent with the location of the eastern nucleus. Since we also observed the strongest HCN~(4--3) and HCO$^+$~(4--3) emission at R21a at the same line-of-sight velocity \citep[see also Appendix~\ref{A3}]{ion13}, it is possible that the main contribution to the CO emission at R21a arises from dense gas (10$^{3.4}$ -- 10$^{5.0}$~cm$^{-3}$) near the eastern nucleus, with a minor contribution from the diffuse gas clouds along the same line of sight observed within the same beam. In contrast to the eastern nucleus, the boxes that cover the western galaxy (R1 -- R11 and R26 -- R29) show moderately dense condition of 10$^{2.0}$ -- 10$^{4.0}$~cm$^{-3}$. This extended and moderately dense gas is associated with the disk-like structure seen in optical images \citep{eva08}, and the star formation traced in Pa$\alpha$ emission and UV/X-ray emission \citep{grm06, tat12}. We note that the strongest off-nuclear Pa$\alpha$ peak (R27 in Table~\ref{table_SF_CO}; SFR = 3.15 $\pm$ 0.05~M$_{\odot}$~yr$^{-1}$) coincides with relatively low gas density ($\sim$ 10$^{3.0}$~cm$^{-3}$). The density of the surrounding region labeled R25a is similar (10$^{3.5}$ -- 10$^{5.0}$~cm$^{-3}$) and this is comparable to the nucleus of the eastern galaxy. The secondary Pa$\alpha$ peak (R29; SFR = 0.92 $\pm$ 0.05~M$_{\odot}$~yr$^{-1}$) is not associated with any molecular line emission.

\begin{figure*}
\begin{center}
\includegraphics[scale=.5]{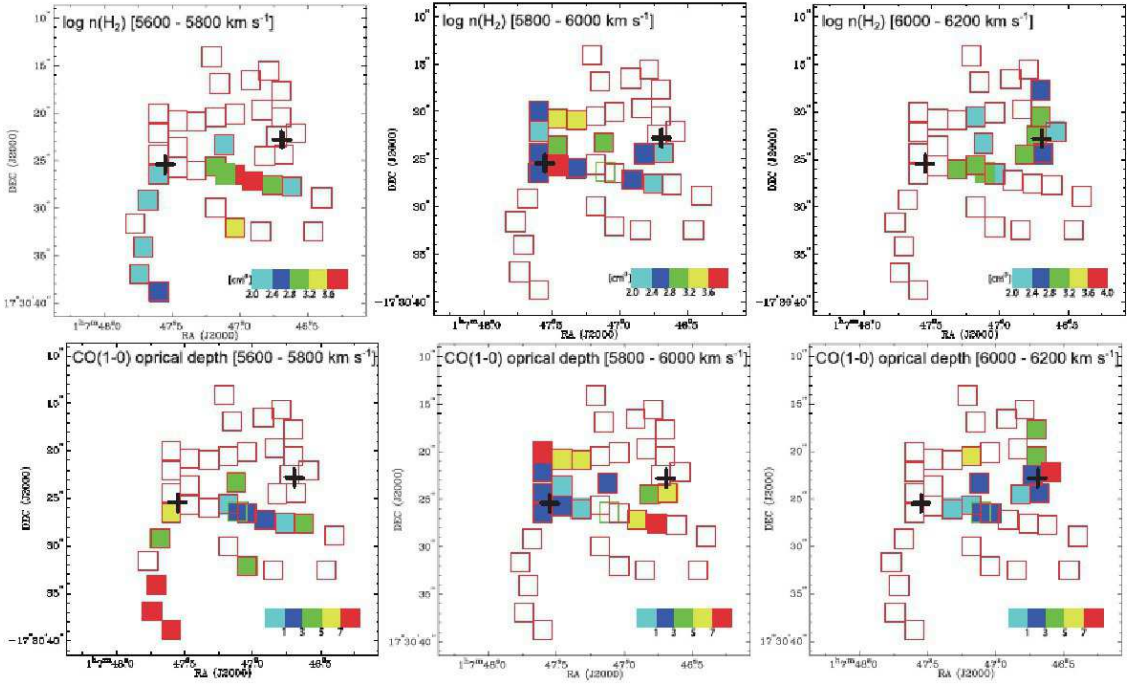}
\caption{The channel maps of the box-averaged RADEX modeling based on the $^{12}$CO~(1--0), $^{13}$CO~(1--0), and $^{12}$CO~(3--2). (top) The best fitted values of logarithmic gas density of the $^{12}$CO~(1--0) emission. The value in color scale ranges from 2.0 to 4.0~cm$^{-3}$. (bottom) The best fitted values of optical depth of the $^{12}$CO~(1--0) emission. The value in color scale ranges from 0 to 9. The black crosses show the positions of the nuclei defined by the peak positions of the Ks-band observation \citep{tat12}. The open squares are regions which we cannot solve the RADEX calculations because of non-detection of the $^{12}$CO~(1--0), $^{13}$CO~(1--0), or $^{12}$CO~(3--2) emission.}
\label{fig_RADEX}
\end{center}
\end{figure*}

It is usually believed that the $^{12}$CO~(1--0) emission is optically thick ($\tau_{\rm{CO}} \gg 1$), while the $^{13}$CO~(1--0) emission is optically thin ($\tau_{\rm{^{13}CO}} \ll 1$) even in luminous mergers \citep{dav13}. In most regions, we find that the optical depth of the $^{12}$CO~(1--0) line is $\gg$ 1 (Figure~\ref{fig_RADEX}). In contrast, the $^{12}$CO~(1--0) opacity at the eastern nucleus and the filament is moderately optically thick ($\tau_{\rm{CO}}$ $\sim$ 1). However, the elevated $R_{12/13}$ at the eastern nucleus (see \S\ref{ratio}) cannot be explained by the relatively low $^{12}$CO~(1--0) opacity alone (the opacity has to be $\tau_{\rm{CO}}$ $\ll$ 0.1; see also \citet{TORA}). Finally, we find that indeed the $^{13}$CO~(1--0) emission is optically thin ($\tau_{\rm{^{13}CO}} \ll 1$) averaged over the whole galaxy, except for the southern dust lane ($\tau_{\rm{^{13}CO}}$ = 0.3 -- 1.5).

From these results, we suggest that the peak of the molecular gas in the central 800~pc of the eastern galaxy is cold ($T_{\rm{kin}}$ = 25 -- 90~K), dense ($n_{\rm{H_2}}$ = 10$^{3.4}$ -- 10$^{5.0}$~cm$^{-3}$), and moderately optically thick ($\tau_{\rm{CO(1-0)}}$ $\sim$ 3), while peaks in the overlap region are warm ($T_{\rm{kin}}$ $>$ 50~K, best-fitted $T_{\rm{kin}}$ is 95 and 175~K at R39a and R39b, respectively), moderately dense ($n_{\rm{H_2}}$ = 10$^{2.3}$ -- 10$^{4.1}$~cm$^{-3}$), and moderately optically thick ($\tau_{\rm{CO(1-0)}}$ $\sim$ 1). The derived density of the eastern galaxy is slightly higher than the range of values found in U/LIRGs using low-$J$ CO emission with$\sim$ kpc resolution \citep[$n_{\rm{H_2}}$ = 10$^{2.3}$ -- 10$^{4.3}$~cm$^{-3}$;][]{d&s98}. In addition, the low opacities predicted from these analyses are consistent with earlier results that investigate the opacities in M82 \citep[$\tau$ = 0.5 -- 4.5;][]{mao00} and U/LIRGs \citep[$\tau$ = 3 -- 10;][]{d&s98}, and the central region of NGC 6240 \citep[$\tau$ = 0.2 -- 2;][]{ion07}. However, the derived temperature of the eastern galaxy is inconsistent with the high values found in nearby starburst galaxies M82, NGC 253, and NGC 6240 \citep{wld92, jac95, s&f00, ion07}.  The disagreement is possibly due to the uncertainties in the [CO]/[$^{13}$CO], or the difference in the observed molecular gas tracers.

\subsubsection{Case 2} \label{2}

The values for $T_{\rm{kin}}$, $n_{\rm{H_2}}$, and the optical depth of HCN~(4--3) and HCO$^+$~(4--3) are shown in Table~\ref{table_E}. The derived parameters for the unresolved component E0, are $T_{\rm{kin}}$ $>$ 100~K, $n_{\rm{H_2}}$ = 10$^{5.0}$ -- 10$^{5.4}$~cm$^{-3}$,  and [HCN]/[HCO$^+$] $>$ 5. The lower limit to the kinetic temperature is higher than those of E1 and E2, mainly due to the unusually high $R_{\rm{HCN/HCO^+}}$ and  $R_{\rm{HCN/CO}}$. In contrast to E0, the derived parameters near E1 show high H$_2$ densities ($n_{\rm{H_2}}$ = 10$^{5.6}$ -- 10$^{5.9}$~cm$^{-3}$).  The overlap region (E2), where the star-formation rate (1.70 $\pm$ 0.05~M$_{\odot}$~yr$^{-1}$) is lower than the eastern nucleus, has densities in the range of $n_{\rm{H_2}}$ = 10$^{5.0}$ -- 10$^{5.6}$~cm$^{-3}$. Finally, the optical depths for the HCO$^+$~(4--3) and HCN~(4--3) lines are calculated for each gas clump, yielding $\tau_{\rm{HCN}}$ $\simeq$ 0.7 and $\tau_{\rm{HCO^+}}$ $\simeq$ 0.2 for E0, $\tau_{\rm{HCN}}$ $\simeq$ 0.2 and $\tau_{\rm{HCO^+}}$ $\simeq$ 0.6 for E1, and $\tau_{\rm{HCN}}$ $\simeq$ 0.4 and $\tau_{\rm{HCO^+}}$ $\simeq$ 0.4 for E2.

The higher linear resolution observations of HCN~(4--3) and HCO$^+$~(4--3) toward NGC~1097 \citep{izm13} revealed that the gas in the central region of NGC 1097 has $T_{\rm{kin}}$ = 70 -- 550~K and $n_{\rm{H_2}}$ = 10$^{4.5}$ -- 10$^{6.0}$~cm$^{-3}$. Moreover, by comparing to LVG models, \citet{krp08} found that HCN and HCO$^+$ emission in AGN-dominated sources appears to emerge from regions with lower H$_2$ densities, higher temperatures, and higher HCN abundance relative to starburst-dominated (SB-dominated) galaxies. Our results obtained toward VV~114 are consistent with these previous results.

\subsection{Filling factor and Column Density under LTE} \label{LTE}

In order to determine the bulk properties of the CO emitting gas, we used an excitation temperature analysis \citep{dav13}. The excitation temperature at each box can be calculated from

\begin{equation}
T_{\rm{ex}} = T_0\:\left(\ln\left[\left(\frac{T_{\rm{b, CO(1-0)}}}{\Phi_{\rm{A}}T_0(1 - e^{-\tau_{\rm{CO}}})} + \frac{1}{e^{T_0/T_{\rm{bg}}} - 1} \right)^{-1} + 1\right] \right)^{-1}
\end{equation}
where $T_0 = h\nu/k$ [= 5.53~K for $^{12}$CO~(1--0) emission], $\nu$ is the frequency of the transition, $h$ is the Planck's constant, $k$ is the Boltzmann's constant, $T_{\rm{b, CO(1-0)}}$ is the brightness temperature of $^{12}$CO~(1--0) emission in Kelvin, $\tau_{\rm{CO}}$ is the optical depth of the $^{12}$CO~(1--0) emission, and $T_{\rm{bg}}$ is the cosmic microwave background temperature (2.73~K). Using $T_{\rm{kin}}$ estimated from the RADEX calculation (\S\ref{chi}), we estimate the beam filling factor $\Phi_{\rm{A}}$,

\begin{equation}
\Phi_{\rm{A}} = \frac{T_{\rm{b, CO(1-0)}}}{T_{\rm{kin}}}
\end{equation}

The optical depth of the $^{12}$CO~(1--0) emission is also estimated from the RADEX calculation in \S\ref{chi}. Assuming that the $^{13}$CO and CO arise from the same molecular cloud, and that the $^{12}$CO~(1--0) is optically thick, we estimate the optical depth of a given molecule using,

\begin{equation}
\tau_{\rm{X}} \simeq \ln\left[\left(1 - \frac{T_{\rm{b}, X}}{T_{\rm{b, CO(1-0)}}} \right)^{-1}\right]
\end{equation}
where $\tau_{\rm{X}}$ is the optical depth of a given transition, and $T_{\rm{b, X}}$ is the observed brightness temperature for transition X. Using $T_{\rm{ex}}$ and $\tau_{\rm{X}}$, we estimate the column density for a given molecule from,

\begin{displaymath}
N_{\rm{X}} = \frac{3k}{8\pi^3\mu^2B(J + 1)}\frac{\exp\left(\frac{2hJ(J + 1)}{kT_{\rm{ex}}} \right)}{\left(1 - \exp\left(-\frac{h\nu}{kT_{\rm{ex}}} \right) \right)}
\end{displaymath}

\begin{equation}
\times\:\frac{\tau_{\rm{X}}}{1 - e^{-\tau_{\rm{X}}}}\frac{1}{J(T_{\rm{ex}}) - J(T_{\rm{bg}})}\int T_{\rm{R}}^{*}dV
\end{equation}

\begin{equation}
J(T) = \frac{h\nu}{k}\frac{1}{\exp(h\nu/kT) - 1}
\end{equation}
where $\mu$ is the dipole moment, $B$ is the rotational constant, $J$ is the lower energy level, and $\int T_{\rm{R}}^{*}dV$ is the integrated intensity \citep{TORA}. The derived column densities are listed in Tables~\ref{table_LTE} and \ref{table_chem}.

\subsection{Dust Mass and ISM Mass Derivation from 340~GHz continuum} \label{dust}

We calculated the dust mass from the 340~GHz (880~$\mu$m) continuum emission (Table~\ref{table_contin}) using \citep{wil08},

\begin{equation}
M_{\rm{dust}} = 74220\:S_{340}\:D_\mathrm{L}^2\frac{e^{\frac{17}{T_d}} - 1}{\kappa_{340}}\:{\rm{M}_{\odot}}
\end{equation}
where $S_{340}$ is the 340~GHz flux in Jy and $D_{\rm{L}}$ is the luminosity distance in Mpc. We assumed a dust emissivity, $\kappa_{340} = 0.9~\rm{cm^2~g^{-1}}$, and the dust temperature $T_{\rm{d}}$ of 39.4~K \citep{wil08}. The box-summed dust masses ranges between 2.0 $\times$ 10$^4$~$\left(\frac{0.9}{\kappa_{340}}\right)$ and 2.8 $\times$ 10$^6$~$\left(\frac{0.9}{\kappa_{340}}\right)$~M$_{\odot}$ (Table~\ref{table_SF_CO}). We note that we used the \citet{d&l84} dust model for $\kappa_{340}$, because the $\kappa_{340}$ derived from observations has a large error \citep{hnn95}.

\citet{scv14} suggested that the submillimeter continuum emission traces the total ISM mass ($M_{\rm{ISM}}$), since the long wavelength Rayleigh-Jeans (RJ) tail of thermal dust emission is often optically thin. In order to compare the $M_{\rm{ISM}}$ with the $M_{\rm{H_2}}$ (see \$\ref{X}) using spatially-resolved data, we calculated the total ISM mass from the 340~GHz continuum emission \citep{scv14}. For $\nu_{\rm{rest}}$ $\lesssim$ 1199~GHz,

\begin{displaymath}
S_{\rm{\nu_{obs}}} = 0.83\frac{M_{\rm{ISM}}}{10^{10}\:\rm{M}_{\odot}}(1 + z)^{4.8}\left(\frac{\nu_{\rm{obs}}}{353~\rm{GHz}}\right)^{3.8}
\end{displaymath}
\begin{equation}
\times\frac{\Gamma_{\rm{RJ}}}{\Gamma_0}\left(\frac{\rm{Gpc}}{D_{\rm{L}}}\right) \rm{mJy}
\end{equation}
where $S_{\rm{\nu_{obs}}}$ is the observed flux, $M_{\rm{ISM}}$ is the ISM mass, $\nu_{\rm{obs}}$ is the observed frequency, and $\Gamma_{\rm{RJ}}$ and $\Gamma_0$ are given by

\begin{equation}
\Gamma_{\rm{RJ}}(T_d, \nu_{\rm{obs}}, z) = \frac{h\nu_{\rm{obs}}(1 + z)/kT_{d}}{e^{h\nu_{\rm{obs}}(1 + z)/kT_d} - 1}
\end{equation}

\begin{equation}
\Gamma_0 = \Gamma_{\rm{RJ}}(T_d, 353~\rm{GHz}, 0).
\end{equation}

The derived box-summed ISM masses of VV~114 range between 5.2 $\times$ 10$^7$ and 7.2 $\times$ 10$^8$ M$_{\odot}$ (Table~\ref{table_SF_CO}). This is comparable to the box-summed H$_2$ masses ($M_{\rm{H_2}}$ = (0.2 -- 4.7) $\times$ 10$^8$~$\left(\frac{\alpha_{\rm{CO}}}{0.8}\right)$ M$_{\odot}$). We find that the $M_{\rm{ISM}}$/$M_{\rm{H_2}}$ ratio is close to unity (0.5 -- 2.0, the average $M_{\rm{ISM}}$/$M_{\rm{H_2}}$ = 0.9 $\pm$ 0.1), while the total $M_{\rm{ISM}}$/$M_{\rm{H_2}}$ ratio is 0.6 $\pm$ 0.1. This means that the spatially-resolved $M_{\rm{ISM}}$ is a good tracer of the ``resolved" H$_2$ mass. However, the total $M_{\rm{ISM}}$ underestimates the H$_2$ mass (even using the $\alpha_{\rm{CO}}$ for ULIRGs to derive the $M_{\rm{H_2}}$) because the global distribution of the 340~GHz continuum emission is significantly different from that of the CO~(1--0) emission (Figures~\ref{fig_mom1} and \ref{fig_contin}). This difference between the 340~GHz continuum and the CO~(1--0) is also seen in recent observations of nearby LIRGs \citep[e.g.,][]{sak14}.

\section{Discussion} \label{dis}

\subsection{Conditions of ``Dense" Gas near the Eastern Nucleus} \label{dense}

Our RADEX modeling yields lower molecular gas density near the AGN ($n_{\rm{H_2}}$ = 10$^{5.0}$ -- 10$^{5.4}$~cm$^{-3}$) compared to the surrounding clumps (10$^{5.6}$ -- 10$^{5.9}$~cm$^{-3}$). Similarly high values are obtained near AGNs in other galaxies \citep{aln02, wil03, krp08}. \citet{krp08} suggest that the gas densities in AGN host galaxies ($<$ 10$^{4.5}$~cm$^{-3}$) are lower than starburst host galaxies (10$^{5.0}$ -- 10$^{6.5}$~cm$^{-3}$), and a common interpretation relies on a clumpy ISM near star-forming regions (which reduces the filling factor) and a continuous ISM near the AGN. Since our current observations ($\sim$ 200 pc resolution) cover a significantly large area and the beam filling factor may be small ($\Phi_{\rm{A}}$ at E0, E1, and E2 are $\lesssim$ 0.03, 0.04 -- 0.06 and 0.01 -- 0.04, respectively), higher resolution observations ($<$ 0\farcs5) are required to confirm this scenario.

\begin{figure*}
\begin{center}
\includegraphics[scale=.7]{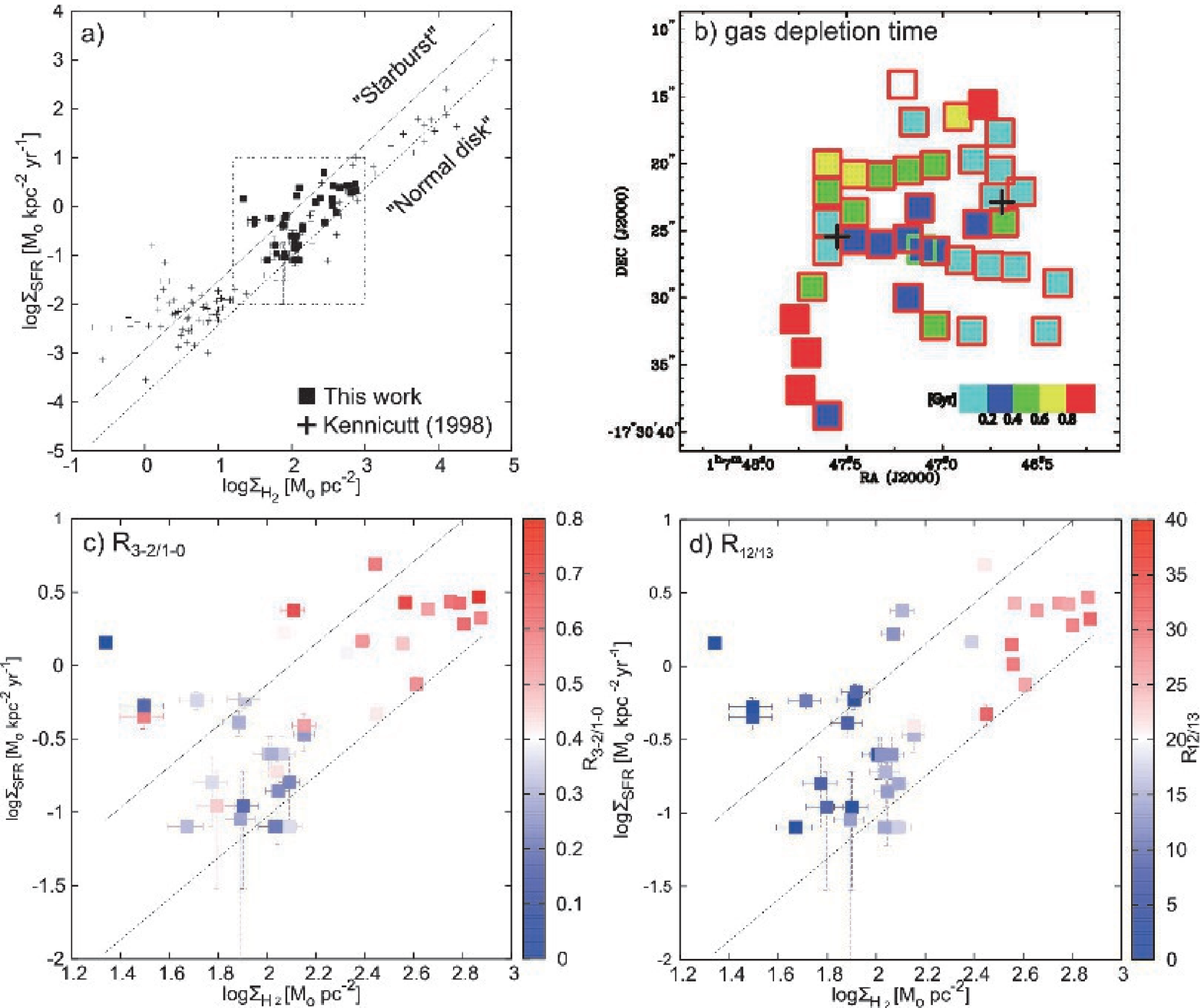}
\caption{(a) The Kennicutt-Scimidt law of VV~114 overlaid on other galaxies. Filled squares show regions of VV~114, while crosses show galaxies in the sample of \citet{ken98}. The dashed line and the dotted line indicate the ``starburst" sequence and ``normal disk" sequence, respectively \citep{dad10}. (b) Distribution of the box-averaged gas depletion time ($\tau_{\rm{gas}}$ = $\Sigma_{\rm{H_2}}/\Sigma_{\rm{SFR}}$). The gas depletion time in color scale ranges from 0 - 1 Gyr. The black crosses show the positions of the nuclei defined by the peak positions of the Ks-band observation \citep{tat12}. The open square is a region which we cannot estimate the gas depletion time because of non-detection of the Pa$\alpha$ emission \citep{tat12}. (c) The Keniccutt-Schmidt law with the $R_{3-2/1-0}$. The ratio in color scale ranges from 0 to 0.8. (d) The Keniccutt-Schmidt law with the $R_{12/13}$. The ratio in color scale ranges from 0 to 40.
}
\label{fig_KS}
\end{center}
\end{figure*}

In addition, our modeling shows higher [HCN]/[HCO$^+$] near the eastern nucleus ($>$ 5) than that in the surrounding clumps ($<$ 4) and the overlap region (1 - 9). The elevated [HCN]/[HCO$^+$] is explained by two mechanisms \citep{krp08}. One is far-UV radiation from OB stars in young starbursts \citep{s&d95}, and the other is strong X-ray radiation from an AGN \citep{mal96}. Because of different penetrating lengths between far-UV and X-ray emission, photon dominated regions (PDRs) are created at the surface of gas clouds and X-rays penetrate deeply into the circumnuclear disk (CND), forming large X-ray dominated regions. As a consequence of this volume versus surface effect, the X-ray radiation from an AGN may produce higher HCN abundances than the UV radiation of starburst activities \citep{krp08}. To some degree, ionization effects from cosmic rays \citep{wld92} such as supernovae or strong shocks are suspected to significantly increase the HCO$^+$ abundance while potentially decreasing the HCN abundance, thus yielding lower $R_{\rm{HCN/HCO^+}}$ in evolved starbursts than in AGNs. The high [HCN]/[HCO$^+$] near the eastern nucleus and low [HCN]/[HCO$^+$] and strong/extended 8~GHz continuum detection at the surrounding clumps \citep{cnd91} are all consistent with a presence of an AGN in the eastern nucleus, surrounded by star-forming dense clumps.

\subsection{Spatially Resolved Kennicutt-Schmidt Law} \label{SFR}

Observational studies of galaxies at global scales have shown that the surface density of SFR and that of cold gas traced in CO~(1--0) obey a power law relation \citep[KS law;][]{scm59, ken98}. ULIRGs are systemically shifted from the normal galaxy population in the $\Sigma_{\rm{SFR}}$ -- $\Sigma_{\rm{H_2}}$ phase \citep{kmg05, dad10, gen10, ler13}. It is suggested that systems lower in IR luminosity (e.g., LIRGs) occupy the region between the ``starburst" sequence and the ``normal disk" sequence in the KS law. Galaxies in the ``starburst" sequence have shorter gas depletion time ($\tau_{\rm{gas}}$ = $\Sigma_{\rm{H_2}}$/$\Sigma_{\rm{SFR}}$ $\sim$ 0.1~Gyr) relative to galaxies in the ``normal disk" sequence \citep[$\tau_{\rm{gas}}$ $\sim$ 1 Gyr;][]{dad10, bou11}. The spatially resolved surface densities of the SFR and the molecular gas mass of VV~114 are shown in Table~\ref{table_SF_CO} and Figure~\ref{fig_KS}. The star-forming regions of VV~114 fill the gap between the ``normal disk" and ``starburst" sequences (Figure~\ref{fig_KS}a). We also show the spatial distribution of $\tau_{\rm{gas}}$ in Figure~\ref{fig_KS}b. The data points close to the ``starburst" sequence are located along the eastern nucleus ($<$ 0.2~$\left(\frac{\alpha_{\rm{CO}}}{0.8}\right)$~Gyr) and the overlap region (= 0.2 -- 0.4~$\left(\frac{\alpha_{\rm{CO}}}{0.8}\right)$~Gyr), while those near the ``normal disk" sequence are located in the NW and SE arms ($>$ 0.8~$\left(\frac{\alpha_{\rm{CO}}}{0.8}\right)$~Gyr). The spatial distribution of $\Sigma_{\rm{SFR}}$ and $\Sigma_{\rm{H_2}}$ are consistent with the distributions of previous optical, UV, and X-ray studies \citep{aln02, lfl02, grm06}. Regions with higher $\Sigma_{\rm{SFR}}$ and $\Sigma_{\rm{H_2}}$ clearly show higher $R_{3-2/1-0}$ and $R_{12/13}$ (Figures~\ref{fig_KS}c and \ref{fig_KS}d).

In summary, transition from the ``normal disk" to ``starburst" sequence may occur when the molecular clouds become excited and dense at the nuclei and the overlap region. Moreover, gas clouds with high $R_{3-2/1-0}$ have high $\Sigma_{\rm{SFR}}$ -- $\Sigma_{\rm{H_2}}$, and this is consistent with past studies which suggest that the $R_{3-2/1-0}$ correlates with the local H$\alpha$ flux \citep{mnm08, fji14}. The $R_{12/13}$ also shows a similar trend, and this is also consistent with the past studies \citep[$>$ 20 in central kpc regions of U/LIRGs, 10 -- 15 in normal starburst galaxies, and $\sim$ 5 in Galactic GMCs;][]{aal97}: The reason for the elevated $R_{12/13}$ in starburst regions of VV~114 will be discussed in detail in \$\ref{isotope}.

\subsection{CO Isotope Ratio Enhancement in the Molecular ``Filament"} \label{isotope}

We suggest from our RADEX modelings that the eastern nucleus and the overlap region have extremely high [CO]/[$^{13}$CO] ($>$ 200), which is at least two times higher than the Galactic value \citep[$\simeq$ 70;][]{w&r94}. The Pa$\alpha$ peaks roughly coincide with the regions where high [CO]/[$^{13}$CO] are expected, suggesting that the increased [CO]/[$^{13}$CO] is related to the star formation activity. Similarly high values are seen in the overlap region of NGC4038/9 \citep{wil03} and the Taffy \citep{zhu07}. \citet{zhu07} suggested that the extreme [CO]/[$^{13}$CO] value in the bridge is explained by three scenarios, 1) selective isotope photodissociation in the diffuse clouds and shocked region, 2) CO enrichment around starburst activities, and/or 3) the destruction and recombination of molecules after shock. We briefly explain each scenario below, but our current data is insufficient for us to identify the exact cause of the high [CO]/[$^{13}$CO] in VV~114.

The first possibility of [CO]/[$^{13}$CO] enhancement is the deficiency in $^{13}$CO. \citet{she92} suggest that selective isotope photodissociation can reduce the $^{13}$CO abundance in diffuse clouds, because CO is self-shielded to a greater extent. Thus, the ISM surrounding young starbursts and/or shocked regions show elevated [CO]/[$^{13}$CO] \citep{zhu07}. The ISM in the nuclei and the overlap region of VV~114 show extremely high [CO]/[$^{13}$CO], presumably due to intense starburst activities and/or large-scale shocks.

The second possibility is that massive stars end their life as supernovae and expel a large amount of $^{12}$C in the interstellar medium. While the elemental abundances (e.g. C and S) are not directly related to the molecular abundances \citep[e.g., CS;][]{cas92}, once the synthesized elements are dispersed in the interstellar medium, molecules \citep[e.g., CO, CS, and CN;][]{hen14} form as soon as the temperature and density conditions are favorable. This occurs with a timescale of a few 10$^5$ yr \citep{l&g89}.

For the overlap region, the destruction and recombination of molecules after shocks (see \S\ref{chemical}) are possible mechanisms to enhance the [CO]/[$^{13}$CO] (the third possibility). The recombination timescale of H$_2$ and CO molecules after shock destruction are shorter than that of $^{13}$CO, since ionized photons from shocked regions lead to selective isotope photodissociation \citep{zhu03}. Shielded regions from the radiation field are needed to form rare $^{13}$CO (Abundant CO can form self-shielded regions). Moreover, the rare isotope molecules generally need a longer time to form, because collisions between molecules and dust grains are less frequent \citep{zhu07}.

\begin{figure}
\begin{center}
\includegraphics[scale=.3]{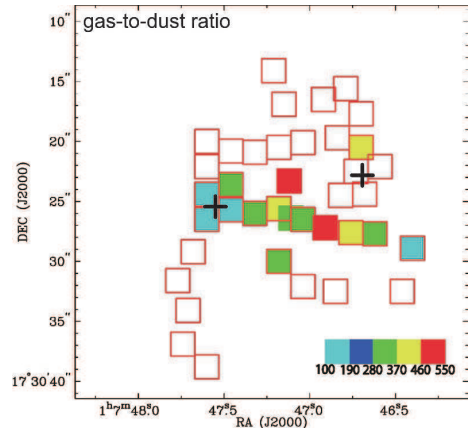}
\caption{Distribution of the box-averaged gas-to-dust ratio map. The black crosses show the positions of the nuclei defined by the peak positions of the Ks-band observation \citep{tat12}. The open squares are regions which we cannot estimate the gas-to-dust ratio because of non-detection of the 340~GHz continuum emission.}
\label{plot_G/D}
\end{center}
\end{figure}

\subsection{Gas-to-Dust Ratio, $M_{\rm{H_2}}/M_{\rm{dust}}$} \label{G/D}

The gas-to-dust ratio, $M_{\rm{H_2}}$/$M_{\rm{dust}}$, provides an important measure of the relative abundance between gas and metallicity. The average $M_{\rm{H_2}}$/$M_{\rm{dust}}$ over the entire galaxy is often derived in single-dish work, and typical $M_{\rm{H_2}}$/$M_{\rm{dust}}$ is 200 -- 300 for local U/LIRGs \citep{c&c03, yao03, sea04}, and 15 -- 231 in high-z sources \citep{s&v05}. \citet{wil08} found $M_{\rm{H_2}}$/$M_{\rm{dust}}$ = 357 $\pm$ 95 from a sample of 13 U/LIRGs, including VV~114, observed at kpc resolution.

We use the gas and dust masses derived in \S\ref{dust} to investigate the distribution of $M_{\rm{H_2}}$/$M_{\rm{dust}}$ (Figure~\ref{plot_G/D}). The smallest value of (128 $\pm$ 16)~$\left(\frac{\alpha_{\rm{CO}}}{0.8}\right)\left(\frac{0.9}{\kappa_{340}}\right)$ occurs in the eastern nucleus, which is similar to the Galactic value \citep[100;][]{hil83}, while higher values of (371 $\pm$ 118)~$\left(\frac{\alpha_{\rm{CO}}}{0.8}\right)\left(\frac{0.9}{\kappa_{340}}\right)$ and (339 $\pm$ 60)~$\left(\frac{\alpha_{\rm{CO}}}{0.8}\right)\left(\frac{0.9}{\kappa_{340}}\right)$ occur in the western nucleus and the overlap region, respectively. The clear differences between the two nuclei may suggest a local gradient in the metallicity. For the overlap region, cold dust associated with diffuse gas clouds cannot avoid the collision. This tends to increase the $M_{\rm{H_2}}$/$M_{\rm{dust}}$, because shocks destruct dust particles preferentially \citep{zhu07}. On the other hand, the low $M_{\rm{H_2}}$/$M_{\rm{dust}}$ in the eastern nucleus may be due to intense starbursts producing dust-rich environments.

\subsection{Fractional Abundances of CS, CH$_3$OH, and CN} \label{chemical}

Table~\ref{table_chem} shows the properties of the detected molecular lines which are not used in the RADEX calculations. Either the dense gas component of VV~114 has extreme variations in excitation among the molecular clumps in the filament (see \S\ref{dense}), or there is widespread chemical differentiation across the filament. The fractional abundances [$N_{\rm{X}}/N_{\rm{H_2}}$] of the different astrochemical species provide evidence of varying chemical influences due to star formation, physical conditions, and dynamics across the galaxy disks. We use the H$_2$ column densities, derived from the RADEX calculations, which are 10$^{20.8}$, 10$^{21.1}$, and 10$^{21.1}$~cm$^{-2}$ at R18 (AGN), R21a (starburst), and R39a (overlap region), respectively. Column densities of each molecules are determined by equation (7) assuming an optically thin emission under LTE. The $T_{\rm{ex}}$ values determined from equation (3) are 38.7 $\pm$ 1.9~K, 58.8 $\pm$ 2.9~K, and 52.6 $\pm$ 2.6~K at R18, R21a, and R39a, respectively. The derived [$N_{\rm{X}}/N_{\rm{H_2}}$] are listed in Table~\ref{table_chem}.

\begin{figure}
\begin{center}
\includegraphics[scale=.3]{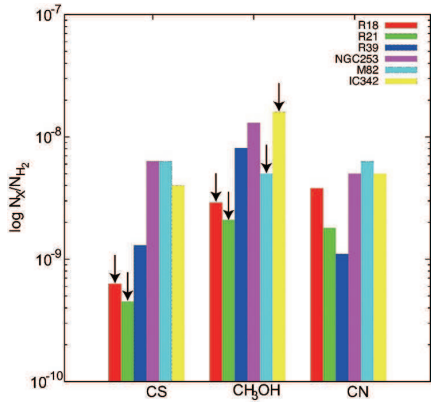}
\caption{Logarithmic fractional abundances relative to H$_2$ ([$X$]/[H$_2$]) of selected extragalactic sources compared to those of specific regions of VV~114, as presented in Table~\ref{table_chem}. The red, green, and blue columns show R18, R21a, and R39a, respectively. The pink, light blue, and yellow columns show NGC~253, M~82, and IC~342, respectively. Arrows represent upper limits.}
\label{plot_molabu}
\end{center}
\end{figure}

In Figure~\ref{plot_molabu}, we show the fractional abundances for CS, CH$_3$OH, and CN in VV~114, and the same ratios for a sample of nearly galaxies, NGC 253, M82, and IC 342, taken from line surveys available in the literature \citep{hen88, mau89, hut97, mar06}. M82 has a relatively old starburst at its core, with an average stellar population age of $\simeq$ 10~Myr \citep{kon09}. This creates strong UV fields, therefore the PDR dominates its chemistry \citep{ala11}. Figure~\ref{plot_molabu} shows that R39a has higher CH$_3$OH abundance than M82, and small CS and CN abundances. A pure PDR similar to M82 may explain the molecular abundances we observe in R18.

The molecular abundances for the overlap region and NGC 253 share similar characteristics. NGC 253 is thought to be in an early stage of starburst evolution, and has young stellar populations in its nucleus \citep[$\simeq$ 6~Myr;][]{fer09}. The chemistry in the nucleus of NGC 253 is dominated by large-scale shocks \citep{ala11}, and we suggest that the overlap region of VV~114 is also dominated by shocks. The low $R_{\rm{HCN/HCO^+}}$ at the overlap region are further evidences for a shock dominated region \citep{krp08}.

\subsection{Merger-driven Tidal Dwarf Galaxy Formation} \label{TDG}

Tidal dwarf galaxies (TDGs) are gas-rich irregular galaxies made out of stellar and gaseous material pulled out by tidal forces from the disks of the colliding parent galaxies into the intergalactic medium. They are found at the ends of long tails and host active star-forming regions \citep{bra00}. \citet{hib01} and \citet{gao01} found the HI gas mass of 4.1 $\times$ 10$^8$ M$_{\odot}$ and the molecular gas mass of 4 $\times$ 10$^6$ M$_{\odot}$ at the edge of the southern tail of NGC 4038/9.

We found an elevated $R_{3-2/1-0}$ (0.36 $\pm$ 0.01), SFR (0.10 $\pm$ 0.05~M$_{\odot}$~yr$^{-1}$), and $M_{\rm{H_2}}$ ($\sim$ 3.8 $\times$ 10$^7$~$\left(\frac{\alpha_{\rm{CO}}}{0.8}\right)$ M$_{\odot}$) at the edge of the southern tidal arm (R38). The derived SFR and $M_{\rm{H_2}}$ of R38 are comparable to those of TDG candidates in other galaxies \citep{bra01}. The gas depletion time of (0.40 $\pm$ 0.22)~$\left(\frac{\alpha_{\rm{CO}}}{0.8}\right)$~Gyr is shorter than the rest of the gas in the tidal arm ($>$ 0.5~$\left(\frac{\alpha_{\rm{CO}}}{0.8}\right)$~Gyr). According to the RADEX modeling, while the ranges of $T_{\rm{kin}}$ and $n_{\rm{H_2}}$ are not confined well, the best fitting values (35~K, 10$^{2.5}$~cm$^{-3}$) are slightly higher than those in the middle of the tidal arm, R36a and R37a (25 -- 30~K, 10$^{2.0}$ -- 10$^{2.2}$~cm$^{-3}$). We suggest that R38 is a forming tidal dwarf galaxy at the edge of the tidal arm of VV~114. Future high sensitivity optical and high resolution HI observations will allow us to constrain the star formation and the atomic gas properties of R38.

\section{CONCLUSION}\label{conclusion}

We investigate the physical conditions of the molecular gas in the mid-stage merger VV~114. We present high-resolution observations of molecular gas and dust continuum emission in this galaxy using ALMA band 3 and band 7. This study includes the first detection of extranuclear CH$_3$OH~(2--1) emission in interacting galaxies. The results can be summarized as follows:

\begin{enumerate}

\item We find that the CO~(1--0) and CO~(3--2) lines show significantly extended structures (i.e., the northern and southern tidal arms), the central filament across the galaxy disks, and double-peaks in the overlap region, while the $^{13}$CO~(1--0) line is only detected at the central filament. The filament is also identified by the strong CN~(1$_{3/2}$ -- 0$_{1/2}$), HCO$^+$~(4--3), 110~GHz, and 340~GHz continuum emission.

\item Higher $R_{\rm{3-2/1-0}}$ (0.5 -- 0.8) and $R_{\rm{12/13}}$ (20 -- 50) are detected at the central filament. These higher ratios indicate that the central filament has highly excited (but not thermalized) molecular ISM, and the eastern nucleus is nearly thermalized when it is observed with a 800~pc beam.

\item The unresolved eastern nucleus has the highest $R_{\rm{HCN/HCO^+}}$ (1.34 $\pm$ 0.09), while the dense gas clumps near the eastern nucleus have significantly lower values ($\sim$ 0.5). The broad HCN~(4--3) and HCO$^+$~(4--3) ($\sim$ 290~km~s$^{-1}$) emission lines seen in the unresolved eastern nucleus suggests an obscured AGN (see also \citetalias{ion13}).

\item Radiative transfer analysis of the CO~(1--0), CO~(3--2), and $^{13}$CO~(1--0) emission enables us to map physical parameters of the ``diffuse" gas of an interacting LIRG with 800~pc scale for the first time. The analysis suggests that ``diffuse" gas clouds in the filament have warmer/denser conditions than those in the galaxy disks. This is consistent with predictions from merger simulations. Our analysis also suggest that the [CO]/[$^{13}$CO] is enhanced in the central filament. The extremely high [CO]/[$^{13}$CO] values are more important than the moderately optically thick $^{12}$CO~(1--0) emission to explain the high $R_{12/13}$ in VV~114.

\item Radiative transfer analysis of the HCN~(4--3), HCO$^+$~(4--3), and $^{12}$CO~(3--2) allow us to compare the dense gas clouds around AGN, starburst activities, and the overlap region. These results show that dense gas clouds around AGN have $n_{\rm{H_2}}$ = 10$^{5.0}$ -- 10$^{5.4}$~cm$^{-3}$ and $T_{\rm{kin}}$ $>$ 100~K with [HCN]/[HCO$^+$] $>$ 5, while gas clumps around starburst activities show $n_{\rm{H_2}}$ = 10$^{5.6}$ -- 10$^{5.9}$~cm$^{-3}$ and $T_{\rm{kin}}$ = 40 --100~K with [HCN]/[HCO$^+$] $<$ 4. In addition, the analysis shows that the overlap region has $n_{\rm{H_2}}$ = 10$^{5.0}$ -- 10$^{5.6}$~cm$^{-3}$ and $T_{\rm{kin}}$ = 5 -- 90~K with [HCN]/[HCO$^+$] = 1 -- 9.

\item The spatially resolved Kennicutt-Schmidt law in VV~114 clearly connects the ``starburst" sequence with the ``normal disk" sequence. Most of the data points near the ``starburst" sequence are found in the nuclei and the overlap region, whereas the data points near the ``normal disk" sequence are found in the tidal arms. We also find the $R_{\rm{3-2/1-0}}$ and $R_{\rm{12/13}}$ are well correlated with the $\Sigma_{\rm{SFR}}$.

\item The $M_{\rm{H_2}}$/$M_{\rm{dust}}$ of (128 $\pm$ 16)~$\left(\frac{\alpha_{\rm{CO}}}{0.8}\right)\left(\frac{0.9}{\kappa_{340}}\right)$ in the eastern nucleus of VV~114 is comparable to the Galactic value, but it is a factor of two higher than that in the overlap region of (339 $\pm$ 60)~$\left(\frac{\alpha_{\rm{CO}}}{0.8}\right)\left(\frac{0.9}{\kappa_{340}}\right)$. Since the 340~GHz emission is spatially correlated with dense gas tracers, the cold dust in VV~114 appears to be closely related to the dense molecular component in the filament. The lowest $M_{\rm{H_2}}$/$M_{\rm{dust}}$ in the eastern nucleus may be due to the dusty starburst.

\item Comparing the CS, CN, and CH$_3$OH emission with other galaxies, we suggest that the overlap region is dominated by large-scale shocks similar to the nucleus of NGC~253. From the abundance analysis and distribution of the line ratios, we postulate that the HCN-rich AGN, the HCO$^+$-rich starbursts, and the CH$_3$OH-rich overlap region are important drivers of the molecular chemistry of VV~114.

\item We find a region with relatively high excitation ($\simeq$ 35~K, $\simeq$ 10$^{2.5}$~cm$^{-3}$) and star formation (SFR = 0.10 $\pm$ 0.05~M$_{\odot}$~yr$^{-1}$) at the edge of the southern tail. This region has a shorter $\tau_{\rm{gas}}$ of (0.40 $\pm$ 0.22)~$\left(\frac{\alpha_{\rm{CO}}}{0.8}\right)$~Gyr than the rest of the southern tail ($>$ 1.35~$\left(\frac{\alpha_{\rm{CO}}}{0.8}\right)$~Gyr), and we suggest that it is a forming tidal dwarf galaxy.

\end{enumerate}

\acknowledgements

The authors thanks the anonymous referee for comments that improved the contents of this paper. TS thanks for Yoichi Tamura, Takuma Izumi, and Akio Taniguchi's help on the RADEX calculation. We used a script developed by Y. Tamura for this calculation (http://www.ioa.s.u-tokyo.ac.jp/\verb|~|ytamura/Wiki/?Science\%2FUsingRADEX). TS and other authors thank ALMA staff for their kind support. TS, J. Ueda, and K. Tateuchi are financially supported by a Research Fellowship from the Japan Society for the Promotion of Science for Young Scientists. D. Iono was supported by the ALMA Japan Research Grant of NAOJ Chile Observaory, NAOJ-ALMA-0011 and JSPS KAKENHI Grant Number 2580016. This paper makes use of the following ALMA data: ADS/JAO.ALMA\#2011.0.00467.S. ALMA is a partnership of ESO (representing its member states), NSF (USA) and NINS (Japan), together with NRC (Canada) and NCS and ASIAA (Taiwan), in cooperation with the Republic of Chile. The Joint ALMA Observatory is operated by ESO, AUI/NRAO, and NAOJ.

\bibliographystyle{yahapj}

\begin{deluxetable*}{lrrrrrrrrrrcrl}
\tabletypesize{\scriptsize}
\tablecaption{Log of ALMA Observations \label{table_obs}}
\tablewidth{0pt}
\tablehead{
UT date & \multicolumn{2}{c}{Spectral windows} & \colhead{} & \multicolumn{3}{c}{Configuration}  & $T_{\rm{sys}}$& MRS & Amplitude caibrator & $T_{\rm{obs}}$\\
\cline{2-3} \cline{5-7} \\
& LSB & USB & & $N_{\rm{ant}}$ & Array & $L_{\rm{baseline}}$ & & & & \\
& [GHz] & [GHz] & & & & [m] & [K] & [arcsec.] & & [min.] \\
(1) & (2) & (3) & & (4) & (5) & (6) & (7) & (8) & (9) & (10)
 }
\startdata
2011 Nov 6 &101.5, 103.5& 114.0, 115.1 &&16 &CMP &18 - 196&\phantom{0}65 - \phantom{0}89& 18 & Uranus &41\\
2012 May 4 &101.5, 103.5& 114.0, 115.1 &&15 &EXT &39 - 402&\phantom{0}48 - \phantom{0}62& \phantom{0}8 & Neptune &40\\
2012 Mar 27 &\phantom{0}97.5, \phantom{0}99.5& 110.2, 111.5 &&17 &EXT &18 - 401&\phantom{0}54 - \phantom{0}73& 19 & Neptune &22\\
2012 Jul 2 &\phantom{0}97.5, \phantom{0}99.5& 110.2, 111.5 &&20 &EXT &16 - 402&\phantom{0}71 - 117& 21 & Neptune &39\\
2012 Nov 5 &331.1, 333.0& 343.5, 345.3 &&14 &CMP &12 - 135&125 - 172& \phantom{0}9 & Uranus &66\\
2012 Nov 5 &331.1, 333.0& 343.5, 345.3 &&14 &CMP &12 - 135&108 - 155& \phantom{0}9 & Uranus &67\\
2012 Nov 5 &331.1, 333.0& 343.5, 345.3 &&14 &CMP &12 - 135&124 - 175& \phantom{0}9 & Callisto &67\\
2012 Jun 1 &342.0, 344.0& 354.5, 356.0 &&18 &EXT &15 - 402&150 - 213& \phantom{0}7 & Uranus &78\\
2012 Jun 2 &342.0, 344.0& 354.5, 356.0 &&20 &EXT &15 - 402&108 - 160& \phantom{0}7 & Uranus &80\\
2012 Jun 3 &342.0, 344.0& 354.5, 356.0 &&20 &EXT &15 - 402&103 - 130& \phantom{0}7 & Uranus &45\\
\enddata
\tablecomments{Column 2 and 3: Central frequencies of the spectral windows (spw). All spw have the frequency coverage of 1.875 GHz. Column 4: Number of available antennas. Column 5: ALMA antenna configuration. CMP is the compact configuration and EXT is the extended configuration. Column 6: Range of projected length of baselines for VV~114. Column 7: DSB system temperature toward VV~114. Column 8: Maximum recoverable scale (MRS) of the configuration. This is defined by $\sim$ 0.6 $\lambda$/(minimum $L_{\rm{baseline}}$). Column 9: Observed calibrators for amplitude correction. Column 10: Total integration time on the galaxy.}
\end{deluxetable*}

\begin{deluxetable*}{lrrrrrrrrrcrl}
\tabletypesize{\scriptsize}
\tablecaption{ALMA Observational Properties \label{table_data}}
\tablewidth{0pt}
\tablehead{
Emission & Band & $\nu_{\rm{rest}}$ & Beam size & P.A. & $\Delta$V & Noise rms & \\
 & & [GHz] & [arcsecond] & [deg] & [km s$^{-1}$] & [mJy beam$^{-1}$] & [mK] \\
 (1) &(2) &(3) &(4) &(5) &(6) &(7) &(8)
 }
\startdata
CH$_3$OH~(2$_k$--1$_k$)     &3 &96.74        &2.03 $\times$ 1.34 &85.7   &30 &1.0 &46\\
CS~(2--1)                                     &3 &97.98                 &2.01 $\times$ 1.37 &83.6   &30 &0.9 &40\\
$^{13}$CO~(1--0)                       &3 &110.20               &1.77 $\times$ 1.20 &85.8   &30 &1.0 &46\\
CN~(1$_{1/2}$--0$_{1/2}$)       &3 &113.14               &1.97 $\times$ 1.27 &$-$85.8   &30 &1.0 &37\\
CN~(1$_{3/2}$--0$_{1/2}$)       &3 &113.49      &1.98 $\times$ 1.29 &$-$84.7  &30 &1.1 &39\\
CO~(1--0)                                    &3 &115.27                &1.97 $\times$ 1.35 &82.3   &10 &2.3 &76\\
CS~(7--6)                                    &7 &342.88                &0.47 $\times$ 0.39 &54.2   &30  &0.7 &38\\
CO~(3--2)                                    &7 &345.80                &1.64 $\times$ 1.17 &112.6 &10 &2.1 &11\\
HCN~(4--3)                                 &7 &354.51                &0.46 $\times$ 0.38 &51.5   &30 &0.8 &42\\
HCO$^+$~(4--3)                        &7 &356.73                &0.45 $\times$ 0.37 &53.4   &30 &0.9 &50\\
Continuum                               &3 &110      &1.89 $\times$ 1.28 &81.8 &\nodata  &0.05 & 2.1\\
Continuum                               &7 &340 &1.33 $\times$ 1.12 &119.6 &\nodata  &0.11 & 0.8\\
Continuum                               &7 &340 &0.45 $\times$ 0.38 &56.2 &\nodata  &0.07 & 4.3\\
\enddata
\tablecomments{Column 1: Identified emission. Column 2: Band which includes the molecular line and continuum emission. Column 3: Rest frequency of the line or mean frequency of the continuum. Column 4: Major and minor axes (FWHM) of the synthesized beam. Column 5: Position angle of the synthesized beam. Column 6: Velocity resolution of our binning images. Column 7 and 8: Noise rms intensity in the data which have velocity resolutions shown in Column 6. The noise in Column 8 is in Rayleigh-Jeans brightness temperature.}
\end{deluxetable*}

\begin{deluxetable*}{lrrrrrrrrrcrl}
\tabletypesize{\scriptsize}
\tablecaption{Radio Continuum Flux of VV~114 \label{table_contin}}
\tablewidth{0pt}
\tablehead{
ID & $S_{8.44}$ & $S_{110}$ & $S_{340}$  \\
 &[mJy] &[mJy] &[mJy] \\
 (1) &(2) &(3) &(4)
 }
\startdata
R7&0.62 $\pm$ 0.08&$<$ 0.18&0.58 $\pm$ 0.14 \\
R8&0.70 $\pm$ 0.08&$<$ 0.18&$<$ 0.42 \\
R9&0.82 $\pm$ 0.08&0.21 $\pm$ 0.06&$<$ 0.42 \\
R10&0.81 $\pm$ 0.08&0.20 $\pm$ 0.06&$<$ 0.42 \\
R11&0.92 $\pm$ 0.08&$<$ 0.18&$<$ 0.42 \\
R17&0.42 $\pm$ 0.08&$<$ 0.18&$<$ 0.42 \\
R18$^{\dag}$&3.37 $\pm$ 0.08&1.48 $\pm$ 0.06&5.17 $\pm$ 0.14 \\
R19&2.40 $\pm$ 0.08&1.03 $\pm$ 0.06&3.03 $\pm$ 0.14 \\
R20&1.14 $\pm$ 0.08&0.54 $\pm$ 0.06&2.41 $\pm$ 0.14 \\
R21$^{\dag\dag}$&5.00 $\pm$ 0.08&1.86 $\pm$ 0.06&8.20 $\pm$ 0.14 \\
R22&1.85 $\pm$ 0.08&0.81 $\pm$ 0.06&4.37 $\pm$ 0.14 \\
R23&1.34 $\pm$ 0.08&0.50 $\pm$ 0.06&3.55 $\pm$ 0.14 \\
R24&0.67 $\pm$ 0.08&$<$ 0.18&1.19 $\pm$ 0.14 \\
R25&1.07 $\pm$ 0.08&0.39 $\pm$ 0.06&3.01 $\pm$ 0.14 \\
R26&0.81 $\pm$ 0.08&0.33 $\pm$ 0.06&1.73 $\pm$ 0.14 \\
R27&1.29 $\pm$ 0.08&0.41 $\pm$ 0.06&1.40 $\pm$ 0.14 \\
R28&0.81 $\pm$ 0.08&0.21 $\pm$ 0.06&0.76 $\pm$ 0.14 \\
R29&0.35 $\pm$ 0.08&$<$ 0.18&0.71 $\pm$ 0.14 \\
R30&$<$ 0.24&$<$ 0.18&0.70 $\pm$ 0.14 \\
R39$^{\dag\dag\dag}$&1.18 $\pm$ 0.08&0.21 $\pm$ 0.06&3.35 $\pm$ 0.14
\enddata
\tablecomments{Column 2: 8.44 GHz continuum flux \citep{cnd91}. Column 3: 110 GHz continuum flux obtained by ALMA/band 3. Column 4: 340 GHz continuum flux obtained by ALMA/band 7.; We only show the statistical error in this table. The systematic error of absolute flux calibration is estimated to be $\sim$ 5\% in band 3 and $\sim$ 10\% in band 7.  \dag represents boxes contained the obscured AGN defined by \citetalias{ion13}.  \dag\dag represents boxes contained the nuclear starbursts defined by \citetalias{ion13}.  \dag\dag\dag represents boxes at the overlap region.}
\end{deluxetable*}

\begin{deluxetable*}{lrrrrrrrrrcrl}
\tabletypesize{\scriptsize}
\tablecaption{CO~(1--0), $^{13}$CO~(1--0), and $^{12}$CO~(3--2) Integrated Intensities \label{table_I_R}}
\tablewidth{0pt}
\tablehead{
ID & $^{12}$CO~(1--0) & $^{12}$CO~(3--2) & $^{13}$CO~(1--0) & $R_{3-2/1-0}$ & $R_{12/13}$ \\
 & [Jy km s$^{-1}$] & [Jy km s$^{-1}$] & [Jy km s$^{-1}$] &&\\
 (1) &(2) &(3) &(4) &(5) &(6)
 }
\startdata
R1&2.12 $\pm$ 0.06&5.92 $\pm$ 0.08&\nodata&0.31 $\pm$ 0.01&\nodata\\
R2&1.41 $\pm$ 0.05&7.47 $\pm$ 0.06&0.29 $\pm$ 0.02&0.59 $\pm$ 0.02&4 $\pm$ 1\\
R3&3.62 $\pm$ 0.09&5.45 $\pm$ 0.09&\nodata&0.17 $\pm$ 0.01&\nodata\\
R4&3.54 $\pm$ 0.08&8.69 $\pm$ 0.10&0.25 $\pm$ 0.04&0.27 $\pm$ 0.01&13 $\pm$ 2\\
R5&3.46 $\pm$ 0.08&9.16 $\pm$ 0.09&0.40 $\pm$ 0.03&0.29 $\pm$ 0.01&7 $\pm$ 1\\
R6&3.66 $\pm$ 0.10&10.71 $\pm$ 0.11&\nodata&0.33 $\pm$ 0.01&\nodata\\
R7&5.26 $\pm$ 0.10&19.64 $\pm$ 0.12&0.40 $\pm$ 0.03&0.41 $\pm$ 0.01&12 $\pm$ 1\\
R8&3.71 $\pm$ 0.09&13.44 $\pm$ 0.12&0.38 $\pm$ 0.03&0.40 $\pm$ 0.01&8 $\pm$ 1\\
R9&9.59 $\pm$ 0.11&33.88 $\pm$ 0.12&0.43 $\pm$ 0.04&0.39 $\pm$ 0.01&20 $\pm$ 2\\
R10&12.69 $\pm$ 0.11&47.74 $\pm$ 0.12&0.37 $\pm$ 0.05&0.42 $\pm$ 0.01&32 $\pm$ 4\\
R11&16.26 $\pm$ 0.13&58.93 $\pm$ 0.14&0.49 $\pm$ 0.04&0.40 $\pm$ 0.01&30 $\pm$ 3\\
R12&4.61 $\pm$ 0.10&13.53 $\pm$ 0.10&\nodata&0.33 $\pm$ 0.01&\nodata\\
R13&5.22 $\pm$ 0.10&16.02 $\pm$ 0.11&0.38 $\pm$ 0.05&0.34 $\pm$ 0.01&12 $\pm$ 2\\
R14&4.70 $\pm$ 0.09&12.22 $\pm$ 0.09&0.35 $\pm$ 0.03&0.29 $\pm$ 0.01&12 $\pm$ 1\\
R15&5.03 $\pm$ 0.09&10.62 $\pm$ 0.09&0.35 $\pm$ 0.03&0.23 $\pm$ 0.01&13 $\pm$ 1\\
R16&5.59 $\pm$ 0.09&11.23 $\pm$ 0.08&0.38 $\pm$ 0.02&0.22 $\pm$ 0.01&14 $\pm$ 1\\
R17&6.44 $\pm$ 0.10&14.33 $\pm$ 0.11&0.38 $\pm$ 0.05&0.25 $\pm$ 0.01&15 $\pm$ 2\\
R18$^{\dag}$&16.54 $\pm$ 0.13&106.39 $\pm$ 0.15&0.60 $\pm$ 0.05&0.71 $\pm$ 0.01&25 $\pm$ 2\\
R19&11.09 $\pm$ 0.12&54.98 $\pm$ 0.14&0.60 $\pm$ 0.05&0.55 $\pm$ 0.01&17 $\pm$ 2\\
R20&18.33 $\pm$ 0.13&93.14 $\pm$ 0.14&0.64 $\pm$ 0.06&0.56 $\pm$ 0.01&26 $\pm$ 2\\
R21$^{\dag\dag}$&33.04 $\pm$ 0.15&225.47 $\pm$ 0.16&1.09 $\pm$ 0.07&0.76 $\pm$ 0.01&28 $\pm$ 2\\
R22&28.67 $\pm$ 0.15&162.41 $\pm$ 0.16&0.90 $\pm$ 0.07&0.63 $\pm$ 0.01&29 $\pm$ 2\\
R23&33.65 $\pm$ 0.16&177.02 $\pm$ 0.17&0.95 $\pm$ 0.07&0.58 $\pm$ 0.01&32 $\pm$ 2\\
R24&16.12 $\pm$ 0.14&68.28 $\pm$ 0.15&0.48 $\pm$ 0.07&0.47 $\pm$ 0.01&31 $\pm$ 4\\
R25&25.23 $\pm$ 0.16&138.20 $\pm$ 0.16&0.86 $\pm$ 0.06&0.61 $\pm$ 0.01&27 $\pm$ 2\\
R26&20.44 $\pm$ 0.14&97.56 $\pm$ 0.15&0.69 $\pm$ 0.06&0.53 $\pm$ 0.01&27 $\pm$ 3\\
R27&12.45 $\pm$ 0.12&67.10 $\pm$ 0.13&0.54 $\pm$ 0.06&0.60 $\pm$ 0.01&21 $\pm$ 2\\
R28&5.78 $\pm$ 0.10&36.14 $\pm$ 0.14&0.35 $\pm$ 0.03&0.69 $\pm$ 0.01&15 $\pm$ 1\\
R29&\nodata&\nodata&\nodata&\nodata&\nodata\\
R30&6.39 $\pm$ 0.12&30.50 $\pm$ 0.12&0.28 $\pm$ 0.03&0.53 $\pm$ 0.01&21 $\pm$ 3\\
R31&4.93 $\pm$ 0.11&19.03 $\pm$ 0.13&0.30 $\pm$ 0.04&0.43 $\pm$ 0.01&15 $\pm$ 2\\
R32&2.34 $\pm$ 0.05&7.34 $\pm$ 0.08&0.21 $\pm$ 0.02&0.35 $\pm$ 0.01&10 $\pm$ 1\\
R33&1.44 $\pm$ 0.05&2.03 $\pm$ 0.05&\nodata&0.16 $\pm$ 0.01&\nodata\\
R34&2.81 $\pm$ 0.08&11.65 $\pm$ 0.10&0.32 $\pm$ 0.03&0.46 $\pm$ 0.01&8 $\pm$ 1\\
R35&5.36 $\pm$ 0.09&9.40 $\pm$ 0.09&0.30 $\pm$ 0.03&0.19 $\pm$ 0.01&16 $\pm$ 2\\
R36&5.55 $\pm$ 0.12&17.95 $\pm$ 0.14&0.31 $\pm$ 0.02&0.36 $\pm$ 0.01&17 $\pm$ 1\\
R37&4.84 $\pm$ 0.11&8.65 $\pm$ 0.11&0.31 $\pm$ 0.03&0.20 $\pm$ 0.01&14 $\pm$ 2\\
R38&2.66 $\pm$ 0.07&8.51 $\pm$ 0.09&0.33 $\pm$ 0.02&0.36 $\pm$ 0.01&7 $\pm$ 1\\
R39$^{\dag\dag\dag}$&27.54 $\pm$ 0.15&155.31 $\pm$ 0.15&0.97 $\pm$ 0.06&0.63 $\pm$ 0.01&26 $\pm$ 2\\
\enddata
\tablecomments{Column 1: These numbers are labeled at the ratio map of Figure~\ref{fig_ratio}. Column 2: Integrated $^{12}$CO~(1--0) intensity at an emission region. Column 3: Integrated $^{12}$CO~(3--2) intensity at an emission region. Column 4: Integrated $^{13}$CO~(3--2) intensity at an emission region. Column 5: The $^{12}$CO~(3--2)/CO~(1--0) integrated intensity ratio. Column 6: The $^{12}$CO~(1--0)/$^{13}$CO~(1--0) integrated intensity ratio.; We only show the statistical error in this table. The systematic error of absolute flux calibration is estimated to be $\sim$ 5\% in band 3 and $\sim$ 10\% in band 7.  \dag represents boxes contained the obscured AGN defined by \citetalias{ion13}.  \dag\dag represents boxes contained the nuclear starbursts defined by \citetalias{ion13}.  \dag\dag\dag represents boxes at the overlap region.}
\end{deluxetable*}

\begin{deluxetable*}{lrrrrrrrrrcrl}
\tabletypesize{\scriptsize}
\tablecaption{HCN~(4--3) and HCO$^+$~(4--3) Integrated Intensities \label{table_I_S}}
\tablewidth{0pt}
\tablehead{
ID & HCN~(4--3) & HCO$^+$~(4--3) & $R_{\rm{HCN/HCO^+}}$ \\
 & [Jy km s$^{-1}$] & [Jy km s$^{-1}$] &\\
 (1) &(2) &(3) &(4) \\
 }
\startdata
S0$^{\dag}$&1.34 $\pm$ 0.09 &0.88$\pm$ 0.10 &1.52 $\pm$ 0.20 \\
S1&0.11 $\pm$ 0.03 &0.34$\pm$ 0.06 &0.34 $\pm$ 0.10 \\
S2&$<$ 0.05 &0.12$\pm$ 0.06 &$<$ 0.43 \\
S3$^{\dag\dag}$&0.49 $\pm$ 0.07 &1.05$\pm$ 0.08 &0.46 $\pm$ 0.07 \\
S4&0.13 $\pm$ 0.06 &0.63$\pm$ 0.08 &0.20 $\pm$ 0.09 \\
S5$^{\dag\dag}$&0.81 $\pm$ 0.07 &2.36$\pm$ 0.09 &0.34 $\pm$ 0.03 \\
S6&0.40 $\pm$ 0.07 &0.92$\pm$ 0.08 &0.43 $\pm$ 0.09 \\
S7&0.19 $\pm$ 0.07 &0.42$\pm$ 0.08 &0.45 $\pm$ 0.18 \\
S8&0.06 $\pm$ 0.04 &0.14$\pm$ 0.06 &0.44 $\pm$ 0.31 \\
S9&0.07 $\pm$ 0.05 &0.13$\pm$ 0.06 &0.54 $\pm$ 0.46 \\
S10&$<$ 0.04 &0.28$\pm$ 0.07 &$<$ 0.14 \\
S11$^{\dag\dag\dag}$&$<$ 0.05 &0.15$\pm$ 0.06 &$<$ 0.33 \\
S12$^{\dag\dag\dag}$&$<$ 0.04 &0.32$\pm$ 0.07 &$<$ 0.12 \\
S13&$<$ 0.04 &0.17$\pm$ 0.06 &$<$ 0.24 \\
S14&$<$ 0.05 &0.19$\pm$ 0.07 &$<$ 0.26
\enddata
\tablecomments{Column 1: These numbers are labeled at the ratio map of Figure~\ref{fig_ratio}. Column 2: Integrated HCN~(4--3) intensity at an emission region. Column 3: Integrated HCO$^+$~(4--3) intensity at an emission region. Column 4: The HCN~(4--3)/HCO$^+$~(4--3) integrated intensity ratio.; We only show the statistical error in this table. The systematic error of absolute flux calibration is estimated to be $\sim$ 5\% in band 3 and $\sim$ 10\% in band 7.  \dag represents boxes contained the obscured AGN defined by \citetalias{ion13}.  \dag\dag represents boxes contained the nuclear starbursts defined by \citetalias{ion13}.  \dag\dag\dag represents boxes at the overlap region.}
\end{deluxetable*}

\begin{deluxetable*}{lrrrrrrrrcrl}
\tabletypesize{\scriptsize}
\tablecaption{Peak Brightness Temperature of $^{12}$CO~(1--0), $^{13}$CO~(1--0), and $^{12}$CO~(3--2) Emission \label{table_T_R}}
\tablewidth{0pt}
\tablehead{
ID & Peak $T_{\rm{b, CO~(1-0)}}$ & Peak $T_{\rm{b, CO~(3-2)}}$ & Peak $T_{\rm{b, ^{13}CO~(1-0)}}$ & Peak $R_{3-2/1-0}$ & Peak $R_{12/13}$ \\
 & [K] & [K] & [K] & &\\
 (1) &(2) &(3) &(4) &(5) &(6)
 }
\startdata
R1&0.40 $\pm$ 0.07&0.13 $\pm$ 0.01&$<$ 0.09&0.32 $\pm$ 0.07&$>$ 4\\
R2&0.69 $\pm$ 0.07&0.25 $\pm$ 0.01&$<$ 0.09&0.37 $\pm$ 0.04&$>$ 8\\
R3&0.61 $\pm$ 0.07&0.07 $\pm$ 0.01&$<$ 0.09&0.12 $\pm$ 0.02&$>$ 7\\
R4&0.98 $\pm$ 0.07&0.30 $\pm$ 0.01&$<$ 0.09&0.30 $\pm$ 0.02&$>$ 11\\
R5&1.19 $\pm$ 0.07&0.42 $\pm$ 0.01&0.06 $\pm$ 0.03&0.35 $\pm$ 0.02&20 $\pm$ 2\\
R6&0.87 $\pm$ 0.07&0.21 $\pm$ 0.01&$<$ 0.09&0.24 $\pm$ 0.02&$>$ 10\\
R7a&1.34 $\pm$ 0.07&0.32 $\pm$ 0.01&$<$ 0.09&0.24 $\pm$ 0.01&$>$ 15\\
R7b&0.45 $\pm$ 0.07&0.36 $\pm$ 0.01&0.06 $\pm$ 0.03&0.79 $\pm$ 0.12&7 $\pm$ 1\\
R8a&0.21 $\pm$ 0.07&$<$ 0.03&$<$ 0.09&$>$ 0.14&$>$ 2\\
R8b&0.96 $\pm$ 0.07&0.24 $\pm$ 0.01&0.08 $\pm$ 0.03&0.25 $\pm$ 0.02&11 $\pm$ 1\\
R9a&0.40 $\pm$ 0.07&0.12 $\pm$ 0.01&$<$ 0.09&0.30 $\pm$ 0.06&$>$ 4\\
R9b&2.60 $\pm$ 0.07&0.99 $\pm$ 0.01&0.08 $\pm$ 0.03&0.38 $\pm$ 0.01&33 $\pm$ 2\\
R10a&0.80 $\pm$ 0.07&0.24 $\pm$ 0.01&0.05 $\pm$ 0.03&0.30 $\pm$ 0.03&17 $\pm$ 2\\
R10b&2.81 $\pm$ 0.07&1.27 $\pm$ 0.01&0.09 $\pm$ 0.03&0.45 $\pm$ 0.01&30 $\pm$ 1\\
R11a&2.11 $\pm$ 0.07&0.82 $\pm$ 0.01&0.12 $\pm$ 0.03&0.39 $\pm$ 0.01&18 $\pm$ 1\\
R11b&3.34 $\pm$ 0.07&1.36 $\pm$ 0.01&0.04 $\pm$ 0.03&0.41 $\pm$ 0.01&89 $\pm$ 7\\
R12a&0.55 $\pm$ 0.07&0.23 $\pm$ 0.01&$<$ 0.09&0.41 $\pm$ 0.06&$>$ 6\\
R12b&1.30 $\pm$ 0.07&0.41 $\pm$ 0.01&$<$ 0.09&0.32 $\pm$ 0.02&$>$ 14\\
R13a&1.12 $\pm$ 0.07&0.34 $\pm$ 0.01&0.06 $\pm$ 0.03&0.30 $\pm$ 0.02&18 $\pm$ 1\\
R13b&0.67 $\pm$ 0.07&0.30 $\pm$ 0.01&$<$ 0.09&0.45 $\pm$ 0.05&$>$ 7\\
R14&1.61 $\pm$ 0.07&0.39 $\pm$ 0.01&0.08 $\pm$ 0.03&0.24 $\pm$ 0.01&20 $\pm$ 1\\
R15&1.77 $\pm$ 0.07&0.42 $\pm$ 0.01&0.08 $\pm$ 0.03&0.24 $\pm$ 0.01&22 $\pm$ 1\\
R16&1.38 $\pm$ 0.07&0.33 $\pm$ 0.01&0.12 $\pm$ 0.03&0.24 $\pm$ 0.01&12 $\pm$ 1\\
R17&1.64 $\pm$ 0.07&0.35 $\pm$ 0.01&0.05 $\pm$ 0.03&0.21 $\pm$ 0.01&36 $\pm$ 3\\
R18$^{\dag}$&3.52 $\pm$ 0.07&1.79 $\pm$ 0.01&0.14 $\pm$ 0.03&0.51 $\pm$ 0.01&25 $\pm$ 1\\
R19a&1.01 $\pm$ 0.07&0.54 $\pm$ 0.01&0.09 $\pm$ 0.03&0.54 $\pm$ 0.04&11 $\pm$ 1\\
R19b&1.99 $\pm$ 0.07&0.98 $\pm$ 0.01&0.09 $\pm$ 0.03&0.49 $\pm$ 0.02&23 $\pm$ 1\\
R20a&3.33 $\pm$ 0.07&1.99 $\pm$ 0.01&0.11 $\pm$ 0.03&0.60 $\pm$ 0.01&30 $\pm$ 1\\
R20b&1.13 $\pm$ 0.07&0.67 $\pm$ 0.01&$<$ 0.09&0.59 $\pm$ 0.04&$>$ 13\\
R21a$^{\dag\dag}$&5.53 $\pm$ 0.07&4.49 $\pm$ 0.01&0.23 $\pm$ 0.03&0.81 $\pm$ 0.01&24 $\pm$ 1\\
R21b$^{\dag\dag}$&1.22 $\pm$ 0.07&0.95 $\pm$ 0.01&$<$ 0.09&0.78 $\pm$ 0.05&$>$ 14\\
R21c$^{\dag\dag}$&0.40 $\pm$ 0.07&0.72 $\pm$ 0.01&$<$ 0.09&1.80 $\pm$ 0.05&$>$ 4\\
R22a&3.60 $\pm$ 0.07&1.93 $\pm$ 0.01&0.12 $\pm$ 0.03&0.54 $\pm$ 0.01&29 $\pm$ 1\\
R22b&3.83 $\pm$ 0.07&2.61 $\pm$ 0.01&0.16 $\pm$ 0.03&0.68 $\pm$ 0.01&24 $\pm$ 1\\
R23a&4.17 $\pm$ 0.07&2.56 $\pm$ 0.01&0.15 $\pm$ 0.03&0.61 $\pm$ 0.01&28 $\pm$ 1\\
R23b&3.39 $\pm$ 0.07&1.87 $\pm$ 0.01&0.12 $\pm$ 0.03&0.55 $\pm$ 0.01&28 $\pm$ 1\\
R24a&1.23 $\pm$ 0.07&0.46 $\pm$ 0.01&0.05 $\pm$ 0.03&0.37 $\pm$ 0.02&23 $\pm$ 2\\
R24b&1.60 $\pm$ 0.07&0.96 $\pm$ 0.01&0.06 $\pm$ 0.03&0.60 $\pm$ 0.03&28 $\pm$ 2\\
R24c&1.78 $\pm$ 0.07&0.58 $\pm$ 0.01&0.06 $\pm$ 0.03&0.32 $\pm$ 0.01&32 $\pm$ 2\\
R25a&4.89 $\pm$ 0.07&3.18 $\pm$ 0.01&0.23 $\pm$ 0.03&0.65 $\pm$ 0.01&21 $\pm$ 1\\
R25b&1.77 $\pm$ 0.07&0.59 $\pm$ 0.01&0.06 $\pm$ 0.03&0.33 $\pm$ 0.01&31 $\pm$ 2\\
R26a&4.49 $\pm$ 0.07&2.57 $\pm$ 0.01&0.14 $\pm$ 0.03&0.57 $\pm$ 0.01&33 $\pm$ 1\\
R26b&1.07 $\pm$ 0.07&0.31 $\pm$ 0.01&0.07 $\pm$ 0.03&0.29 $\pm$ 0.02&15 $\pm$ 1\\
R27a&1.96 $\pm$ 0.07&1.31$\pm$ 0.01&0.04 $\pm$ 0.03&0.67 $\pm$ 0.02&47 $\pm$ 4\\
R27b&1.53 $\pm$ 0.07&0.75 $\pm$ 0.01&0.13 $\pm$ 0.03&0.49 $\pm$ 0.02&12 $\pm$ 1\\
R28a&0.76 $\pm$ 0.07&0.37 $\pm$ 0.01&0.05 $\pm$ 0.03&0.48 $\pm$ 0.05&16 $\pm$ 2\\
R28b&0.86 $\pm$ 0.07&0.54 $\pm$ 0.01&$<$ 0.09&0.62 $\pm$ 0.05&$>$ 10\\
R29&$<$ 0.21&$<$ 0.03&$<$ 0.09&\nodata&\nodata\\
R30a&0.77 $\pm$ 0.07&0.39 $\pm$ 0.01&$<$ 0.09&0.51 $\pm$ 0.05&$>$ 9\\
R30b&1.39 $\pm$ 0.07&0.69 $\pm$ 0.01&$<$ 0.09&0.50 $\pm$ 0.03&$>$ 15\\
R31a&0.72 $\pm$ 0.07&0.33 $\pm$ 0.01&0.06 $\pm$ 0.03&0.46 $\pm$ 0.03&13 $\pm$ 1\\
R31b&0.55 $\pm$ 0.07&0.24 $\pm$ 0.01&$<$ 0.09&0.44 $\pm$ 0.06&$>$ 6\\
R32&0.55 $\pm$ 0.07&0.29 $\pm$ 0.01&$<$ 0.09&0.52 $\pm$ 0.07&$>$ 6\\
R33&0.35 $\pm$ 0.07&0.14 $\pm$ 0.01&$<$ 0.09&0.41 $\pm$ 0.09&$>$ 4\\
R34&0.83 $\pm$ 0.07&0.37 $\pm$ 0.01&0.05 $\pm$ 0.03&0.45 $\pm$ 0.04&18 $\pm$ 2\\
R35&1.42 $\pm$ 0.07&0.30 $\pm$ 0.01&$<$ 0.09&0.21 $\pm$ 0.01&$>$ 16\\
R36a&0.65 $\pm$ 0.07&0.24 $\pm$ 0.01&0.07 $\pm$ 0.03&0.37 $\pm$ 0.04&9 $\pm$ 1\\
R36b&0.47 $\pm$ 0.07&0.22 $\pm$ 0.01&$<$ 0.09&0.47 $\pm$ 0.07&$>$ 5\\
R37a&0.79 $\pm$ 0.07&0.15 $\pm$ 0.01&0.06 $\pm$ 0.03&0.19 $\pm$ 0.02&14 $\pm$ 1\\
R37b&0.71 $\pm$ 0.07&0.18 $\pm$ 0.01&$<$ 0.09&0.26 $\pm$ 0.03&$>$ 8\\
R38&0.38 $\pm$ 0.07&0.16 $\pm$ 0.01&0.04 $\pm$ 0.03&0.42 $\pm$ 0.08&9 $\pm$ 2\\
R39a$^{\dag\dag\dag}$&4.91 $\pm$ 0.07&3.15 $\pm$ 0.01&0.24 $\pm$ 0.03&0.64 $\pm$ 0.01&21 $\pm$ 1\\
R39b$^{\dag\dag\dag}$&1.96 $\pm$ 0.07&0.81 $\pm$ 0.01&0.06 $\pm$ 0.03&0.41 $\pm$ 0.02&31 $\pm$ 2\\
\enddata
\tablecomments{Column 1: These numbers are labeled at the ratio map of Figure~\ref{fig_ratio}. Column 2: Peak $^{12}$CO~(1--0) brightness temperature at an emission region. Column 3: Peak $^{12}$CO~(3--2) brightness temperature at an emission region. Column 4: Peak $^{13}$CO~(3--2) brightness temperature at an emission region. Column 5: The $^{12}$CO~(3--2)/CO~(1--0) brightness temperature ratio. Column 6: The $^{12}$CO~(1--0)/$^{13}$CO~(1--0) brightness temperature ratio.; We only show the statistical error in this table. The systematic error of absolute flux calibration is estimated to be $\sim$ 5\% in band 3 and $\sim$ 10\% in band 7.  \dag represents boxes contained the obscured AGN defined by \citetalias{ion13}.  \dag\dag represents boxes contained the nuclear starbursts defined by \citetalias{ion13}.  \dag\dag\dag represents boxes at the overlap region.}
\end{deluxetable*}

\begin{deluxetable*}{lrrrrrrrrrcrl}
\tabletypesize{\scriptsize}
\tablecaption{Peak Brightness Temperature of HCN~(4--3) and HCO$^+$~(4--3) \label{table_T_S}}
\tablewidth{0pt}
\tablehead{
ID & $T_{\rm{b, HCN~(4-3)}}$ & $T_{\rm{b, HCO^+(4-3)}}$ & $R_{\rm{HCN/HCO^+}}$ \\
 & [K] & [K] &\\
 (1) &(2) &(3) &(4) \\
 }
\startdata
S0$^{\dag}$&0.32 $\pm$ 0.05&0.27 $\pm$ 0.05&1.18 $\pm$ 0.29\\
S1&$<$ 0.15&0.20 $\pm$ 0.05&$<$ 0.75\\
S2&$<$ 0.15&0.16 $\pm$ 0.05&$<$ 0.94\\
S3$^{\dag\dag}$&0.20 $\pm$ 0.05&0.65 $\pm$ 0.05&0.31 $\pm$ 0.08\\
S4&0.10 $\pm$ 0.05&0.51 $\pm$ 0.05&0.19 $\pm$ 0.10\\
S5$^{\dag\dag}$&0.32 $\pm$ 0.05&0.91 $\pm$ 0.05&0.35 $\pm$ 0.06\\
S6&0.14 $\pm$ 0.05&0.55 $\pm$ 0.05&0.26 $\pm$ 0.09\\
S7&$<$ 0.15&0.24 $\pm$ 0.05&$<$ 0.63\\
S8&$<$ 0.15&0.14 $\pm$ 0.05&$<$ 1.07\\
S9&$<$ 0.15&0.17 $\pm$ 0.05&$<$ 0.88\\
S10&0.10 $\pm$ 0.05&0.17 $\pm$ 0.05&0.60 $\pm$ 0.34\\
S11$^{\dag\dag\dag}$&$<$ 0.15&0.18 $\pm$ 0.05&$<$ 0.83\\
S12$^{\dag\dag\dag}$&$<$ 0.15&0.24 $\pm$ 0.05&$<$ 0.63\\
S13&$<$ 0.15&0.20 $\pm$ 0.05&$<$ 0.75\\
S14&$<$ 0.15&0.23 $\pm$ 0.05&$<$ 0.65
\enddata
\tablecomments{Column 1: These numbers are labeled at the ratio map of Figure~\ref{fig_ratio}. Column 2: Peak HCN~(4--3) brightness temperature at an emission region. Column 3: Peak HCO$^+$~(4--3) brightness temperature at an emission region. Column 4: The HCN~(4--3)/HCO$^+$~(4--3) brightness temperature ratio.; We only show the statistical error in this table. The systematic error of absolute flux calibration is estimated to be $\sim$ 5\% in band 3 and $\sim$ 10\% in band 7.  \dag represents boxes contained the obscured AGN defined by \citetalias{ion13}.  \dag\dag represents boxes contained the nuclear starbursts defined by \citetalias{ion13}.  \dag\dag\dag represents boxes at the overlap region.}
\end{deluxetable*}

\begin{deluxetable*}{lrrrrrrrrrcrl}
\tabletypesize{\scriptsize}
\tablecaption{Peak Brightness Temperature of $^{12}$CO~(3--2), HCN~(4--3), and HCO$^+$~(4--3) \label{table_E}}
\tablewidth{0pt}
\tablehead{
ID & Peak $T_{\rm{b, CO~(3-2)}}$ & Peak $T_{\rm{b, HCN~(4-3)}}$ & Peak $T_{\rm{b, HCO^+(4-3)}}$ & Peak $R_{\rm{HCO^+/CO}}$ & Peak $R_{\rm{HCN/HCO^+}}$ \\
 & [K] & [K] & [K] & &\\
 (1) &(2) &(3) &(4) &(5) &(6)
 }
\startdata
E0$^{\dag}$&2.94 $\pm$ 0.04&0.06 $\pm$ 0.01&0.04 $\pm$ 0.01&0.014 $\pm$ 0.004&1.55 $\pm$ 0.49\\
E1$^{\dag\dag}$&5.82 $\pm$ 0.04&0.11 $\pm$ 0.01&0.32 $\pm$ 0.01&0.056 $\pm$ 0.002&0.36 $\pm$ 0.03\\
E2$^{\dag\dag\dag}$&4.71 $\pm$ 0.04&0.04 $\pm$ 0.01&0.08 $\pm$ 0.01&0.017 $\pm$ 0.002&0.52 $\pm$ 0.14
\enddata
\tablecomments{Column 1: These numbers are regions where were convolved to the 1\farcs2 $\times$ 1\farcs0 resolution (P.A. = 119 deg.). Column 2: Peak $^{12}$CO~(3--2) brightness temperature at an emission region. Column 3: Peak HCN~(4--3) brightness temperature at an emission region. Column 4: Peak HCO$^+$~(4--3) brightness temperature at an emission region. Column 5: The HCO$^+$~(4--3)/CO~(3--2) brightness temperature ratio. Column 6: The HCN~(4--3)/HCO$^+$~(4--3) brightness temperature ratio.; We only show the statistical error in this table. The systematic error of absolute flux calibration is estimated to be $\sim$ 5\% in band 3 and $\sim$ 10\% in band 7.  \dag represents boxes contained the obscured AGN defined by \citetalias{ion13}.  \dag\dag represents boxes contained the nuclear starbursts defined by \citetalias{ion13}.  \dag\dag\dag represents boxes at the overlap region.}
\end{deluxetable*}

\begin{deluxetable*}{lrrrrrrrrrcrl}
\tabletypesize{\scriptsize}
\tablecaption{Parameters Used for RADEX Modelings \label{table_radex_parm}}
\tablewidth{0pt}
\tablehead{
Case & $T_{\rm{kin}}$ & $\log$ $n_{\rm{H_2}}$ & $\log$ $N({\rm{H_2}})$ & abundance ratio & box size \\
& [K] & [cm$^{-3}$] & [cm$^{-2}$] && [pc] \\
(1) & (2) & (3) & (4) & (5) & (6)
 }
\startdata
1 &5 - 300 (5) &2 - 5 (0.1) &18 - 22 (0.1) &70 & 800\\
2 &5 - 400 (5) &3 - 7 (0.1) &21.2, 21.6, 21.5 &1 - 10 (1) & 320\\
\enddata
\tablecomments{Column 2, 3, 4, 5: Fitting ranges (steps) of $T_{\rm{kin}}$, $n_{\rm{H_2}}$, $N(\rm{H_2})$, and abundance ratios. The abundance ratios mean the [CO]/[$^{13}$CO] and [HCN]/[HCO$^+$] in case 1 and 2, respectively. We fixed the [CO]/[$^{13}$CO] in case 1 \citep[Galactic value;][]{w&r94} and the $N(\rm{H_2})$ (E0, E1, and E2, respectively) in case 2. Column 6: Box sizes.; We consider the statistical error and the systematic error in these calculations. The systematic error of absolute flux calibration is estimated to be $\sim$ 5\% in band 3 and $\sim$ 10\% in band 7.}
\end{deluxetable*}

\begin{deluxetable*}{lrrrrrrrrrcrl}
\tabletypesize{\scriptsize}
\tablecaption{Gas, Dust, and Star-forming Properties (R1 - R39) \label{table_SF_CO}}
\tablewidth{0pt}
\tablehead{
ID & $M_{\rm{H_2}}$ & $L_{\rm{Pa\alpha}}$ & SFR &$\tau_{\rm{gas}}$ & $S_{340}$ & $M_{\rm{dust}}$ & $M_{\rm{H_2}}/M_{\rm{dust}}$ & $M_{\rm{ISM}}$\\
 & [$\times$10$^7$ M$_{\odot}$] &[$\times$10$^{38}$ erg s$^{-1}$] &[M$_{\odot}$ yr$^{-1}$] & [Gyr] & [mJy] & [$\times$10$^{4}$ M$_{\odot}$] & & [$\times$10$^7$ M$_{\odot}$] \\
 (1) &(2) &(3) &(4) &(5) &(6) &(7) &(8) &(9)
 }
\startdata
R1&3.0 $\pm$ 0.6&$<$ 24.6&$<$ 0.15&$>$ 0.20&$<$ 0.51&$<$ 17&$>$ \phantom{0}180&$<$ 3.6\\
R2&2.0 $\pm$ 0.4&46.0 $\pm$ 8.2&0.29 $\pm$ 0.05&0.07 $\pm$ 0.02&$<$ 0.51&$<$ 17&$>$ \phantom{0}119&$<$ 3.6\\
R3&5.1 $\pm$ 0.8&11.0 $\pm$ 8.2&0.07 $\pm$ 0.05&0.75 $\pm$ 0.57&$<$ 0.51&$<$ 17&$>$ \phantom{0}306&$<$ 3.6\\
R4&5.0 $\pm$ 0.7&9.3 $\pm$ 8.2&0.06 $\pm$ 0.05&0.88 $\pm$ 0.78&$<$ 0.51&$<$ 17&$>$ \phantom{0}299&$<$ 3.6\\
R5&4.9 $\pm$ 0.7&41.8 $\pm$ 8.2&0.26 $\pm$ 0.05&0.19 $\pm$ 0.05&$<$ 0.51&$<$ 17&$>$ \phantom{0}293&$<$ 3.6\\
R6&5.2 $\pm$ 0.9&60.7 $\pm$ 8.2&0.38 $\pm$ 0.05&0.14 $\pm$ 0.03&$<$ 0.51&$<$ 17&$>$ \phantom{0}310&$<$ 3.6\\
R7&7.5 $\pm$ 0.9&172.1 $\pm$ 8.2&1.07 $\pm$ 0.05&0.07 $\pm$ 0.01&0.61 $\pm$ 0.18&20 $\pm$ \phantom{0}6&371 $\pm$ \phantom{0}118&5.2 $\pm$ 1.4\\
R8&5.3 $\pm$ 0.8&69.8 $\pm$ 8.2&0.43 $\pm$ 0.05&0.12 $\pm$ 0.02&$<$ 0.51&$<$ 17&$>$ \phantom{0}314&$<$ 3.6\\
R9&13.6 $\pm$ 1.2&125.1 $\pm$ 8.2&0.78 $\pm$ 0.05&0.18 $\pm$ 0.02&$<$ 0.51&$<$ 17&$>$ \phantom{0}811&$<$ 3.6\\
R10&18.0 $\pm$ 1.3&49.0 $\pm$ 8.2&0.30 $\pm$ 0.05&0.59 $\pm$ 0.11&$<$ 0.51&$<$ 17&$>$ 1074&$<$ 3.6\\
R11&23.1 $\pm$ 1.6&106.9 $\pm$ 8.2&0.66 $\pm$ 0.05&0.35 $\pm$ 0.04&$<$ 0.51&$<$ 17&$>$ 1376&$<$ 3.6\\
R12&6.5 $\pm$ 0.9&25.5 $\pm$ 8.2&0.16 $\pm$ 0.05&0.41 $\pm$ 0.14&$<$ 0.51&$<$ 17&$>$ \phantom{0}390&$<$ 3.6\\
R13&7.4 $\pm$ 1.0&25.9 $\pm$ 8.2&0.16 $\pm$ 0.05&0.46 $\pm$ 0.16&$<$ 0.51&$<$ 17&$>$ \phantom{0}441&$<$ 3.6\\
R14&6.7 $\pm$ 0.8&25.5 $\pm$ 8.2&0.16 $\pm$ 0.05&0.42 $\pm$ 0.14&$<$ 0.51&$<$ 17&$>$ \phantom{0}398&$<$ 3.6\\
R15&7.1 $\pm$ 0.9&14.7 $\pm$ 8.2&0.09 $\pm$ 0.05&0.78 $\pm$ 0.44&$<$ 0.51&$<$ 17&$>$ \phantom{0}425&$<$ 3.6\\
R16&7.9 $\pm$ 0.9&15.9 $\pm$ 8.2&0.10 $\pm$ 0.05&0.80 $\pm$ 0.42&$<$ 0.51&$<$ 17&$>$ \phantom{0}473&$<$ 3.6\\
R17&9.1 $\pm$ 1.0&35.2 $\pm$ 8.2&0.22 $\pm$ 0.05&0.42 $\pm$ 0.11&$<$ 0.51&$<$ 17&$>$ \phantom{0}545&$<$ 3.6\\
R18$^{\dag}$&23.5 $\pm$ 1.6&277.4 $\pm$ 8.2&1.72 $\pm$ 0.05&0.14 $\pm$ 0.01&5.57 $\pm$ 0.58&183 $\pm$ 19&128 $\pm$ \phantom{00}16&47.1 $\pm$ 4.9\\
R19&15.7 $\pm$ 1.3&150.8 $\pm$ 8.2&0.94 $\pm$ 0.05&0.17 $\pm$ 0.02&2.70 $\pm$ 0.32&89 $\pm$ 11&177 $\pm$ \phantom{00}25&22.9 $\pm$ 2.6\\
R20&26.0 $\pm$ 1.7&78.1 $\pm$ 8.2&0.48 $\pm$ 0.05&0.54 $\pm$ 0.07&2.40 $\pm$ 0.29&79 $\pm$ 10&329 $\pm$ \phantom{00}46&20.3 $\pm$ 2.4\\
R21$^{\dag\dag}$&46.9 $\pm$ 2.7&303.2 $\pm$ 8.2&1.88 $\pm$ 0.05&0.25 $\pm$ 0.02&8.55 $\pm$ 0.87&281 $\pm$ 29&167 $\pm$ \phantom{00}19&72.3 $\pm$ 7.3\\
R22&40.7 $\pm$ 2.4&197.1 $\pm$ 8.2&1.22 $\pm$ 0.05&0.33 $\pm$ 0.02&4.08 $\pm$ 0.44&134 $\pm$ 15&303 $\pm$ \phantom{00}37&34.5 $\pm$ 3.7\\
R23&47.8 $\pm$ 2.7&217.5 $\pm$ 8.2&1.35 $\pm$ 0.05&0.35 $\pm$ 0.02&3.60 $\pm$ 0.40&119 $\pm$ 13&403 $\pm$ \phantom{00}50&30.5 $\pm$ 3.3\\
R24&22.9 $\pm$ 1.7&145.5 $\pm$ 8.2&0.90 $\pm$ 0.05&0.25 $\pm$ 0.02&1.28 $\pm$ 0.21&42 $\pm$ \phantom{0}7&543 $\pm$ \phantom{00}99&10.8 $\pm$ 1.7\\
R25&35.8 $\pm$ 2.2&281.2 $\pm$ 8.2&1.74 $\pm$ 0.05&0.21 $\pm$ 0.01&3.10 $\pm$ 0.35&102 $\pm$ 12&351 $\pm$ \phantom{00}46&26.2 $\pm$ 2.9\\
R26&29.0 $\pm$ 1.9&250.1 $\pm$ 8.2&1.55 $\pm$ 0.05&0.19 $\pm$ 0.01&1.66 $\pm$ 0.24&54 $\pm$ \phantom{0}8&533 $\pm$ \phantom{00}84&14.0 $\pm$ 1.9\\
R27&17.7 $\pm$ 1.4&507.8 $\pm$ 8.2&3.15 $\pm$ 0.05&0.06 $\pm$ 0.01&1.42 $\pm$ 0.22&47 $\pm$ \phantom{0}7&379 $\pm$ \phantom{00}66&12.0 $\pm$ 1.7\\
R28&8.2 $\pm$ 1.0&246.3 $\pm$ 8.2&1.53 $\pm$ 0.05&0.05 $\pm$ 0.01&0.73 $\pm$ 0.18&24 $\pm$ \phantom{0}6&344  $\pm$ \phantom{00}97&6.1 $\pm$ 1.4\\
R29&$<$ 1.4&147.8 $\pm$ 8.2&0.92 $\pm$ 0.05&$<$ 0.02&0.76 $\pm$ 0.19&25 $\pm$ \phantom{0}6&$<$ \phantom{00}55&6.4 $\pm$ 1.4\\
R30&9.1 $\pm$ 1.1&40.2 $\pm$ 8.2&0.25 $\pm$ 0.05&0.36 $\pm$ 0.09&0.79 $\pm$ 0.19&26 $\pm$ \phantom{0}6&348 $\pm$ \phantom{00}92&6.7 $\pm$ 1.4\\
R31&7.0 $\pm$ 1.0&19.0 $\pm$ 8.2&0.12 $\pm$ 0.05&0.59 $\pm$ 0.27&$<$ 0.51&$<$ 17&$>$ \phantom{0}418&$<$ 3.6\\
R32&3.3 $\pm$ 0.5&60.3 $\pm$ 8.2&0.37 $\pm$ 0.05&0.09 $\pm$ 0.02&$<$ 0.51&$<$ 17&$>$ \phantom{0}198&$<$ 3.6\\
R33&2.0 $\pm$ 0.4&55.0 $\pm$ 8.2&0.34 $\pm$ 0.05&0.06 $\pm$ 0.02&$<$ 0.51&$<$ 17&$>$ \phantom{0}122&$<$ 3.6\\
R34&4.0 $\pm$ 0.7&10.9 $\pm$ 8.2&0.07 $\pm$ 0.05&0.59 $\pm$ 0.45&$<$ 0.51&$<$ 17&$>$ \phantom{0}238&$<$ 3.6\\
R35&7.6 $\pm$ 0.9&$<$ 24.6&$<$ 0.15&$>$ 0.51&$<$ 0.51&$<$ 17&$>$ \phantom{0}453&$<$ 3.6\\
R36&7.9 $\pm$ 1.1&$<$ 24.6&$<$ 0.15&$>$ 0.52&$<$ 0.51&$<$ 17&$>$ \phantom{0}469&$<$ 3.6\\
R37&6.9 $\pm$ 1.0&$<$ 24.6&$<$ 0.15&$>$ 0.46&$<$ 0.51&$<$ 17&$>$ \phantom{0}410&$<$ 3.6\\
R38&3.8 $\pm$ 0.7&15.2 $\pm$ 8.2&0.10 $\pm$ 0.05&0.40 $\pm$ 0.22&$<$ 0.51&$<$ 17&$>$ \phantom{0}225&$<$ 3.6\\
R39$^{\dag\dag\dag}$&39.1 $\pm$ 2.3&273.6 $\pm$ 8.2&1.70 $\pm$ 0.05&0.23 $\pm$ 0.02&3.50 $\pm$ 0.58&115 $\pm$ 19&339 $\pm$ \phantom{00}60&29.6 $\pm$ 3.2\\
\enddata
\tablecomments{Column 2: The molecular gas mass derived using the conversion factor $\alpha_{\rm{CO}}$ = 0.8 (K km s$^{-1}$ pc$^2$)$^{-1}$. Column 3: The Pa$\alpha$ flux \citep{tat12}. Column 4: The star formation rate derived using the conversion factor SFR/$L_{\rm{Pa\alpha}}$ = 6.2 $\times$ 10$^{-41}$ [erg s$^{-1}$/(M$_{\odot}$ yr$^{-1}$)$^{-1}$]. Column 5: The gas depletion time (= $\Sigma_{\rm{H_2}}/\Sigma_{\rm{SFR}}$). Column 6: The 340 GHz continuum flux. Column 7: The dust mass derived using the equation (7). We adopt the \citet{d&l84} dust model for $\kappa_{340}$ to derive the $M_{\rm{dust}}$. Column 9: The ISM mass derived using the equation (8).; We consider the statistical error and the systematic error in this table. The systematic error of absolute flux calibration is estimated to be $\sim$ 5\% in band 3 and $\sim$ 10\% in band 7.  \dag represents boxes contained the obscured AGN defined by \citetalias{ion13}.  \dag\dag represents boxes contained the nuclear starbursts defined by \citetalias{ion13}.  \dag\dag\dag represents boxes at the overlap region.}
\end{deluxetable*}

\begin{deluxetable*}{lrrrrrrrrrcrl}
\tabletypesize{\scriptsize}
\tablecaption{Gas, Dust, and Star-forming Properties (S0 - S14) \label{table_SF_HCN}}
\tablewidth{0pt}
\tablehead{
ID & $M_{\rm{dense}}$ & $L_{\rm{Pa\alpha}}$ & SFR &$\tau_{\rm{gas}}$ & $S_{340}$ & $M_{\rm{dust}}$ & $M_{\rm{dense}}/M_{\rm{dust}}$\\
 & [$\times$10$^6$ M$_{\odot}$] &[$\times$10$^{38}$ erg s$^{-1}$] &[M$_{\odot}$ yr$^{-1}$] & [Myr] & [mJy] & [$\times$10$^{4}$ M$_{\odot}$] &\\
 (1) &(2) &(3) &(4) &(5) &(6) &(7) &(8)
 }
\startdata
S0$^{\dag}$&38.3 $\pm$ 4.7 &322 $\pm$ 12 &2.00 $\pm$ 0.07 &19.2 $\pm$ 2.5 &1.71 $\pm$ 0.18 &56.1 $\pm$ 6.0&68 $\pm$ 11 \\
S1&3.2 $\pm$ 1.8 &240 $\pm$ 12 &1.49 $\pm$ 0.07 &2.2 $\pm$ 1.2 &0.25 $\pm$ 0.07 &8.1 $\pm$ 2.3&40 $\pm$ 25 \\
S2&$<$ 4.3 &$<$ 36 &$<$ 0.21 &\nodata &0.10 $\pm$ 0.07 &3.1 $\pm$ 2.2&$<$ 137 \\
S3$^{\dag\dag}$&13.9 $\pm$ 2.6 &663 $\pm$ 12 &4.11 $\pm$ 0.07 &3.4 $\pm$ 0.6 &1.28 $\pm$ 0.14 &42.1 $\pm$ 4.7&33 $\pm$  \phantom{0}7 \\
S4&3.6 $\pm$ 2.2 &446 $\pm$ 12 &2.76 $\pm$ 0.07 &1.3 $\pm$ 0.8 &0.63 $\pm$ 0.09 &20.6 $\pm$ 3.0&17 $\pm$ 11 \\
S5$^{\dag\dag}$&23.2 $\pm$ 3.6 &606 $\pm$ 12 &3.75 $\pm$ 0.07 &6.2 $\pm$ 0.7 &2.56 $\pm$ 0.26 &84.4 $\pm$ 8.7&27 $\pm$  \phantom{0}5 \\
S6&11.5 $\pm$ 2.6 &268 $\pm$ 12 &1.66 $\pm$ 0.07 &6.9 $\pm$ 1.6 &0.90 $\pm$ 0.11 &29.7 $\pm$ 3.7&39 $\pm$  10 \\
S7&5.4 $\pm$ 2.3 &138 $\pm$ 12 &0.85 $\pm$ 0.07 &6.4 $\pm$ 2.7 &0.40 $\pm$ 0.08 &13.0 $\pm$ 2.5&42 $\pm$ 19 \\
S8&1.8 $\pm$ 1.6 &158 $\pm$ 12 &0.98 $\pm$ 0.07 &1.8 $\pm$ 1.6 &0.32 $\pm$ 0.07 &10.5 $\pm$ 2.4&17 $\pm$ 16 \\
S9&2.0 $\pm$ 1.8 &106 $\pm$ 12 &0.66 $\pm$ 0.07 &3.1 $\pm$ 2.8 &0.28 $\pm$ 0.07 &9.2 $\pm$ 2.3&22 $\pm$ 21 \\
S10&$<$ 4.3 &203 $\pm$ 12 &1.26 $\pm$ 0.07 &$<$ 3.4 &0.34 $\pm$ 0.07 &11.3 $\pm$ 2.4&$<$ 38 \\
S11$^{\dag\dag\dag}$&$<$ 4.3 &242 $\pm$ 12 &1.50 $\pm$ 0.07 &$<$ 2.9 &0.21 $\pm$ 0.07 &7.0 $\pm$ 2.3&$<$ 61 \\
S12$^{\dag\dag\dag}$&$<$ 1.4 &323 $\pm$ 12 &2.00 $\pm$ 0.07 &$<$ 2.1 &0.46 $\pm$ 0.08 &15.3 $\pm$ 2.6&$<$  28 \\
S13&$<$ 4.3 &291 $\pm$ 12 &1.81 $\pm$ 0.07 &$<$ 2.4 &0.46 $\pm$ 0.08 &15.2 $\pm$ 2.6&$<$  28 \\
S14&$<$ 1.4 &219 $\pm$ 12 &1.36 $\pm$ 0.07 &$<$ 3.2 &0.27 $\pm$ 0.07 &9.0 $\pm$ 2.3&$<$ 48 \\
\enddata
\tablecomments{Column 2: The molecular gas mass derived using the conversion factor $\alpha_{\rm{HCN}}$ = 10/0.63 (K km s$^{-1}$ pc$^2$)$^{-1}$. Column 3: The Pa$\alpha$ flux \citep{tat12}. Column 4: The star formation rate derived using the conversion factor SFR/$L_{\rm{Pa\alpha}}$ = 6.2 $\times$ 10$^{-41}$ [erg s$^{-1}$/(M$_{\odot}$ yr$^{-1}$)$^{-1}$]. Column 5: The gas depletion time (= $\Sigma_{\rm{dense}}/\Sigma_{\rm{SFR}}$). Column 6: The 340 GHz continuum flux. Column 7: The dust mass derived using the equation (8). We adopt the \citet{d&l84} dust model for $\kappa_{340}$ to derive the $M_{\rm{dust}}$.; We consider the statistical error and the systematic error in this table. The systematic error of absolute flux calibration is estimated to be $\sim$ 5\% in band 3 and $\sim$ 10\% in band 7.  \dag represents boxes contained the obscured AGN defined by \citetalias{ion13}.  \dag\dag represents boxes contained the nuclear starbursts defined by \citetalias{ion13}.  \dag\dag\dag represents boxes at the overlap region.}
\end{deluxetable*}

\begin{deluxetable*}{lrrrrrrrrrrrl}
\tabletypesize{\scriptsize}
\tablecaption{RADEX Results of Case 1 \label{table_RADEX}}
\tablewidth{0pt}
\tablehead{
ID & \multicolumn{2}{c}{$T_{\rm{kin}}$} & & \multicolumn{2}{c}{$\log\:$ $n_{\rm{H_2}}$} & & \multicolumn{2}{c}{$\log\:$$N$(H$_2$)} & min. $\chi^2$ \\
 & $\chi^2$ $<$ 7.81 & min. $\chi^2$  && $\chi^2$ $<$ 7.81 & min. $\chi^2$ && $\chi^2$ $<$ 7.81 & min. $\chi^2$ &\\
 &[K]&[K]&&[cm$^{-3}$]&[cm$^{-3}$]&&[cm$^{-2}$]&[cm$^{-2}$]&  \\
 (1) &(2) &(3) &&(4) &(5) &&(6) &(7) &(8)
 }
\startdata
R5&$>$ 5 &90 &&$<$ 3.7 &2.4 &&20.4 - 21.2 &20.7 &0.0001508 \\
R7b&5 - 120 &25 &&$>$ 2.8 &4.1 &&$>$ 20.8 &21.4 &0.0008348 \\
R8b&5 - 30 &25 &&$<$ 2.4 &2.0 &&20.9 - 21.8 &21.3 &0.04646 \\
R9b&$>$ 5 &50 &&2.2 - 4.0 &3.0 &&20.1 - 20.7 &20.4 &0.004822 \\
R10a&5 - 120 &65 &&$<$ 3.3 &2.1 &&20.4 - 21.8 &20.9 &0.001529 \\
R10b&5 - 20 &10 &&3.6 - 4.6 &4.4 &&19.9 - 20.4 &20.1 &0.00948 \\
R11a&$>$ 20 &75 &&$<$ 3.2 &2.5 &&20.5 - 21.0 &20.8 &0.000315 \\
R11b&\nodata &15 &&2.6 - 4.6 &3.9 &&$<$ 20.4 &18.1 &0.001707 \\
R13a&5 - 200 &85 &&$<$ 3.5 &2.2 &&20.3 - 21.4 &20.7 &0.002488 \\
R14&5 - 200 &15 &&$<$ 3.4 &3.2 &&20.2 - 20.6 &20.4 &0.004464 \\
R15&5 - 140 &35 &&$<$ 3.4 &2.6 &&20.2 - 20.8 &20.5 &0.001343 \\
R16&5 - 30 &15 &&2.4 - 3.2 &2.8 &&20.6 - 21.0 &20.8 &0.03139 \\
R17&5 - 180 &135 &&$<$ 3.6 &2.0 &&19.6 - 21.2 &20.2 &0.003854 \\
R18$^{\dag}$&$>$ 35 &180 &&2.4 - 3.3 &2.7 &&20.6 - 21.0 &20.8 &0.000186 \\
R19a&$>$ 25 &85 &&$<$ 3.2 &2.5 &&20.9 - 21.5 &21.2 &0.0001408 \\
R19b&$>$ 20 &100 &&2.2 - 3.5 &2.8 &&20.5 - 21.0 &20.8 &0.002962 \\
R20a&$>$ 30 &100 &&2.7 - 3.6 &3.1 &&20.5 - 21.1 &20.8 &0.001267 \\
R21a$^{\dag\dag}$&25 - 90 &50 &&3.4 - 5.0 &3.7 &&20.8 - 21.5 &21.1 &0.01197 \\
R22a&$>$ 25 &70 &&2.6 - 3.6 &3.1 &&20.5 - 20.9 &20.7 &0.01047 \\
R22b&$>$ 90 &265 &&2.8 - 3.3 &3.0 &&21.0 - 21.6 &21.3 &0.00147 \\
R23a&$>$ 40 &140 &&2.7 - 3.5 &3.1 &&20.6 - 21.1 &20.9 &0.001661 \\
R23b&$>$ 35 &140 &&2.5 - 3.5 &2.9 &&20.5 - 21.0 &20.8 &0.002198 \\
R24a&$>$ 5 &160 &&$<$ 4.9 &2.4 &&20.2 - 21.3 &20.6 &0.000268 \\
R24b&$>$ 15 &275 &&2.5 - 3.6 &2.8 &&20.6 - 21.4 &21.0 &0.001229 \\
R24c&$>$ 5 &115 &&$<$ 3.9 &2.5 &&20.0 - 20.8 &20.4 &0.0001756 \\
R25a&5 - 40 &20 &&3.5 - 5.0 &3.9 &&20.5 - 20.8 &20.7 &0.002611 \\
R25b&$>$ 5 &285 &&$<$ 3.9 &2.2 &&20.2 - 21.0 &20.4 &0.0001641 \\
R26a&$>$ 40 &265 &&2.6 - 3.4 &2.8 &&20.5 - 21.1 &20.8 &0.0002885 \\
R26b&10 - 100 &50 &&$<$ 3.3 &2.2 &&20.6 - 21.4 &20.9 &0.00003488 \\
R27a&$>$ 5 &255 &&$>$ 2.9 &3.0 &&$<$ 21.6 &20.6 &0.0004384 \\
R27b&$>$ 50 &160 &&$<$ 2.6 &2.0 &&21.0 - 21.7 &21.3 &0.001546 \\
R28a&$>$ 5 &110 &&$<$ 3.9 &2.5 &&20.6 21.6 &21.0 &0.00007702 \\
R31a&$>$ 20 &115 &&$<$ 3.0 &2.0 &&20.8 - 22.0 &21.3 &0.001847 \\
R34&$>$ 5 &130 &&$<$ 3.4 &2.4 &&20.5 - 21.6 &20.9 &0.0006237 \\
R36a&10 - 40 &30 &&$<$ 2.5 &2.0 &&$>$ 21.1 &21.6 &0.00002201 \\
R37a&5 - 30 &25 &&$<$ 2.7 &2.0 &&20.7 - 21.8 &21.1 &0.03561 \\
R38&$>$ 5 &35 &&$<$ 3.7 &2.5 &&$>$ 20.8 &21.2 &0.0005419 \\
R39a$^{\dag\dag\dag}$&$>$ 50 &95 &&2.7 - 3.4 &3.1 &&20.9 - 21.3 &21.1 &0.006462 \\
R39b$^{\dag\dag\dag}$&$>$ 5 &175 &&2.3 - 4.1 &2.6 &&20.0 - 20.9 &20.5 &0.005283 \\
\enddata
\tablecomments{Column 2 - 7: RADEX parameters noted above. Column 8: Value of $\chi^2$ associated with the fit. Note that for each position, (3), (5), and (7) are the best-fit parameters and (2), (4), and (6) are estimated within a confidence of 95~\%.; We consider the statistical error and the systematic error in this table. The systematic error of absolute flux calibration is estimated to be $\sim$ 5\% in band 3 and $\sim$ 10\% in band 7.  \dag represents boxes contained the obscured AGN defined by \citetalias{ion13}.  \dag\dag represents boxes contained the nuclear starbursts defined by \citetalias{ion13}.  \dag\dag\dag represents boxes at the overlap region.}
\end{deluxetable*}

\begin{deluxetable*}{lrrrrrrrrrrrl}
\tabletypesize{\scriptsize}
\tablecaption{RADEX Results of Case 2 \label{table_RADEX_HCN}}
\tablewidth{0pt}
\tablehead{
ID & $\log\:$$N$(H$_2$) & \multicolumn{2}{c}{$T_{\rm{kin}}$} & & \multicolumn{2}{c}{$\log\:$ $n_{\rm{H_2}}$} & & \multicolumn{2}{c}{[HCN]/[HCO$^+$]} & min. $\chi^2$ \\
 && $\chi^2$ $<$ 7.81& min. $\chi^2$  && $\chi^2$ $<$ 7.81 & min. $\chi^2$ && $\chi^2$ $<$ 0.35 & min. $\chi^2$ &\\
 &[cm$^{-2}$]&[K]&[K]&&[cm$^{-3}$]&[cm$^{-3}$]&&[cm$^{-2}$]&[cm$^{-2}$]&  \\
 (1) &(2) &(3) &(4) &&(5) &(6) &&(7) &(8) &(9)\\
 }
\startdata
E0$^{\dag}$&21.2 &$>$ 100 &270 &&5.0 - 5.4 &5.3 &&$>$ 5 &8 & 0.00563\\
E1$^{\dag\dag}$&21.6 &40 - 100 &70 &&5.6 - 5.9 &5.8 &&$<$ 4 &2 & 0.01073\\
E2$^{\dag\dag\dag}$&21.5 &5 - 90 &40 &&5.0 - 5.6 &5.1 &&1 - 9 &2 &0.0003303 \\
\enddata
\tablecomments{Column 2: Adopted $N(\rm{H_2})$ which derived from CO~(1-0) data and $X{_{\rm{CO}}}$. Column 3 - 8: RADEX parameters noted above. Column 9: Value of $\chi^2$ associated with the fit. Note that for each position, (4), (6), and (8) are the best-fit parameters and (3), (5), and (7) are estimated within the confidence level of 95~\% for $T_{\rm{kin}}$ and $n_{\rm{H_2}}$, and 95~\% (2~$\sigma$) for [HCN]/[HCO$^+$].; We consider the statistical error and the systematic error in this table. The systematic error of absolute flux calibration is estimated to be $\sim$ 5\% in band 3 and $\sim$ 10\% in band 7.  \dag represents boxes contained the obscured AGN defined by \citetalias{ion13}.  \dag\dag represents boxes contained the nuclear starbursts defined by \citetalias{ion13}.  \dag\dag\dag represents boxes at the overlap region.}
\end{deluxetable*}

\begin{deluxetable*}{lrrrrrrrrrcrl}
\tabletypesize{\scriptsize}
\tablecaption{Gas Properties under the LTE Assumption \label{table_LTE}}
\tablewidth{0pt}
\tablehead{
ID&$\Phi_{\rm{A}}$ (= $\rm{T_{b, CO(1-0)}}/T_{kin}$)&Adopted $\Phi_{\rm{A}}$&$T_{\rm{ex}}$&$\tau_{\rm{CO(1-0)}}$&$\tau_{\rm{^{13}CO(1-0)}}$\\ 
 & & &[K] & & &\\ 
 (1) &(2) &(3) &(4) &(5) &(6)\\ 
}
\startdata
R5&$<$ 0.24 &0.10 &15.3 $\pm$ 0.9&3.59 &0.14 \\ 
R7b&$<$ 0.09 &0.09 &8.3 $\pm$ 0.8&3.97 &0.02 \\ 
R8b&0.03 - 0.19 &0.10 &13.0 $\pm$ 0.9&25.28 &1.57 \\ 
R9b&$<$ 0.52&0.10 &29.5 $\pm$ 1.5&1.27 &0.02 \\ 
R10a&0.01 - 0.16 &0.10 &11.4 $\pm$ 0.8&6.72 &0.35 \\ 
R10b&0.14 - 0.56 &0.14 &23.5 $\pm$ 1.1&1.24 &0.03 \\ 
R11a&$<$ 0.11 &0.10 &24.6 $\pm$ 1.3&3.88 &0.19 \\ 
R11b&\nodata &0.10 &36.9 $\pm$ 1.8&0.01 &0.01 \\ 
R13a&0.01 - 0.22 &0.10 &14.6 $\pm$ 0.9&6.87 &0.37 \\ 
R14&0.01 - 0.32 &0.10 &19.6 $\pm$ 1.1&5.89 &0.35 \\ 
R15&0.01 - 0.35 &0.10 &21.2 $\pm$ 1.1&6.44 &0.27 \\ 
R16&0.05 - 0.28 &0.10 &17.2 $\pm$ 1.0&9.00 &0.31 \\ 
R17&0.01 - 0.33 &0.10 &19.9 $\pm$ 1.1&2.59 &0.06 \\ 
R18$^{\dag}$&$<$ 0.10 &0.10 &38.7 $\pm$ 1.9&1.34 &0.03 \\ 
R19a&$<$ 0.04 &0.04 &28.7 $\pm$ 2.2&6.99 &0.76 \\ 
R19b&$<$ 0.10 &0.10 &23.4 $\pm$ 1.2&1.68 &0.05 \\ 
R20a&$<$ 0.11 &0.11 &36.8 $\pm$ 1.8&0.57 &0.01 \\ 
R21a$^{\dag\dag}$&0.06 - 0.22 &0.10 &58.8 $\pm$ 2.9&2.75 &0.03 \\ 
R22a&$<$ 0.14 &0.10 &39.5 $\pm$ 1.9&0.90 &0.01 \\ 
R22b&$<$ 0.04 &0.04 &99.3 $\pm$ 5.1&0.42 &0.01 \\ 
R23a&$<$ 0.10 &0.10 &45.2 $\pm$ 2.2&0.61 &0.01 \\ 
R23b&$<$ 0.10 &0.10 &37.4 $\pm$ 1.8&0.87 &0.01 \\ 
R24a&$<$ 0.25 &0.10 &15.7 $\pm$ 0.9&4.00 &0.21 \\ 
R24b&$<$ 0.11 &0.10 &19.5 $\pm$ 1.1&1.02 &0.01 \\ 
R24c&$<$ 0.36 &0.10 &21.3 $\pm$ 1.1&1.83 &0.05 \\ 
R25a&0.12 - 0.98 &0.12 &44.3 $\pm$ 2.1&2.85 &0.03 \\ 
R25b&$<$ 0.35 &0.10 &21.2 $\pm$ 1.1&2.16 &0.07 \\ 
R26a&$<$ 0.11 &0.10 &48.4 $\pm$ 2.4&1.56 &0.02 \\ 
R26b&0.01 - 0.11 &0.10 &14.1 $\pm$ 0.9&6.33 &0.30 \\ 
R27a&$<$ 0.39 &0.10 &23.1 $\pm$ 1.2&0.09 &0.01 \\ 
R27b&$<$ 0.03 &0.03 &54.5 $\pm$ 3.5&9.17 &0.92 \\ 
R28a&$<$ 0.15 &0.10 &11.0 $\pm$ 0.8&4.21 &0.31 \\ 
R31a&$<$ 0.04 &0.04 &21.5 $\pm$ 2.0&4.61 &0.09 \\ 
R34&$<$ 0.17 &0.10 &11.7 $\pm$ 0.8&4.23 &0.29 \\ 
R36a&0.02 - 0.07 &0.07 &12.7 $\pm$ 1.1&20.17 &1.53 \\ 
R37a&0.03 - 0.16 &0.10 &11.3 $\pm$ 0.8&19.92 &1.04 \\ 
R38&$<$ 0.08 &0.08 &8.0 $\pm$ 0.9&10.07 &0.67 \\ 
R39a$^{\dag\dag\dag}$&$<$ 0.10 &0.10 &52.6 $\pm$ 2.6&1.42 &0.01 \\ 
R39b$^{\dag\dag\dag}$&$<$ 0.39 &0.10 &23.1 $\pm$ 1.2&1.29 &0.02 \\ 
\enddata
\tablecomments{Column 2: The beam filling factor estimated from RADEX modeling and the brightness temperature. Column 3: The adopted beam filling factor. Column 4: The excitation temperature of a spectral line calculated from the equation (3). Column 5 and 6: the optical depth of $^{12}$CO~(1--0) and $^{13}$CO~(1--0) emission from RADEX modeling, respectively. Column 7: The total column density of $^{13}$CO.; We consider the statistical error and the systematic error in this table. The systematic error of absolute flux calibration is estimated to be $\sim$ 5\% in band 3 and $\sim$ 10\% in band 7.  \dag represents boxes contained the obscured AGN defined by \citetalias{ion13}.  \dag\dag represents boxes contained the nuclear starbursts defined by \citetalias{ion13}.  \dag\dag\dag represents boxes at the overlap region.}
\end{deluxetable*}

\begin{deluxetable*}{llrrrrrrrrcrl}
\tabletypesize{\scriptsize}
\tablecaption{Properties of CS, CH$_3$OH, and CN \label{table_chem}}
\tablewidth{0pt}
\tablehead{
Molecule &ID &$T_{\rm{b}}$ &$\Delta v$ &$N_{\rm{X}}$ &[$X$]/[H$_2$]\\
 & &[K] &[km s$^{-1}$] &[cm$^{-2}$] &\\
 (1) &(2) &(3) &(4) &(5) &(6)
 }
\startdata
CS &R18$^{\dag}$ &$<$ 0.04 &\nodata &$<$ 4.0 $\times$ 10$^{11}$ &$<$ 6.3 $\times$ 10$^{-10}$\\
 &R21a$^{\dag\dag}$ &$<$ 0.04 &\nodata &$<$ 5.6 $\times$ 10$^{11}$ &$<$ 4.5 $\times$ 10$^{-10}$\\
 &R39a$^{\dag\dag\dag}$ &0.13 $\pm$ 0.04 &44 $\pm$ 11 &1.6 $\times$ 10$^{12}$ &1.3 $\times$ 10$^{-9}$\\
CH$_3$OH &R18$^{\dag}$ &$<$ 0.04 &\nodata &$<$ 1.9 $\times$ 10$^{12}$ &$<$ 2.9 $\times$ 10$^{-9}$\\
 &R21a$^{\dag\dag}$ &$<$ 0.04 &\nodata &$<$ 2.6 $\times$ 10$^{12}$ &$<$ 2.1 $\times$ 10$^{-9}$\\
 &R39a$^{\dag\dag\dag}$ &0.17 $\pm$ 0.04 &43 $\pm$ 5 &1.0 $\times$ 10$^{13}$ &8.1 $\times$ 10$^{-9}$\\
CN &R18$^{\dag}$ &0.23 $\pm$ 0.04 &63 $\pm$ 5 &2.4 $\times$ 10$^{12}$ &3.8 $\times$ 10$^{-9}$\\
 &R21a$^{\dag\dag}$ & 0.15 $\pm$ 0.04 &66 $\pm$ 3 &2.2 $\times$ 10$^{12}$ &1.8 $\times$ 10$^{-9}$\\
 &R39a$^{\dag\dag\dag}$ & 0.14 $\pm$ 0.04 &49 $\pm$ 9 &1.4 $\times$ 10$^{12}$ &1.1 $\times$ 10$^{-9}$
\enddata
\tablecomments{Column 1: The molecular line considered.  Column 3: Peak brightness temperature of considered line. Column 4: Line width of considered line. Column 5: the derived total box-averaged column density of this species. Column 6: Fractional abundance relative to H$_2$ of this species.; We consider the statistical error and the systematic error in this table. The systematic error of absolute flux calibration is estimated to be $\sim$ 5\% in band 3 and $\sim$ 10\% in band 7.  \dag represents boxes contained the obscured AGN defined by \citetalias{ion13}.  \dag\dag represents boxes contained the nuclear starbursts defined by \citetalias{ion13}.  \dag\dag\dag represents boxes at the overlap region.}
\end{deluxetable*}

\clearpage

\appendix

\section{Images and spectra of VV114}

\subsection{The Pa$\alpha$ and $K_{\rm{s}}$ band image with miniTAO/ANIR observation} \label{A1}

\begin{figure*}[bth]
\begin{center}
\includegraphics[scale=.36]{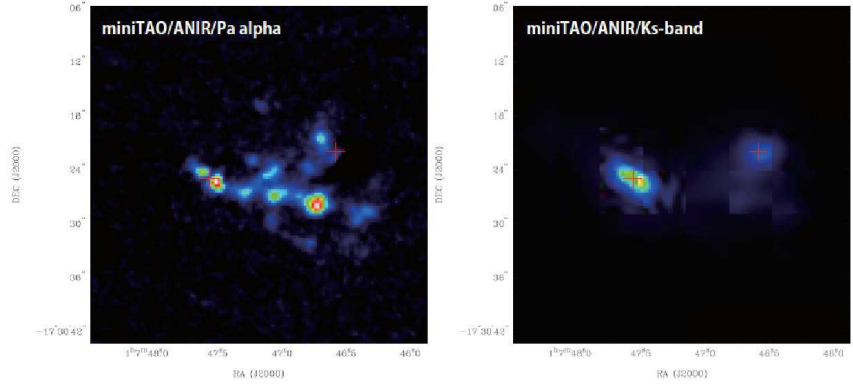}
\caption{The Pa$\alpha$ and $K_{\rm{s}}$ band images of VV~114 with miniTAO/ANIR observation \citep{tat12}.}
\label{fig_previous}
\end{center}
\end{figure*}

\clearpage

\subsection{Channel maps of each line emission} \label{A2}

\begin{figure*}[bth]
\begin{center}
\includegraphics[scale=0.6]{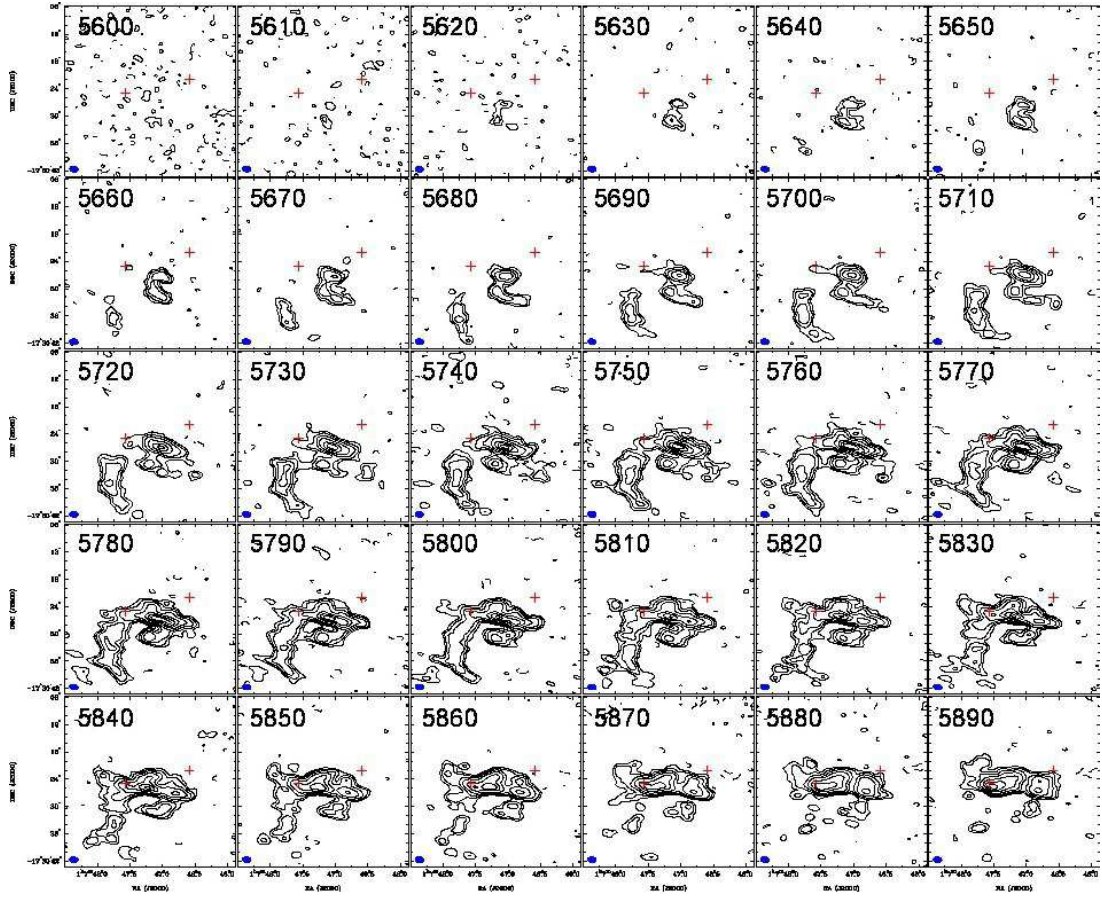}
\caption{The velocity channel maps of the $^{12}$CO~(1--0) line emission of VV~114. Two crosses in each channel show the positions of the nuclei defined by the peak positions of the Ks-band observation \citep{tat12}. The velocity width of each channel is 10 km s$^{-1}$. The beam size is plotted in the bottom-left corner of each channel. The contours represent flux intensity levels: -4.6, 4.6, 9.2, 18.4, 36.8, 73.6, 110.4 and 147.2 mJy beam$^{-1}$.}
\end{center}
\end{figure*}

\addtocounter{figure}{-1}
\begin{figure*}
\begin{center}
\includegraphics[scale=0.6]{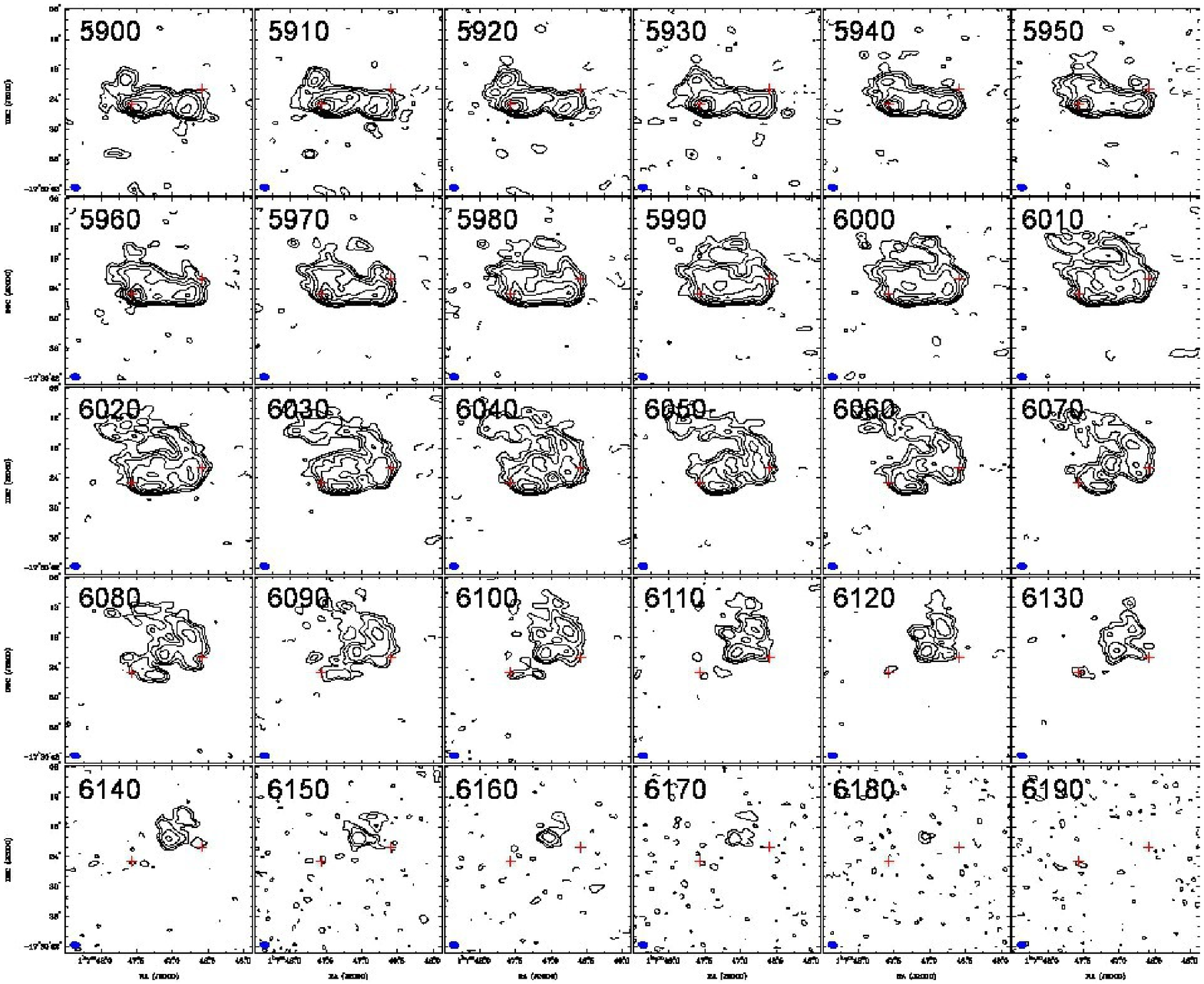}
\caption{Continuied.}
\end{center}
\end{figure*}

\begin{figure*}
\begin{center}
\includegraphics[scale=0.6]{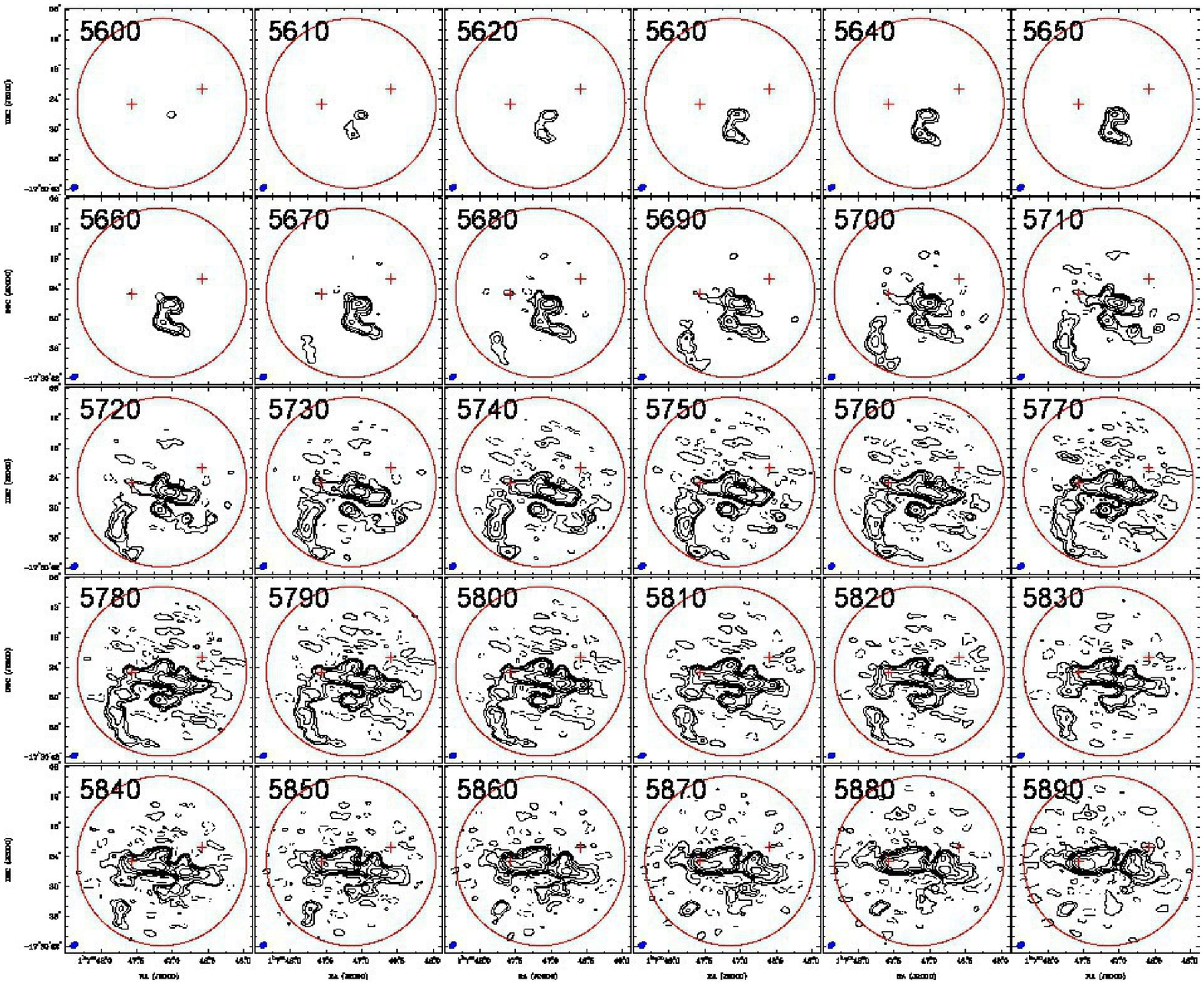}
\caption{The same as Figure~14 but for the CO~(3--2) line emission of VV~114. The approximate field of view of ALMA 7-point mosaic at this frequency is indicated by the large red circle. The contours represent flux intensity levels: -12.6, 12.6, 25.2, 50.4, 100.8, 201.6, and 403.2 mJy beam$^{-1}$.}
\end{center}
\end{figure*}

\addtocounter{figure}{-1}
\begin{figure*}
\begin{center}
\includegraphics[scale=0.6]{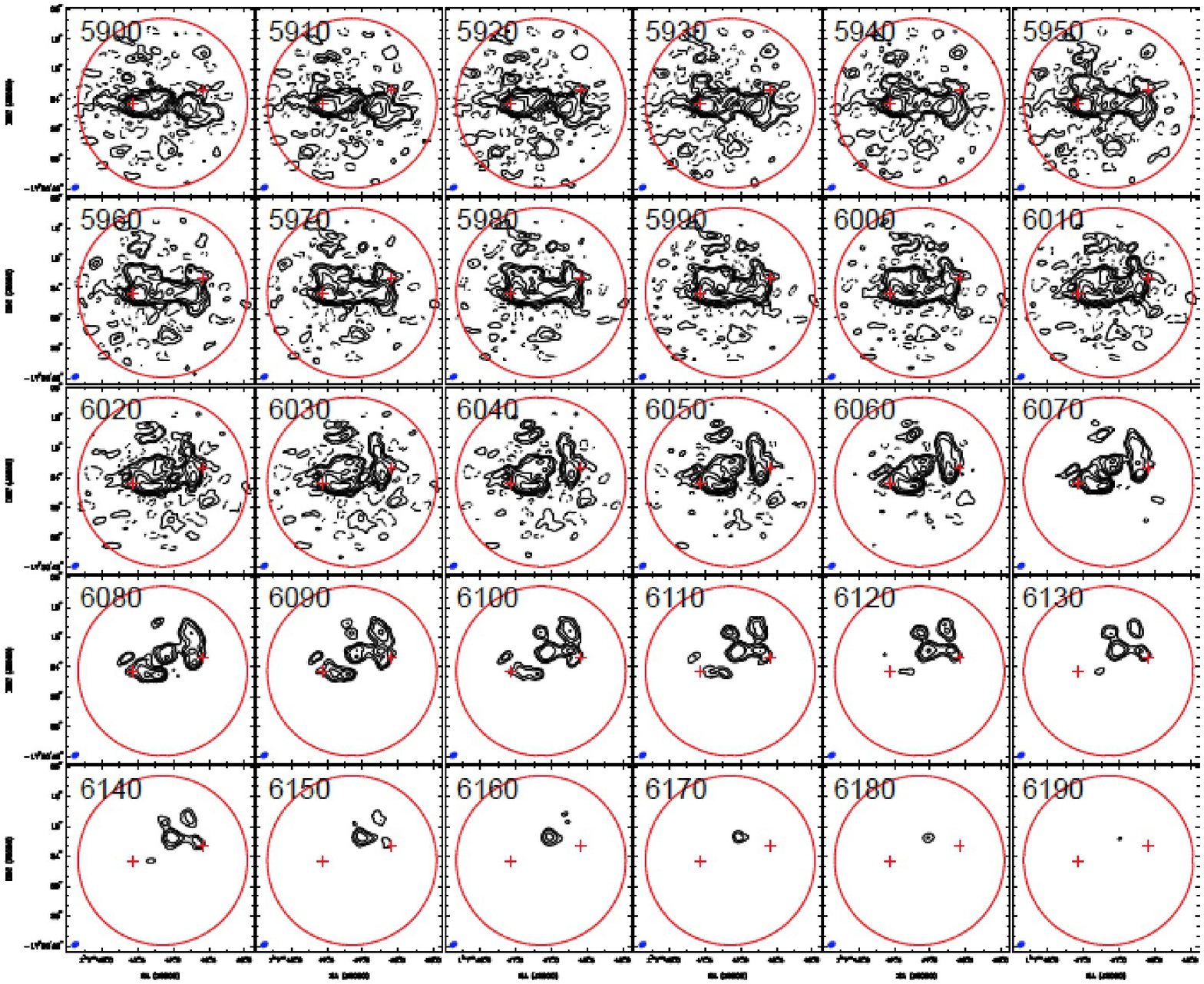}
\caption{Continuied.}
\end{center}
\end{figure*}

\begin{figure*}
\begin{center}
\includegraphics[scale=.9]{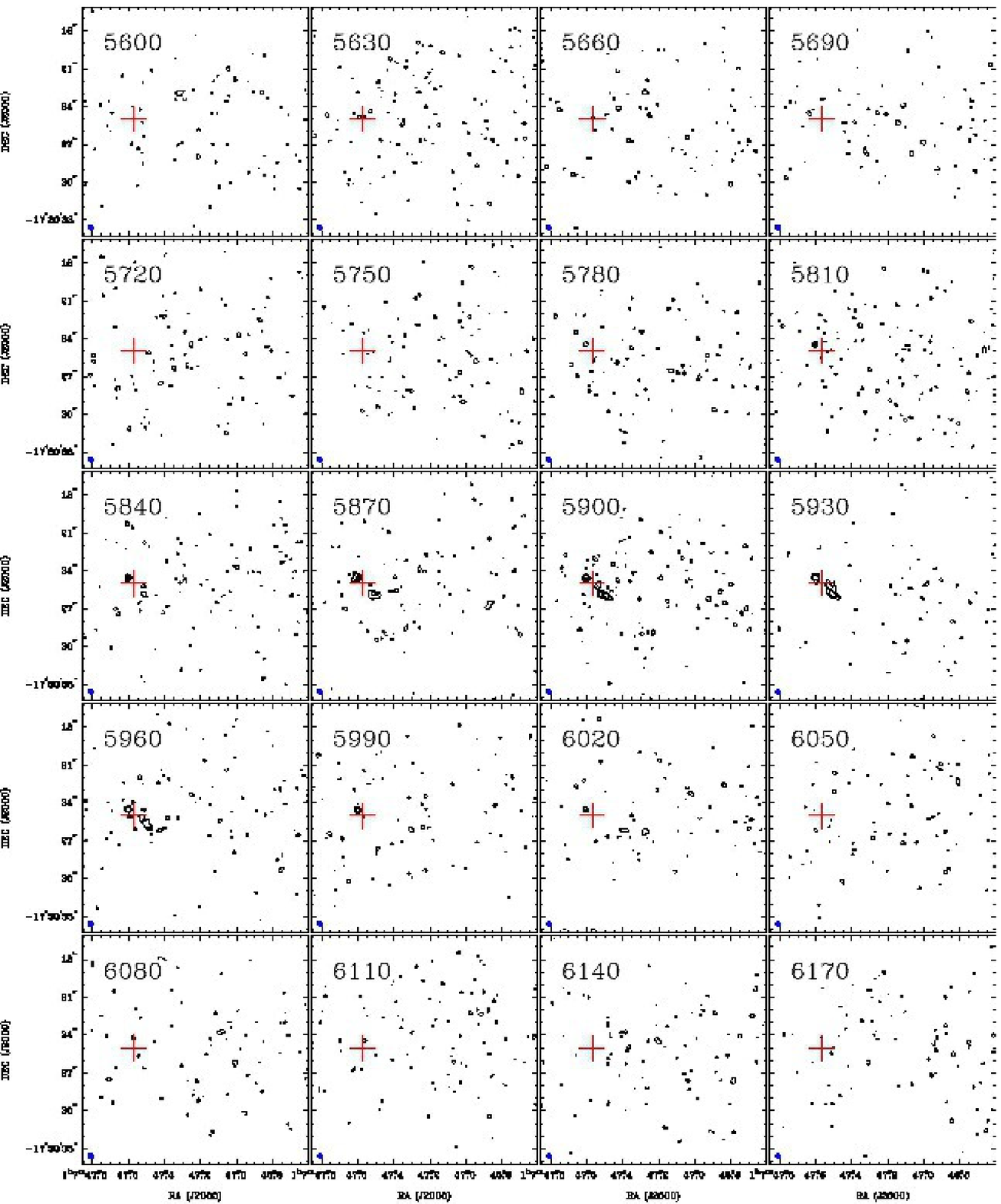}
\caption{The same as Figure~14 but for the HCN~(4--3) line emission of VV~114. The velocity width of each channel is 30 km s$^{-1}$. The cross in each channel shows the position of the eastern nucleus defined by the peak position of the Ks-band observation \citep{tat12}. The contours represent flux intensity levels: -2.4, 2.4, 4.8, 9.6, and 19.2 mJy beam$^{-1}$.}
\end{center}
\end{figure*}

\begin{figure*}
\begin{center}
\includegraphics[scale=.9]{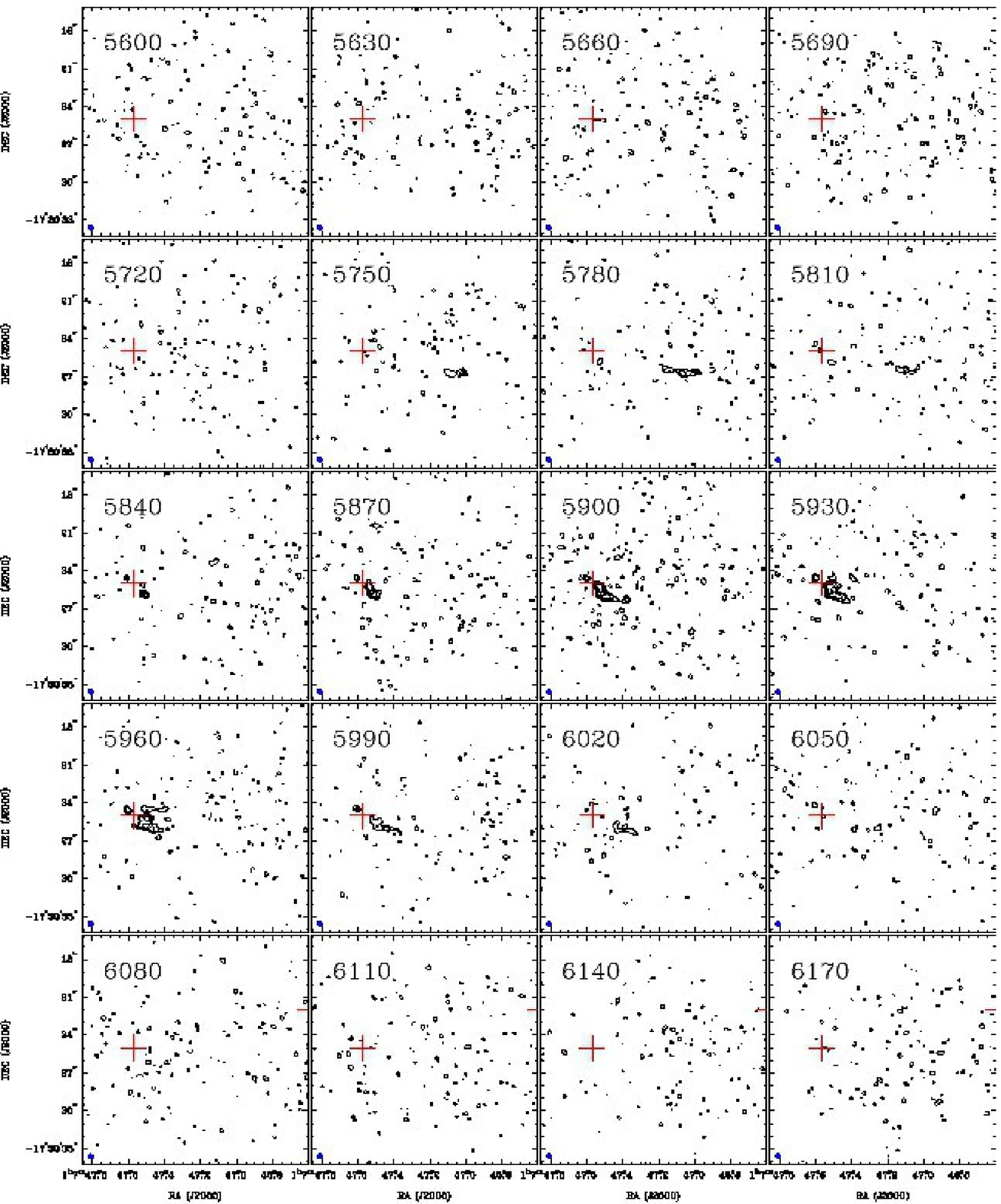}
\caption{The same as Figure~14 but for the HCO$^+$~(4--3) line emission of VV~114. The velocity width of each channel is 30 km s$^{-1}$. The cross in each channel shows the position of the eastern nucleus defined by the peak position of the Ks-band observation \citep{tat12}. The contours represent flux intensity levels: -2.4, 2.4, 4.8, 9.6, and 19.2 mJy beam$^{-1}$.}
\end{center}
\end{figure*}

\clearpage

\subsection{Box-summed spectra of each line emission} \label{A3}

\begin{figure*}[bth]
\begin{center}
\includegraphics[scale=.5]{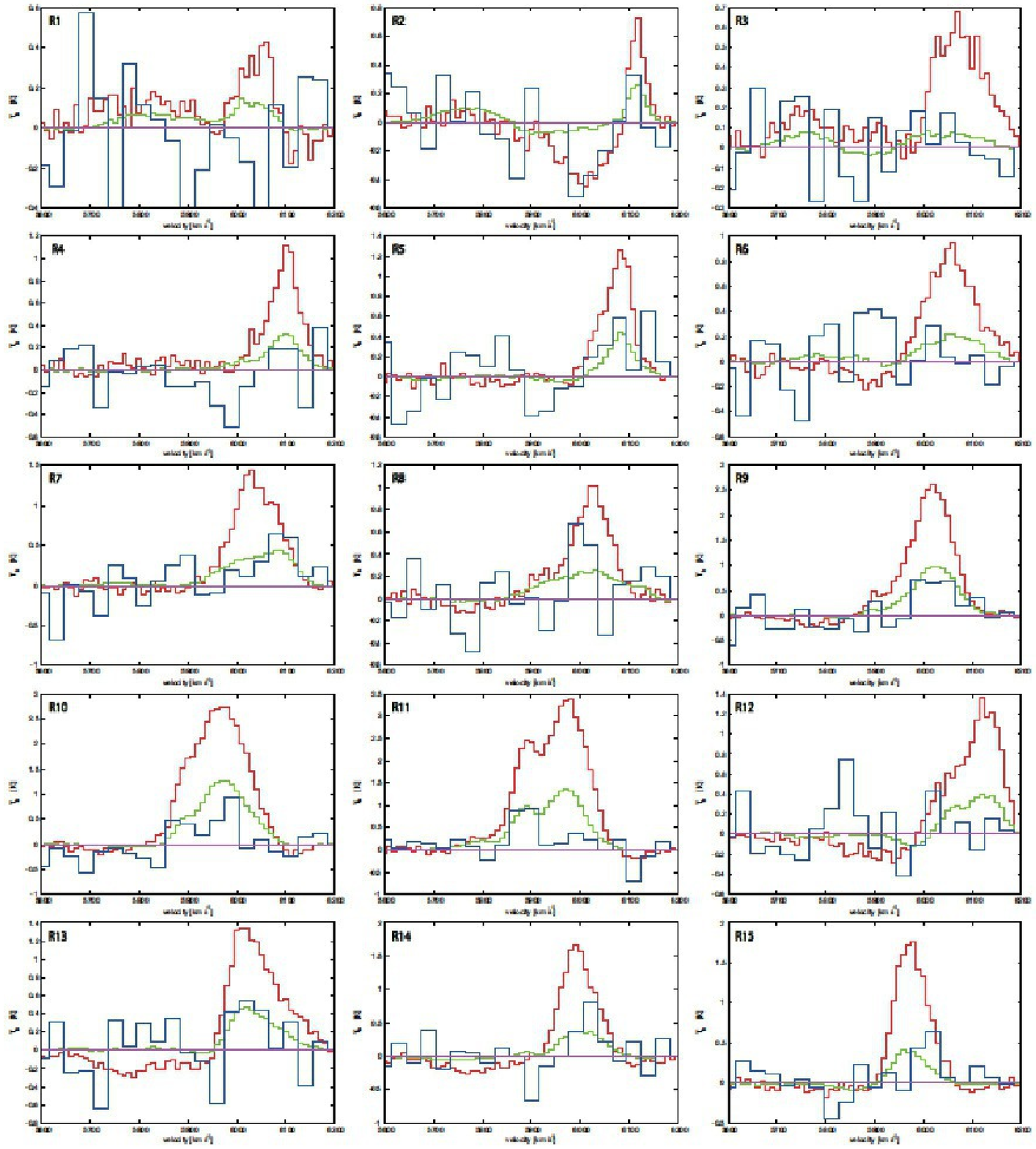}
\caption{2\farcs0 box-summed spectra of $^{12}$CO~(1--0) (red line), $^{13}$CO~(1--0) $\times$ 10 (green broken line), and $^{12}$CO~(3--2) (blue broken line) at the each box, labeled R1 - R39 of Fig \ref{fig_ratio}. The spectra are taken from the ALMA data cubes after correcting the cubes for the primary beam attenuation and convolving them to 2\farcs0 $\times$ 1\farcs5 resolution (P.A. = 83 deg).}
\label{fig_spec_CO}
\end{center}
\end{figure*}

\addtocounter{figure}{-1}
\begin{figure*}
\begin{center}
\includegraphics[scale=.5]{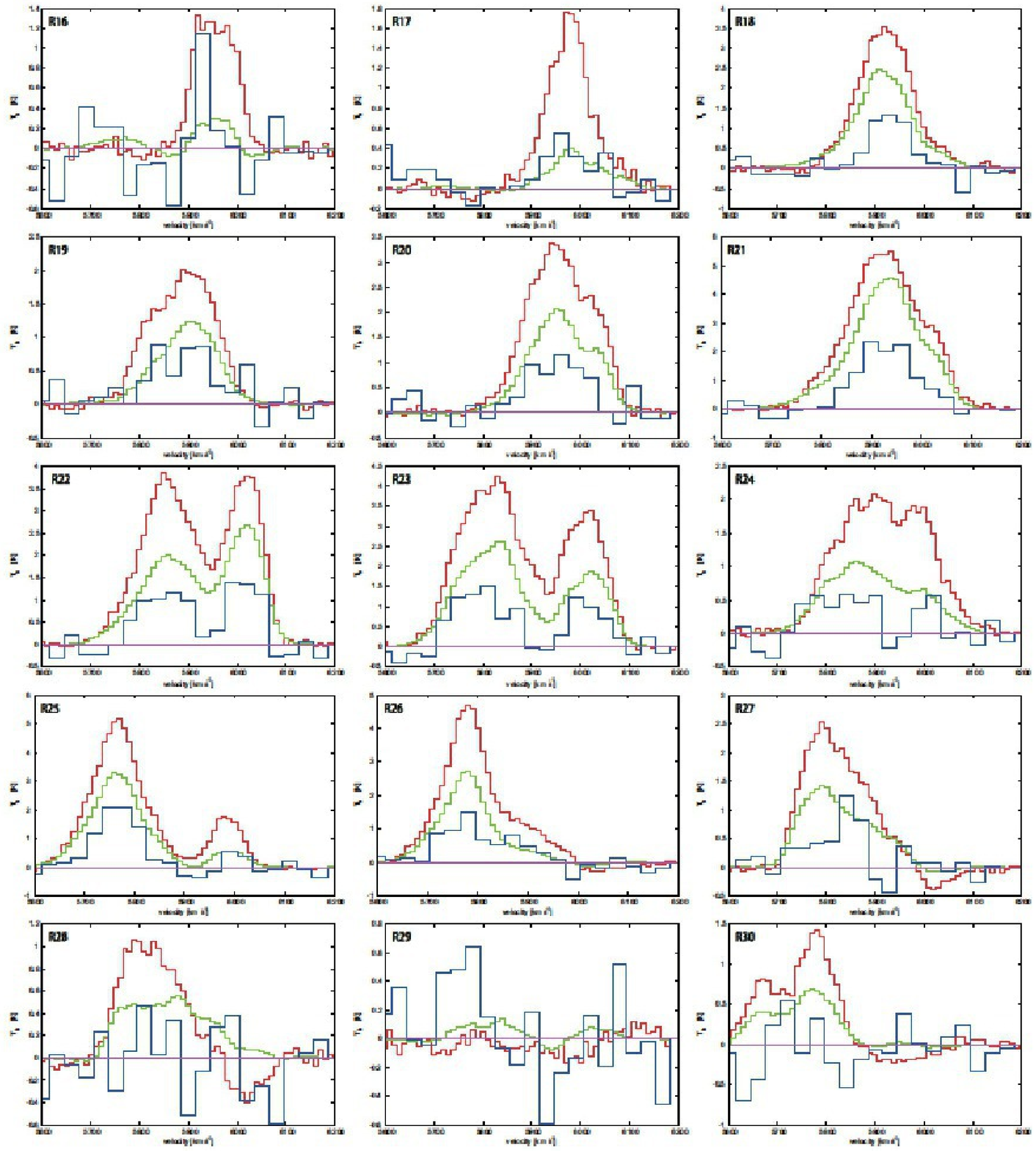}
\caption{continued.}
\end{center}
\end{figure*}

\addtocounter{figure}{-1}
\begin{figure*}
\begin{center}
\includegraphics[scale=.5]{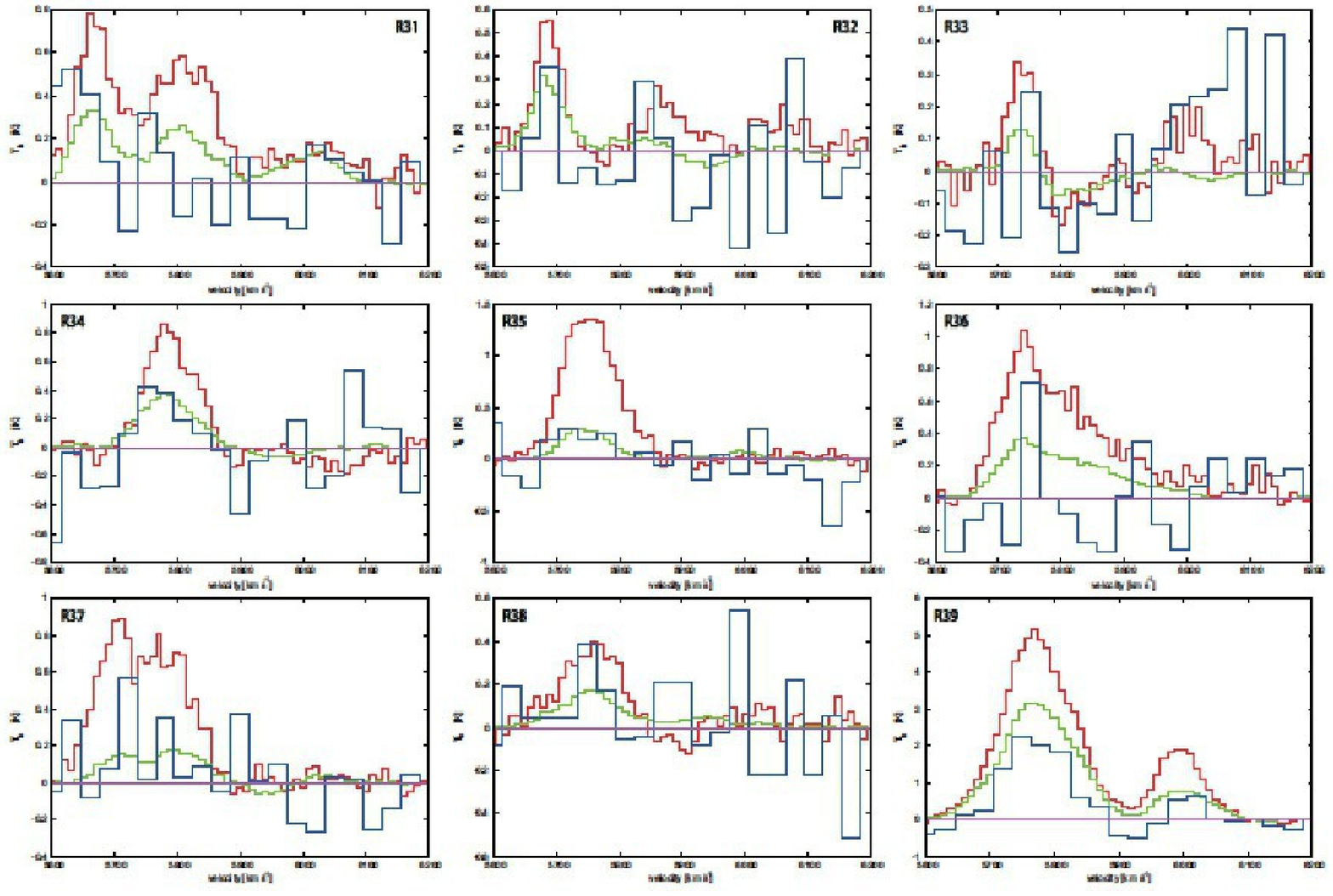}
\caption{continued.}
\end{center}
\end{figure*}

\begin{figure*}
\begin{center}
\includegraphics[scale=.5]{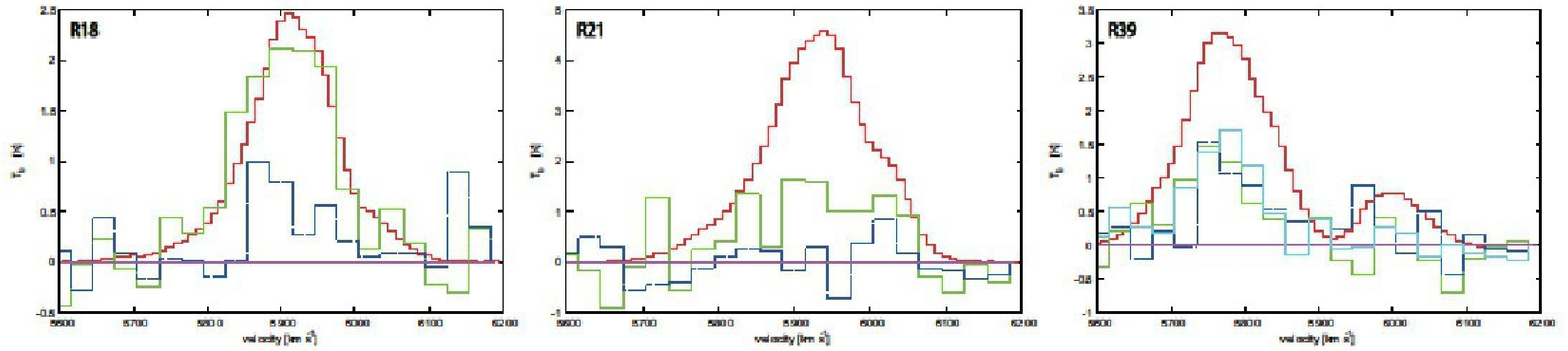}
\caption{2\farcs0 box-summed spectra of $^{12}$CO~(3--2) (red line), CN~(1$_{3/2}$--0$_{1/2}$) $\times$ 10 (green broken line), CS~(2--1) $\times$ 10 (blue broken line), and CH$_3$OH~(2$_k$--1$_k$) $\times$ 10 (light blue broken line) at R18, R21, and R39, of Fig \ref{fig_ratio}. The spectra are taken from the ALMA data cubes after correcting the cubes for the primary beam attenuation and convolving them to 2\farcs0 $\times$ 1\farcs5 resolution (P.A. = 83 deg).}
\label{fig_spec_CN}
\end{center}
\end{figure*}

\begin{figure*}
\begin{center}
\includegraphics[scale=.5]{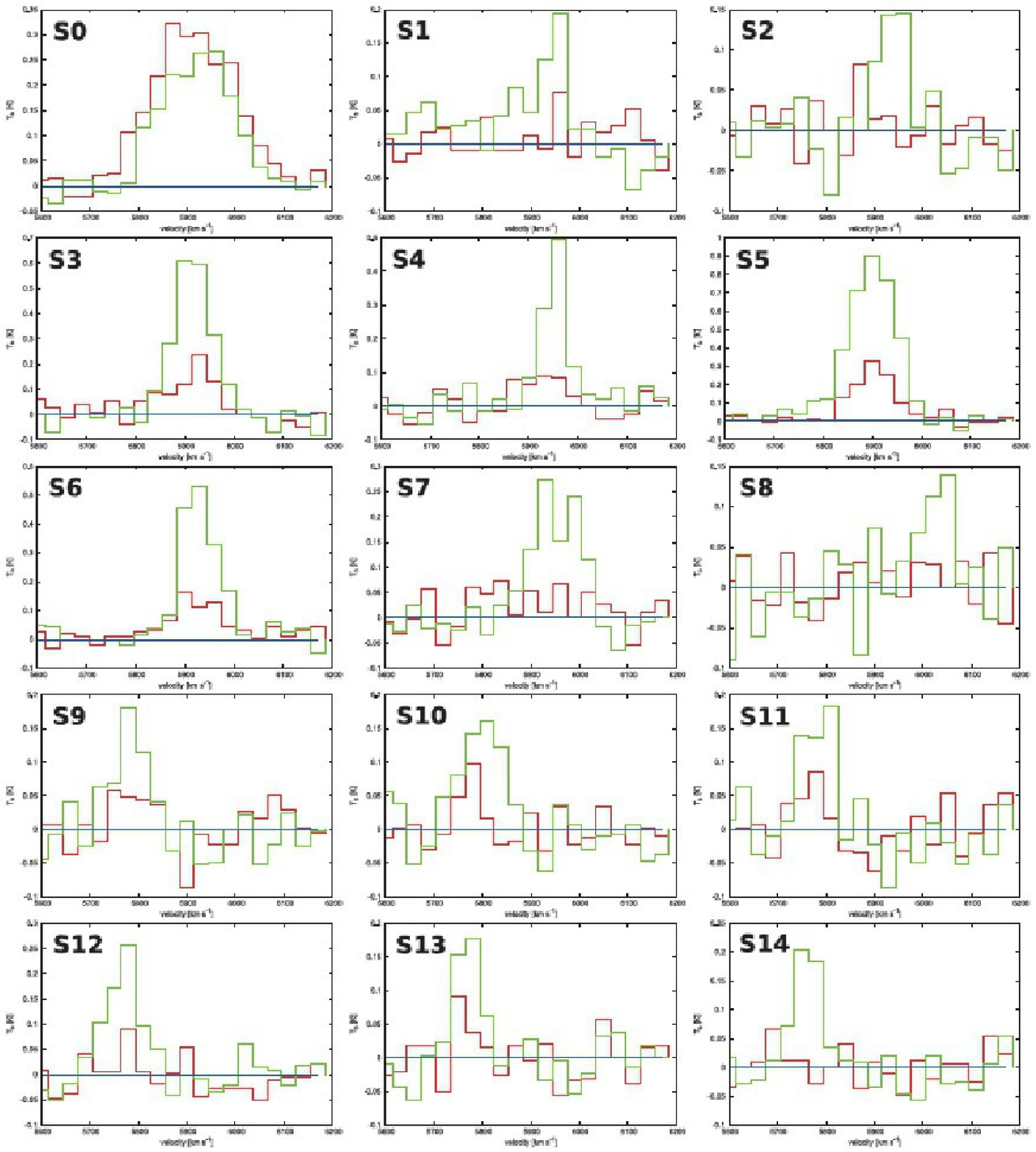}
\caption{0\farcs5 box-summed spectra of HCN~(4--3) (red line) and HCO$^+$~(4--3) (green broken line) at the each box, labeled S0 - S14, of Fig \ref{fig_ratio}.}
\label{fig_spec_HCN}
\end{center}
\end{figure*}

\end{document}